\documentclass[11pt]{article}
\pdfoutput=1
\usepackage{jcapmod}
\usepackage{shorthand}

\usepackage{mathtools}
\usepackage{booktabs}
\usepackage[english]{babel}
\usepackage{amsmath,amssymb,amsbsy,amstext, amsthm, simplewick, amsfonts}
\usepackage{hyperref}
\usepackage{graphicx}
\usepackage[small]{caption}
\usepackage{slashed}
\usepackage{siunitx}
\usepackage{upgreek}
\usepackage{framed}
\usepackage{wrapfig}
\usepackage{multirow}
\usepackage{bbm}
\usepackage[svgnames,dvipsnames,x11names]{xcolor}
\usepackage[utf8x]{inputenc}
\usepackage{selinput}
\usepackage{bm}
\usepackage{float}
\usepackage{geometry}
\usepackage{yfonts}
\usepackage{caption}
\usepackage{subcaption}
\usepackage{sidecap}
\usepackage{longtable}
\usepackage{anyfontsize}
\usepackage{dsfont}
\usepackage{tikz}
\usepackage{relsize}
\usepackage{tcolorbox}

\usepackage[framemethod=default]{mdframed}
\newmdenv[skipabove=7pt,
skipbelow=7pt,
rightline=false,
leftline=false,
topline=false,
bottomline=false,
backgroundcolor=gray!10,
linecolor=gray,
innerleftmargin=5pt,
innerrightmargin=5pt,
innertopmargin=5pt,
innerbottommargin=5pt,
leftmargin=0cm,
rightmargin=0cm,
linewidth=4pt]{eBox}
\newmdenv[skipabove=7pt,
skipbelow=7pt,
rightline=false,
leftline=false,
topline=false,
bottomline=false,
backgroundcolor=gray!10,
linecolor=gray,
innerleftmargin=5pt,
innerrightmargin=5pt,
innertopmargin=-5pt,
innerbottommargin=5pt,
leftmargin=0cm,
rightmargin=0cm,
linewidth=4pt]{eBox2}
\usepackage[framemethod=default]{mdframed}
\newmdenv[skipabove=7pt,
skipbelow=7pt,
rightline=true,
leftline=true,
topline=true,
bottomline=true,
backgroundcolor=gray!15,
linecolor=gray,
innerleftmargin=5pt,
innerrightmargin=5pt,
innertopmargin=5pt,
innerbottommargin=5pt,
leftmargin=0cm,
rightmargin=0cm,
linewidth=0.75pt]{eBox3}

\definecolor{blue3}{RGB}{31, 119, 180}
\definecolor{red3}{RGB}{	214, 39, 40}
\definecolor{orange3}{RGB}{255, 127, 14}
\definecolor{green3}{RGB}{44, 160, 44}
\definecolor{repBlue}{RGB}{31, 119, 180}
\definecolor{repRed}{RGB}{	214, 39, 40}
\definecolor{repGreen}{RGB}{44, 160, 44}
\definecolor{violet2}{RGB}{102,0,204}
\definecolor{Blue}{RGB}{214, 39, 40}
\definecolor{Red}{RGB} {31, 119, 180}

\usepackage{colortbl}
\definecolor{lightgreen}{cmyk}{0.2, 0, 0.2, 0.2}
\definecolor{lightgray2}{cmyk}{0.1,0.1,0,0.1}
\definecolor{Red2}{RGB}{214, 39, 40}
\definecolor{Blue2}{RGB} {31, 119, 180}
\definecolor{Orange2}{RGB}{255, 127, 14}
\definecolor{Green2}{RGB}{44, 160, 44}
\definecolor{greyish2}{rgb}{.96,.96,.96}

\makeatletter
\newlength{\apb@width}
\newcommand{\autoparbox}[2][c]{\settowidth{\apb@width}{#2}\parbox[#1]{\apb@width}{#2}}

\makeatother


\def\hs{\hskip 1pt}

\def\beq{\begin{equation}}
\def\eeq{\end{equation}}
\def\LA{\langle}
\def\RA{\rangle}

\def\k{\vec k}

\newcommand{\rd}{{\rm d}}
\newcommand{\vp}{\varphi}

\newcommand{\kxi}[2]{ {\vec k}_{#1} \cdot {\vec \xi}_{#2}}

\newcommand{\xixi}[2]{ {\vec \xi}_{#1} \cdot {\vec \xi}_{#2}}

\renewcommand{\AA}[2]{\langle #1 #2 \rangle}
\newcommand{\AB}[2]{\langle #1 \bar{#2} \rangle}
\newcommand{\BA}[2]{\langle \bar{#1} #2 \rangle}
\usetikzlibrary{arrows,calc,decorations.pathmorphing,decorations.markings,shapes.misc}
\pgfkeys{/tikz/.cd,
  circle color/.initial=black,
  circle color/.get=\circlecolor,
  circle color/.store in=\circlecolor,
}
\tikzset{/pgf/decoration/.cd,
    number of sines/.initial=10,
    angle step/.initial=20,
}
\newdimen\tmpdimen\pgfdeclaredecoration{complete sines}{initial}
{
    \state{initial}[
        width=+0pt,
        next state=move,
        persistent precomputation={
            \pgfmathparse{\pgfkeysvalueof{/pgf/decoration/angle step}}%
            \let\anglestep=\pgfmathresult%
            \let\currentangle=\pgfmathresult%
            \pgfmathsetlengthmacro{\pointsperanglestep}%
                {(\pgfdecoratedremainingdistance/\pgfkeysvalueof{/pgf/decoration/number of sines})/360*(\anglestep)}%
        }] {}
    \state{move}[width=+\pointsperanglestep, next state=draw]{
        \pgfpathmoveto{\pgfpointorigin}
    }
    \state{draw}[width=+\pointsperanglestep, switch if less than=1.25*\pointsperanglestep to final, % <- bit of a hack
        persistent postcomputation={
        \pgfmathparse{mod(\currentangle+\anglestep, 360)}%
        \let\currentangle=\pgfmathresult%
    }]{%
        \pgfmathsin{+\currentangle}%
        \tmpdimen=\pgfdecorationsegmentamplitude%
        \tmpdimen=\pgfmathresult\tmpdimen%
        \divide\tmpdimen by2\relax%
        \pgfpathlineto{\pgfqpoint{0pt}{\tmpdimen}}%
    }
    \state{final}{
        \ifdim\pgfdecoratedremainingdistance>0pt\relax
            \pgfpathlineto{\pgfpointdecoratedpathlast}
        \fi
   }
}
\allowdisplaybreaks[1]
\begin{document}

%TITLE PAGE=============================

\newgeometry{top=2cm, bottom=2cm, left=2cm, right=2cm}

\begin{titlepage}
\setcounter{page}{1} \baselineskip=15.5pt 
\thispagestyle{empty}

\begin{center}
{\fontsize{18}{18} \bf Linking the Singularities of Cosmological Correlators}\\[14pt]
\end{center}

\vskip 20pt
\begin{center}
\noindent
{\fontsize{12}{18}\selectfont 
Daniel Baumann,$^{1,2}$  Wei-Ming Chen,$^{1,2}$ Carlos Duaso Pueyo,$^{1}$\\  [4pt]
Austin Joyce,$^{1}$ Hayden Lee,$^{3}$ and Guilherme L.~Pimentel\hskip 1pt$^{4,1}$}
\end{center}

\begin{center}
  \vskip8pt
\textit{$^1$ Institute of Physics, University of Amsterdam, Amsterdam, 1098 XH, The Netherlands}

  \vskip8pt
\textit{$^2$  Center for Theoretical Physics, National Taiwan University, Taipei 10617, Taiwan}

\vskip 8pt
\textit{$^3$  Department of Physics, Harvard University, Cambridge, MA 02138, USA}

\vskip 8pt
\textit{$^4$ Lorentz Institute for Theoretical Physics, Leiden University, Leiden, 2333 CA, The Netherlands}
\end{center}

%=========================================
\vspace{0.4cm}
\begin{center}{\bf Abstract}
\end{center}
\noindent
Much of the structure of cosmological correlators is controlled by their singularities, which in turn 
are fixed in terms of flat-space scattering amplitudes. An important challenge is to interpolate between the singular limits to determine the full correlators at arbitrary kinematics. This is particularly relevant because the singularities of correlators are not directly observable, but can only be accessed by analytic continuation. In this paper, we study rational correlators---including those of gauge fields, gravitons, and the inflaton---whose only singularities at tree level are poles and whose behavior away from these poles is strongly constrained by unitarity and locality.
We describe how  unitarity translates into a set of cutting rules that consistent correlators must satisfy, and explain how this can be used to bootstrap correlators given information about their singularities. We also derive recursion relations that allow the iterative construction of more complicated correlators from simpler building blocks. In flat space, all energy singularities are simple poles, so that the combination of unitarity constraints and recursion relations provides an efficient way to bootstrap the full correlators. In many cases, these flat-space correlators can then be transformed into their more complex de Sitter counterparts.  As an example of this procedure, we derive the correlator associated to graviton Compton scattering in de Sitter space, though the methods are much more widely applicable.

\end{titlepage}
\restoregeometry

\newpage
\setcounter{tocdepth}{3}
\setcounter{page}{2}

\linespread{1.6}
\tableofcontents
\linespread{1.1}

\newpage
%=======================================
\section{Introduction}
\label{sec:intro}

Singularities aren't real, but they can be useful. Despite the fact that truly singular configurations cannot be measured, singularities play an important role in our mathematical descriptions of nature. Consequently, we can often infer a great deal about the behavior of a physical system by examining the singularities of our model.

\vskip4pt
An elementary example is provided by the $1/r$ potential of a point charge, where the singularity at $r=0$ is an artifact of the point particle approximation.
Nevertheless, given the locations of a set of charges, we can infer the electric field everywhere in space. More abstractly, we can trace the surprising utility of singularities back to their role in mathematics, where they are central in a variety of settings. For example, differential equations and their solutions can be classified by the nature of their singularities. Similarly, the properties of complex functions are largely controlled by their singularities.   
Given that our physical models often involve these mathematical concepts, it is not surprising that these features have counterparts in our description of nature.

\vskip4pt
In quantum field theory, the singularity structure of observables reflects fundamental physical properties like causality and unitarity.
One of the key insights of the bootstrap approach to scattering amplitudes is that singularities serve as a useful input, instead of simply being the output of an explicit computation~\cite{Cheung:2017pzi}. From this perspective, much of the structure of scattering amplitudes is actually controlled by their singularities. For example, at tree level, amplitudes have poles when intermediate particles go on-shell, and on these poles they must factorize into a product of lower-point amplitudes with positive coefficients. Physically, these singularities reflect the fact that an intermediate particle becomes long-lived and propagates a long distance in spacetime. Factorization into independent subprocesses is then a consequence of locality and unitarity. Amplitudes can have multiple poles and an important challenge is connecting these poles to obtain the function for general kinematics.  For theories involving massless particles, consistently extending the scattering amplitudes away from their singularities is extremely restrictive, in many cases uniquely fixing the S-matrix, and in other cases ruling out possible interactions~\cite{Benincasa:2007xk,Schuster:2008nh,McGady:2013sga}.

\vskip4pt 
In this paper, we apply a similar philosophy to cosmology, where the fundamental observables are spatial correlation functions.  
These correlations in the distribution of structure in the universe can be traced back in time to a spatial surface at the beginning of the hot Big Bang. However,  to explain the observed correlations, we require that this moment was not the beginning of time, but rather the end of an earlier high-energy epoch, typically imagined to be a period of inflationary expansion~\cite{Guth:1980zm, Linde:1981mu, Albrecht:1982wi, Baumann:2009ds}.  It is then natural to ask how the time evolution of this pre-Big Bang phase manifests in the static correlations on the reheating surface.  The usual approach to this question is to explicitly follow the time evolution of fields propagating and interacting in the inflationary spacetime, where the standard rules of quantum field theory ensure that the final observables are consistent with cherished principles like locality, causality, and unitarity. However, it is important to stress that we don't actually have direct observational access to the dynamics during the inflationary epoch, and so we are inspired to ask how this time evolution is encoded directly at the level of the observable quantities. This motivates an approach to cosmological correlators that makes no explicit reference to time evolution, taking principles like locality and unitarity as fundamental and deriving the form of the correlators from consistency requirements alone. 
This bootstrap philosophy has already revealed some hidden simplicity of cosmological correlators---for example, correlators describing the production and decay of massive particles have a unified underlying structure~\cite{Arkani-Hamed:2018kmz,Baumann:2019oyu,Baumann:2020dch}---along with many other new insights~\cite{Arkani-Hamed:2017fdk,Arkani-Hamed:2018bjr,Benincasa:2018ssx,Benincasa:2019vqr,Albayrak:2018tam,Sleight:2019mgd, Sleight:2019hfp,Albayrak:2019asr,Albayrak:2019yve,Hillman:2019wgh,Sleight:2020obc,Albayrak:2020isk,Benincasa:2020aoj,Goodhew:2020hob,Armstrong:2020woi,Albayrak:2020fyp,Meltzer:2021bmb,1866373}.\footnote{Cosmological correlators in de Sitter space are also very closely related to the study of conformal field theory in momentum space, where there has also been much recent progress~\cite{Coriano:2013jba, Coriano:2019nkw, Maglio:2019grh, Bzowski:2013sza, Bzowski:2019kwd, Isono:2018rrb, Isono:2019wex, Bautista:2019qxj, Gillioz:2019lgs, Gillioz:2019iye,  Gillioz:2020mdd,1866362}.} Aside from being of structural interest, these new calculational techniques can also help to realize the promise of using the inflationary epoch as a ``cosmological collider"~\cite{Chen:2009zp, Baumann:2011nk, Assassi:2012zq, Chen:2012ge, Pi:2012gf, Noumi:2012vr, Baumann:2012bc, Assassi:2013gxa, Gong:2013sma, Arkani-Hamed:2015bza, Lee:2016vti, Kehagias:2017cym, Kumar:2017ecc, An:2017hlx, An:2017rwo, Baumann:2017jvh, Kumar:2018jxz, Goon:2018fyu, Anninos:2019nib,Kim:2019wjo,Alexander:2019vtb, Hook:2019zxa,Kumar:2019ebj, Liu:2019fag,Wang:2019gbi,Chen:2018uul,Chen:2018brw}.

\begin{figure}[t!]
   \centering
            \includegraphics[width=.9\textwidth]{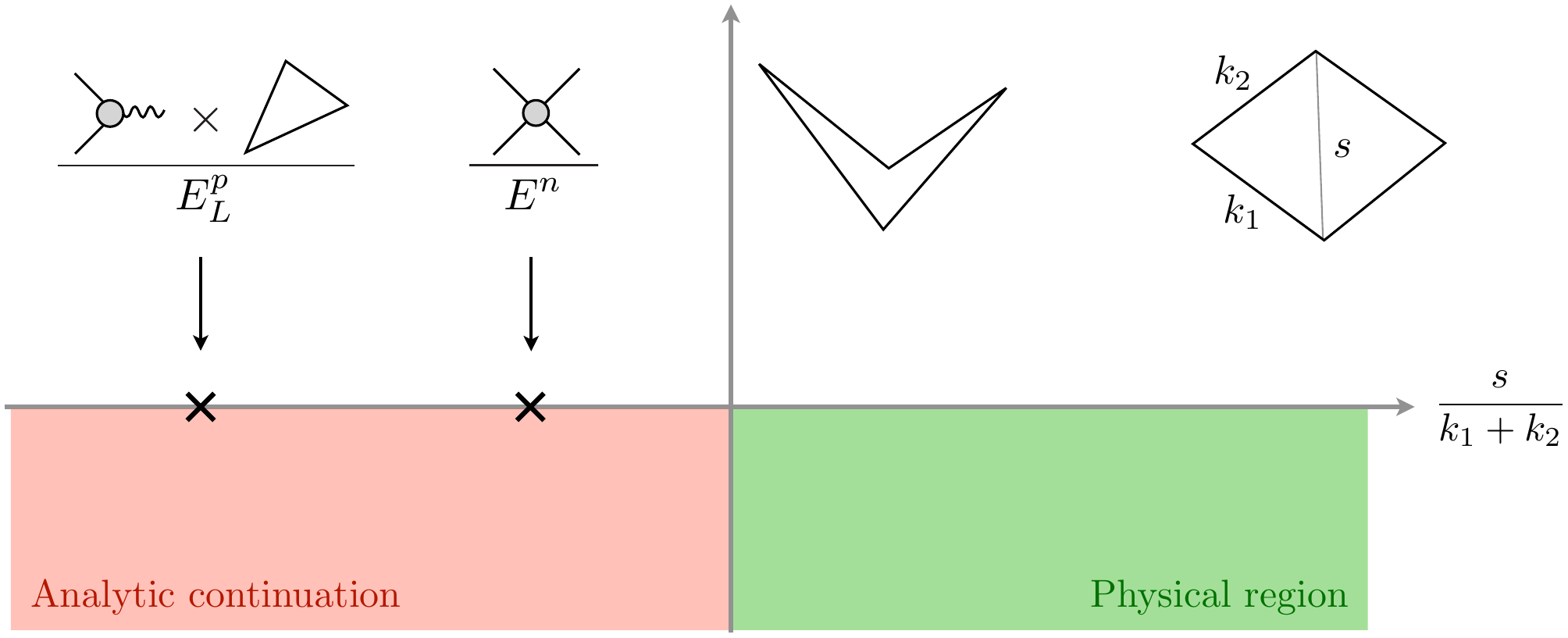}
   \caption{A one-dimensional slice through a four-point correlation function. Singularities in the unphysical region (where some of the energies have been analytically continued to negative values) determine the form of the correlator in the physical region. }
  \label{fig:AnalyticStructure}
\end{figure}

\vskip4pt 
There has been a lot of recent progress in understanding the singularity 
structure of cosmological correlators~\cite{Arkani-Hamed:2017fdk,Arkani-Hamed:2018bjr,Benincasa:2018ssx,Baumann:2020dch}.
Essentially all correlators have a singularity when the energies of the external particles add up to zero. This ``total energy singularity" arises from interactions in the early-time (or short-distance) limit, which behave as though they are in flat spacetime.
Correspondingly, the coefficient of this singularity is the scattering amplitude for the same process~\cite{Maldacena:2011nz,Raju:2012zr}. Importantly, these singularities cannot be accessed for physical configurations, but must instead be reached by analytically continuing some of the energies to negative values. The precise details of this analytic continuation (i.e.~which energies are chosen to be negative) select which particles are in/outgoing in the corresponding scattering amplitude.  
Correlators arising from the bulk exchange of particles have additional ``partial energy singularities" when the energies flowing into any subgraph add up to zero.  At these singularities, the correlator factorizes into a product of a lower-point correlator and a lower-point scattering amplitude~\cite{Arkani-Hamed:2017fdk,Arkani-Hamed:2018bjr,Benincasa:2018ssx,Baumann:2020dch}.  
To obtain the full correlator at general kinematics, we must find a way to connect its singular limits. Remarkably, although the singularities only arise for unphysical energy configurations, they still determine the correlators at physical energy values (see Fig.~\ref{fig:AnalyticStructure}).

\vskip4pt
It is natural to think of the singular points of a correlator as boundary conditions and seek a differential equation that describes deformations of the kinematics away from these loci.
The existence of such a differential equation is a consequence of local and causal time evolution in the bulk spacetime. Changing the kinematical shape of the boundary correlator corresponds to changing the relative time delay between the local bulk interactions involved in a particle exchange. In this sense, the boundary correlations retain a memory of the bulk dynamics.
A particularly natural way to derive the differential equation governing the boundary correlators is to exploit the symmetries of the bulk spacetime, which constrain the evolution. In de Sitter space, these symmetries translate into conformal Ward identities satisfied by the boundary correlators. This approach was pursued in~\cite{Arkani-Hamed:2018kmz,Baumann:2019oyu,Baumann:2020dch} to derive four-point correlators corresponding to a variety of bulk processes. One of the main advantages of this method is that it is possible to treat correlators involving heavy particles analytically. In this case, the singularities are in fact branch points, with the branch cut extending to the boundary of the physical region~\cite{Arkani-Hamed:2018kmz}. It was shown that this branch cut in the unphysical region forces the appearance of a characteristic oscillatory feature in the physical region. These oscillations are a key signature of particle production in the time-dependent background, but here arise as a solution of the bootstrap constraints imposed on completely static boundary correlators~\cite{Arkani-Hamed:2015bza,Arkani-Hamed:2018kmz}. 

\vskip4pt
Many of the correlators of interest in cosmology---especially those involving gauge fields, gravitons, or inflatons---have a much simpler singularity structure than the generic case: all singular points are poles and the correlators are therefore rational functions of their energy variables. This  opens up new avenues to extend correlators away from their singular loci. It also suggests that it is possible to combine together simple building blocks into more complicated processes.  In this paper, we explore several approaches to doing precisely this. We will still require additional information in order to extend correlators away from their singular configurations, but the simplified structure of the singularities means that we can reconstruct the full answer utilizing coarser information about the bulk dynamics. 
 
 \begin{itemize}
 
 \item {\it Cutting:} An additional piece of information that can be used to extend correlators away from their singular limits comes from the unitarity of bulk time evolution. This imposes specific constraints on correlation functions, which were first introduced in the 
form of a ``cosmological optical theorem" in \cite{Goodhew:2020hob} (see also~\cite{Meltzer:2020qbr, Cespedes:2020xqq}) and have recently been extended
to a systematic set of ``cutting rules" in~\cite{Melville:2021lst,Goodhew:2021oqg}. These cuts encode slightly different information about the bulk Green's function from the differential equations described above, relating part of a correlator to lower-point objects, but importantly for general kinematics. 
(The details of the cutting rules will be given in Section~\ref{sec:cuttingsec} and Appendix~\ref{app:cutproof}.) 
These relations provide important constraints on the analytic structure of correlators.
For example, the residues of all partial energy singularities (including those of subleading poles in the case of de Sitter correlators) are fixed in this way~\cite{Jazayeri:2021fvk}.  In this work, we give further examples of the utility of the cosmological cutting rules and show how they can be used to bootstrap complex correlators efficiently, particularly when the final correlators are rational functions.

\item {\it Gluing:} As we have argued, the structure of correlators at physical configurations is controlled in large part by their singularities at unphysical locations (see Fig.~\ref{fig:AnalyticStructure}). In fact, in many cases, correlators are completely specified by these singularities, allowing us to ``glue" together higher-point functions when we know the relevant lower-point data. This gluing can be formalized through recursion relations. In the case of scattering amplitudes, the celebrated BCFW recursion relations~\cite{Britto:2005fq} are an inspiring example of this philosophy. While the precise analog of BCFW does not yet exist for correlators, interesting energy recursion relations were developed for a class of scalar theories in~\cite{Arkani-Hamed:2017fdk} and extended in~\cite{Jazayeri:2021fvk}. Utilizing a similar approach, we will show that in many cases rational correlators both in flat space and in de Sitter space can be constructed recursively. 

\item {\it Lifting:} 
The combination of unitarity and recursion is especially powerful for correlators in flat space, where all energy singularities are just simple poles. In contrast, the singularities of de Sitter correlators are typically higher-order poles, and additional information is required to fix the subleading singularities. 
We will exploit the relative simplicity of the flat-space correlators by ``transmuting" them into the corresponding de Sitter objects using certain derivative (and sometimes integral) operations. 
This makes manifest that the more complicated singularities in de Sitter space capture the same physical information as those in flat space. At three points, this transmutation procedure allows us to construct correlators involving massless particles of arbitrary spin in de Sitter space.
At four points, the relevant transmutation operations have a natural interpretation in terms of the cuts of correlators---we can derive the transmutation operations by demanding that they transform the cut of a flat-space correlator into the cut of the corresponding de Sitter correlator. 
 \end{itemize}
 
\noindent
One of the advantages of these approaches to the problem is that they do not
rely explicitly on de Sitter symmetry.\footnote{There is an implicit reliance on de Sitter symmetries at the background level, as these are responsible for the simplicity of the mode functions of massless fields, leading to rational correlators.}
This makes it easy to input a breaking of the de Sitter symmetries at the level of interactions by using Lorentz-violating amplitudes as the residues of the energy singularities, connecting to recent work classifying such amplitudes~\cite{Grall:2020ibl,Pajer:2020wnj,Stefanyszyn:2020kay}. Extending away from these singular loci, we then end up with the correlator of a de Sitter boost-breaking theory~\cite{Pajer:2020wxk,Jazayeri:2021fvk}. The transmutation procedure is useful in systematizing this idea, allowing us to generate de Sitter boost-breaking correlators from  simple flat-space seeds. As an example of this approach, we will show how to construct correlators in the EFT of inflation~\cite{Cheung:2007st} (see Appendix \ref{app:EFTapp}).

\vskip4pt
The ideas of cutting, gluing, and lifting correlators are interlinked, and are most powerful when combined to efficiently bootstrap correlators. In particular, cutting rules provide information about the singularities---useful for recursion relations---as well as giving a way to systematically construct transmutation operators that lift the cuts of flat-space correlators to their de Sitter counterparts.
As an example of the utility of these tools, in Section~\ref{sec:bootstrapping} we synthesize these ideas to bootstrap the graviton Compton scattering correlator in de Sitter space. 
Conceptually, this approach to correlation functions is useful because it makes manifest how correlation functions are assembled in terms of simpler building blocks. This reorganization of the information is suggestive of further hidden structures, waiting to be discovered.

\vskip10pt
\noindent
{\bf Outline:} The outline of the paper is as follows: In Section~\ref{sec:preliminaries}, we review the perturbative calculation of wavefunction coefficients and explain the origin and nature of their energy singularities. Expert readers may skip this section or just skim it to become familiar with our notation. In Section~\ref{sec:cuttingsec}, we derive unitarity cutting rules---both in flat space and de Sitter---and then apply them to bootstrap a number of illustrative correlators. In Section~\ref{sec:gluing}, we describe energy recursion relations for rational correlators and apply them to several simple examples. In Section~\ref{sec:lifting}, we show how flat-space correlators can be transmuted to de Sitter space. In~Section~\ref{sec:bootstrapping}, we combine the tools introduced in the previous sections to study a more complex example: the correlator associated with the Compton scattering of gravitons. We first bootstrap this correlator in flat space and then transmute the answer to de Sitter space. Our conclusions are presented in Section~\ref{sec:conclusions}.

\vskip 4pt
The appendices contain additional technical details and reference material: In Appendix~\ref{app:pols}, we collect results on the polarization sums arising in the exchange of particles with spin.
In Appendix~\ref{app:cutproof}, we provide details of the cutting rules for general graph topologies. In Appendix~\ref{app:Residues}, we derive a relation between the residues of energy singularities and the coefficients of the corresponding Laurent expansions, relevant for energy recursion in de Sitter space. In Appendix~\ref{app:YMapp}, we use the methods introduced in this paper to derive the four-point function of Yang--Mills theory.
Finally, in Appendix~\ref{app:EFTapp}, we show how our approach can be applied to the boost-breaking interactions in the EFT of inflation~\cite{Creminelli:2006xe,Cheung:2007st}.

\vskip 4pt
\noindent
While this paper was in preparation several interesting papers appeared~\cite{Melville:2021lst,Jazayeri:2021fvk,Goodhew:2021oqg}, that explore related ideas from a complementary perspective.

 \vskip10pt
 \noindent
{\bf Notation and conventions:} 
We use natural units, $\hbar = c \equiv 1$.  Our metric signature is $({-}\hs{+}\hs{+}\hs{+})$.
We use Greek letters for spacetime indices, $\mu =0,1,2,3$, and Latin letters for spatial indices, $i=1,2,3$. Physical time is $t$ and conformal time is $\eta$.
Spatial vectors are denoted by $\vec x$ and their components by~$x^i$.
The corresponding three-momenta are $\vec k$, with magnitudes $k = |\vec k|$ and unit vectors written as $\hat k = \vec k/k$.
We use Latin letters from the beginning of the alphabet to label the momenta of the different legs of a correlation function, i.e.~$\vec k_a$ is the momentum of the $a$-th leg. The sum of $k_a$ and $k_b$ is often written as $k_{ab} \equiv k_a + k_b$.  
Three-dimensional polarization vectors are denoted by $\vec \xi_a$. We (anti-)symmetrize indices with unit weight, e.g.~$A_{(ij)} \equiv \tfrac{1}{2}(A_{ij}+A_{ji})$.

\vskip 4pt
We will mostly focus on
tree-level four-point correlators defined on the spacelike boundary of four-dimensional spacetime. 
It is therefore useful to assign the following kinematic data:
\vspace{-4.5pt}
\begin{equation*}
\scalebox{1.0}{
 %\raisebox{-33pt}{
\begin{tikzpicture}[line width=1. pt, scale=2]

%\draw[gray,fill=lightgray] (0,0) circle (.15cm);
%\draw[gray,fill=lightgray] (1,0) circle (.15cm);
\draw[fill=black] (0,0) -- (1,0);
\draw[fill=black] (0,0) -- (1,0);
\draw[line width=1.pt] (0,0) -- (-0.25,0.55);
\draw[line width=1.pt] (0,0) -- (0.25,0.55);
\draw[line width=1.pt] (1,0) -- (0.75,0.55);
\draw[line width=1.pt] (1,0) -- (1.25,0.55);

\node[scale=1,above] at (-0.25,0.55) {$k_1$};
\node[scale=1,above] at (0.25,0.55) {$k_2$};
\node[scale=1,above] at (0.75,0.55) {$k_3$};
\node[scale=1,above] at (1.25,0.55) {$k_4$};

\draw[lightgray, line width=2.pt] (-0.6,0.55) -- (1.6,0.55);
\draw[fill=black] (0,0) circle (.03cm);
\draw[fill=black] (1,0) circle (.03cm);
\node[scale=1] at (0,-.15) {$E_L$};
\node[scale=1] at (1,-.15) {$E_R$};
\node[scale=1] at (0.5,-.12) {$s$};
\end{tikzpicture}
%} 
}
\vspace{-5.5pt}
\end{equation*}
where $s\equiv  |\k_1+\k_2|$ is the magnitude of the exchanged momentum and $E_L\equiv k_{12}+s$, $E_R \equiv k_{34}+s$ are the energies flowing into the left and right vertices, respectively. The latter variables are natural for $s$-channel exchange, and we will sometimes also write them as  $E_{L,R}^{(s)}$.  Similar quantities can be defined in the $t$ and $u$-channels by permutation. For example, we have $t \equiv |\k_1+\k_4|$ and $u \equiv |\k_1+\k_3|$, along with $E_L^{(t)}\equiv k_{14}+t$, $E_R^{(t)} \equiv k_{23}+t$ and $E_L^{(u)}\equiv k_{13}+u$, $E_R^{(u)} \equiv k_{24}+u$.

\vskip 4pt
We will use the following conventions for scattering amplitudes:
All four-momenta $p^\mu_a$ are ingoing and polarization vectors are denoted by $\epsilon_a^\mu$. 
The Mandelstam variables are $S \equiv -(p_1+p_2)^2$, $T \equiv -(p_1+p_4)^2$ and $U \equiv - (p_1+p_3)^2$, which we capitalize to avoid confusion with the exchange momenta $s,t,u$ used in cosmological correlators. It will also sometimes be convenient to define the variables ${\cal S}\equiv -E_L^{(s)}E_R^{(s)}$, ${\cal T}\equiv -E_L^{(t)}E_R^{(t)}$, ${\cal U}\equiv -E_L^{(u)}E_R^{(u)}$, which reduce to $S, T, U$ on the locus of total energy conservation, $E=0$.

\newpage
\section{The Wavefunction and its Singularities}
\label{sec:preliminaries}

The object of our central interest is the cosmological wavefunctional, so we briefly introduce its elementary properties, with a particular emphasis on its singularity structure, which will play an important role in the following. 
More details can be found in~\cite{Anninos:2014lwa,Goon:2018fyu,Baumann:2020dch,Goodhew:2020hob}.

\subsection{Cosmological Wavefunction}
\label{sec:basic}

Consider a bulk field $\phi$, whose spatial profile at a fixed time $t_*$ is $\varphi(\vec x) = \phi(\vec x, t_*)$. The probability density for spatial field configurations is given by the square of a wavefunctional, $\Psi[ \varphi(\vec x),t_*] \equiv \langle \varphi(\vec x) \rvert 0\rangle$, which is the projection of the vacuum state onto the basis of Heisenberg picture eigenstates of $\phi$.
 Equal-time correlation functions of the field are then defined as
\beq
\langle\vp(\vec x_1)\cdots \vp(\vec x_n)\rangle =\frac{ \mathlarger\int{\cal D} \vp\,\vp(\vec x_1)\cdots \vp(\vec x_n) \left\lvert\Psi[\vp]\right\rvert^2}{\mathlarger\int{\cal D} \vp \left\lvert\Psi[\vp]\right\rvert^2} \, .
\label{eq:inincorr}
\eeq
The wavefunctional and (equal-time) correlation functions therefore contain the same information, which can be made completely explicit in perturbation theory.

\vskip4pt
It is convenient to work  in momentum space and expand the wavefunctional as 
\beq
\begin{aligned}
\Psi[\vp] \,\simeq \ \exp\left(- \sum_{n=2}^\infty \frac{1}{n!}\int\frac{{\rm d}^3k_1\cdots {\rm d}^3k_n}{(2\pi)^{3n}}\, \Psi_n(\vec k_N) \, \vp_{\vec k_1}\cdots\vp_{\vec k_n}
\right) ,
\end{aligned}
\label{eq:WFFourier}
\eeq
where the kernel functions $\Psi_n$ are called {\it wavefunction coefficients}\hskip 1pt\footnote{Viewing the wavefunctional~\eqref{eq:WFFourier} as a generating function, the wavefunction coefficients bear a formal similarity to correlation functions. This is not accidental. Indeed, wavefunction coefficients can be interpreted as correlation functions of operators, $O$, which are dual to the bulk fields $\phi$~\cite{Maldacena:2002vr}. Because of this---and because the information contained in wavefunction coefficients and equal-time correlation functions are essentially the same---we will often blur the distinction between the two objects and abuse terminology by referring to the wavefunction coefficients as ``correlators." In another abuse of notation, we will often call the operators dual to massless spin-1 and spin-2 fields $J_i$ and $T_{ij}$, respectively, even in flat space.}  and
$\vec k_N \equiv \{\vec k_1, \ldots, \vec k_n \}$ denotes the set of all momenta.
As a consequence of translation invariance, the wavefunction coefficients contain a momentum-conserving delta function. It is often useful to extract this delta function and write
\be
\Psi_n(\vec k_N) = (2\pi)^3  \delta(\vec k_1+\cdots+\vec k_n)\, \psi_n(\vec k_N) \, ,
\label{equ:O}
\ee
where the coefficients $\psi_n$ are the objects that we will 
study in the remainder of the paper. They are the outputs of time evolution in the bulk spacetime and encode properties of the bulk physics---like locality and unitarity---in the spatial dependence of correlations. 

\subsubsection*{Perturbation theory}
The wavefunction at a given time (which we can set to $t_* = 0$ without loss of generality) is formally computed by the following path integral
\beq
\Psi[\vp] \ =  \hspace{-0.4cm} \int\limits_{\substack{\phi(0) \,=\,\vp\\ \hspace{-0.4cm}\phi(-\infty)\,=\,0}} 
\hspace{-0.5cm} \raisebox{-.05cm}{ ${\cal D} \phi\, e^{iS[\phi]}\,,$ }
\eeq
which defines a sum over field configurations with the specified boundary conditions. The standard $i\varepsilon$ prescription, which selects the vacuum, is implicit in the early-time boundary condition, while the late-time Dirichlet boundary condition 
makes this expression a functional of the field profile $\vp(\vec{x},t_*=0)$. Our interest is in computing the wavefunction in perturbation theory, where the path integral can be approximated by its saddle point, $\Psi[ \vp] \approx \exp\left(iS[ \vp_{\rm cl}]\right)$, with $\vp_{\rm cl}$ being the boundary profile of the classical solution to the equations of motion, $\phi_{\rm cl}$.

\vskip4pt
The relevant classical solution, and hence the wavefunction coefficients, can be constructed in a diagrammatic expansion in perturbation theory. In particular, the classical equation of motion admits the following formal solution 
\beq
\phi_{\rm cl}(\vec k,t) = {\cal K}(k,t)\hskip 1pt \vp_{\vec k}+\int{\rm d} t'\,{\cal G}(k; t,t') \frac{\delta S_{\rm int}}{\delta \phi_{\vec k}(t')}\,. \label{equ:formaleomsol}
\eeq
 Substituting this solution back into the action yields the wavefunction to whatever order in perturbation theory is desired. The perturbative computation of the wavefunction coefficients therefore requires the two distinct Green's functions appearing in~\eqref{equ:formaleomsol}. The  bulk-to-boundary propagator, ${\cal K}(k,t)$, solves the (homogeneous) linearized equation of motion, oscillates with positive frequency in the far past, and approaches unity as $t\to 0$. The bulk-to-bulk propagator, ${\cal G}(k; t, t')$, solves the inhomogeneous equation $\sqrt{-g}\hskip 1pt(\square-m^2)\hskip 1pt {\cal G}(k;t,t') = i\delta(t-t')$, subject to the boundary condition that it vanishes when either of its time arguments is taken to zero. The combined boundary conditions on the two Green's functions guarantee that the solution~\eqref{equ:formaleomsol} will satisfy the correct Dirichlet boundary conditions.

\vskip4pt
With these preliminaries out of the way, we can now state the Feynman rules for the computation of the wavefunction coefficients, $\psi_n$:
\begin{itemize}
\item Draw all diagrams with a fixed number of lines, $n$, ending on the boundary.
\vspace{-3pt}
\item Assign a vertex factor, $iV$, to each bulk interaction.
\vspace{-3pt}
\item Assign a bulk-to-bulk propagator, ${\cal G}$, to each internal line.
\vspace{-3pt}
\item Assign a bulk-to-boundary propagator, ${\cal K}$, to each external line.
\vspace{-3pt}
\item  Integrate over the time insertions of all bulk vertices.
\vspace{-3pt}
\item For diagrams involving loops, integrate over the loop momenta.
\end{itemize}
The vertex factors, $iV$, can be derived from the action in the same way as for the S-matrix Feynman rules, with the caveat that we are using Fourier space only for the spatial variables. This  leads to a time dependence for each vertex which must then be integrated over. In fact, much of the complication of computing wavefunction coefficients comes from these time integrals, motivating the search for alternative calculational approaches.

\vskip 4pt
In addition to the wavefunction $\Psi$, it will be convenient to introduce the {\it conjugate wavefunction}, $\bar \Psi$. The corresponding conjugate wavefunction coefficients, $\bar\psi_n$, are computed using the same Feynman rules as for the wavefunction coefficients, except that we assign the {\it anti}-bulk-to-bulk propagator to internal lines, $\bar{\cal G} = {\cal G}^*$, and replace each vertex factor with $-iV$, 
but use the same bulk-to-boundary propagators as for the ordinary wavefunction.
An important feature of the conjugate wavefunction coefficients is that in a {\it unitary} theory, they are related to the complex conjugate of the ordinary wavefunction coefficients, but with analytically continued external energies~\cite{Goodhew:2020hob,Cespedes:2020xqq}\footnote{As we will discuss more in detail in Section~\ref{sec:dscutting}, this analytic continuation to negative energies 
%corresponds to 
is performed by rotating clockwise in the complex plane. Moreover, when dealing with external spinning particles, the wavefunction coefficients can have an explicit dependence on spatial momenta $\k_a$, contracted with polarization vectors. In such cases, we also flip the signs of external momenta $\k_a \mapsto -\k_a$, which can be understood as analytically continuing the energies $k_a \mapsto -k_a$, while keeping the directions of $\hat k_a$ fixed in the real domain. The polarization vectors can be either stripped off in the process or kept unchanged under the replacement rule $[\xi_a^\pm(-\hat k_a)]^* \mapsto \xi_a^\pm(\hat k_a)$; see also~\cite{Goodhew:2021oqg}. With this being understood, in what follows, we will only show energies in the argument of $\psi$ for brevity.} 
\beq
\bar\psi_n(k_1,\cdots, k_n) = \psi^*_n(-k_1,\cdots, -k_n) \qquad\quad ({\rm unitarity})\,.
\label{eq:unitarityflat}
\eeq
This relation will play an important role in the cutting rules for the wavefunction coefficients in Section~\ref{sec:cuttingsec}.

\subsubsection*{The flat-space wavefunction}
To get some intuition for both the construction and the properties of the wavefunction, it is useful to study explicit examples in a simplified setting. An illuminating example is provided by the theory of a single scalar field in flat spacetime with polynomial self-interactions
\beq
S = \int\rd^4x\left(-\frac{1}{2}(\partial\phi)^2 + \sum_{n=3}^\infty\frac{g_n}{n!}\phi^n\right) .
\label{equ:S}
\eeq
This theory is simple enough to allow a detailed study of the structure of bulk perturbation theory, but it captures much of the physics,  
so that the generalization to de Sitter space will be straightforward. Many features of this simplified model have been studied in~\cite{Arkani-Hamed:2017fdk,Arkani-Hamed:2018bjr,Benincasa:2018ssx,Benincasa:2020aoj}.

\vskip 4pt
In this theory, the relevant bulk-to-boundary and bulk-to-bulk propagators are~\cite{Arkani-Hamed:2017fdk}
\begin{align}
{\cal K}(k,t) &= e^{ikt}\,, \label{eq:bboundaryflat}\\
{\cal G}(k;t,t') &= \frac{1}{2k}\left(e^{i k(t'-t)}\theta(t-t')+e^{i k(t-t')}\theta(t'-t)-e^{ik(t+t')}\right) ,
\label{eq:bbulkflat}
\end{align}
where $k = |\vec k|$ is the energy of a mode with momentum $\k$. Note that the bulk-to-bulk propagator~\eqref{eq:bbulkflat} contains an additional non-time-ordered piece compared to the ordinary Feynman propagator. This term enforces that the Green's function vanishes when one of its time arguments is on the slice where the wavefunction lives (here $t = 0$).

\vskip4pt
With these perturbative building blocks in hand, we now turn to computing some example wavefunction coefficients in a theory with only a $\phi^3$ interaction.\footnote{The simplicity of the bulk-to-boundary propagator~\eqref{eq:bboundaryflat} means that the wavefunction coefficients do not depend in an explicit way on the number of external lines, but rather depend only on the total energy flowing from a vertex to the boundary. This implies that correlators only really depend on the structure of internal lines, so that the correlators in theories with more complicated polynomial interactions are essentially the same.}
The simplest wavefunction coefficient in this theory is the three-point function 
\vspace{.1cm}
\beq
\psi_3(k_1,k_2,k_3) =
\ ~ \raisebox{-35pt}{
\begin{tikzpicture}[line width=1. pt, scale=2]
\draw[lightgray, line width=1.pt] (0,0) -- (-0.4,0.75);
\draw[lightgray, line width=1.pt] (0,0) -- (0.4,0.75);
\draw[lightgray, line width=1.pt] (0,0) -- (0.0,0.75);
\draw[lightgray, line width=2.pt] (-0.6,0.75) -- (.6,0.75);
\draw[fill=black] (0,0) circle (.03cm);
\node[scale=.9] at (0,-.15) {$K\equiv k_1+k_2+k_3$};
\end{tikzpicture}
} 
~=~
ig\int_{-\infty}^0\rd t\,e^{iKt}= \frac{g}{K}\,,
\label{eq:flat3pt}
\eeq
where we have set $g_3 \equiv g$ and defined the total energy $K\equiv k_1+k_2+k_3$. 
The next-simplest wavefunction coefficient is the four-point function, which has a single internal line: 
\vspace{.1cm}
\beq
 \psi_4(k_{12},k_{34},s) \ \equiv~ \raisebox{-33pt}{
\begin{tikzpicture}[line width=1. pt, scale=2]
\draw[fill=black] (0,0) -- (1,0);
\draw[lightgray, line width=1.pt] (0,0) -- (-0.3,0.75);
\draw[lightgray, line width=1.pt] (0,0) -- (0.3,0.75);
\draw[lightgray, line width=1.pt] (1,0) -- (0.7,0.75);
\draw[lightgray, line width=1.pt] (1,0) -- (1.3,0.75);
\draw[lightgray, line width=2.pt] (-0.5,0.75) -- (1.6,0.75);
\draw[fill=black] (0,0) circle (.03cm);
\draw[fill=black] (1,0) circle (.03cm);
\node[scale=1] at (0,-.15) {$k_{12}$};
\node[scale=1] at (1,-.15) {$k_{34}$};
\node[scale=1] at (0.5,-.12) {$s$};
\end{tikzpicture}
} 
%\hspace{-10pt}
= \  -g^2\int_{-\infty}^0\rd t \hskip 1pt \rd t'\,e^{ik_{12}t}\,{\cal G}(s; t, t')\,e^{ik_{34}t'}\,,
\label{eq:2sitegraph}
\eeq
where $k_{12}\equiv k_1+k_2$ is the sum of energies flowing from the left vertex to the boundary (and similarly for $k_{34}$), while $s \equiv \lvert\vec k_1+\vec k_2\rvert$ is the energy of the exchanged particle.
Evaluating the integrals explicitly, we obtain
\beq
\psi_4(k_{12},k_{34},s) = \frac{g^2}{E E_L E_R}\,,
\label{eq:flatspacescalar4pt}
\eeq
where we have defined $E_L\equiv k_{12}+s$ and $E_R \equiv k_{34}+s$ as the energies flowing into the left and right vertices, and $E \equiv k_1+k_2+k_3+k_4$ is the total energy involved in the process. It is straightforward (though somewhat tedious) to compute higher-order examples with more exchanges, some of which can be found in Appendix~\ref{app:cutproof}. 

\vskip4pt
The explicit examples~\eqref{eq:flat3pt} and~\eqref{eq:flatspacescalar4pt} have a few interesting features, which turn out to be general. First of all, both answers are singular when the total energy involved in the process adds up to zero: $K\to 0$ for~\eqref{eq:flat3pt} and $E\to 0$ for~\eqref{eq:flatspacescalar4pt}. The residue of~\eqref{eq:flat3pt} is~$g$, which is exactly the three-point scattering amplitude $A_3$. More nontrivially, the residue of the total energy singularity in~\eqref{eq:flatspacescalar4pt} is also a scattering amplitude: 
\beq
\lim_{E\to 0}\psi_4(k_{12},k_{34}) = -\frac{1}{E} \frac{g^2}{S}\,=\, \frac{A_4}{E}\,,
\eeq
where $S  = k_{12}^2-s^2$ is the Mandelstam variable, so that $A_4$ is the corresponding
flat-space scattering amplitude. This is not an accident, but a manifestation of a more general phenomenon~\cite{Maldacena:2011nz,Raju:2012zr}. In addition to this total energy singularity, the exchange correlator~\eqref{eq:flatspacescalar4pt} also has singularities when either $E_L$ or $E_R$ vanish. At this location, we have
\beq
\lim_{E_L\to 0}\psi_4(k_{12},k_{34}) = \frac{g}{E_L} \frac{g}{(k_{34}+s)(k_{34}-s)} \,=\, \frac{A_3\times\tl\psi_3}{E_L}\,,
\eeq
and similarly for $E_R \to 0$.
The residue of this partial energy singularity is the product of a three-point scattering amplitude $A_3=g$ and a {\it shifted wavefunction} 
\beq
\tl\psi_3(k_3,k_4,s)  \equiv \frac{1}{2s}\Big(\psi_3(k_3,k_4,-s) -\psi_3(k_3,k_4,s) \Big)\,,
\label{eq:flatspaceshiftedwf1}
\eeq 
which is also the first manifestation of a more general phenomenon~\cite{Arkani-Hamed:2017fdk,Arkani-Hamed:2018bjr,Benincasa:2018ssx,Baumann:2020dch}.

\vskip4pt
In fact, these simple correlators are completely fixed by their singularities. In particular, the result~\eqref{eq:flatspacescalar4pt} is the unique function that has the correct factorization singularities.\footnote{Interestingly, for this simple example, the total energy singularity is required in order for the correlator to have the correct factorization singularities~\cite{Baumann:2020dch}. In more complicated examples, the total energy singularity is an additional input.} This is a very basic example of the general theme that we will explore in the rest of the paper, namely, that correlators can often be fixed by extending them away from their singular points in the only consistent manner.

\subsubsection*{De Sitter perturbation theory}
Perturbation theory in de Sitter space is essentially the same as in flat space, the only real difference being the form of the Green's functions and vertex factors. Recall that the de Sitter line element is
\beq
\rd s^2 = \frac{1}{H^2\eta^2}\left(-\rd\eta^2 + \rd\vec x^{\hskip 1pt 2}\right) ,
\label{eq:dsmetric}
\eeq
where $\eta$ is conformal time.
The bulk-to-boundary propagator for a scalar field of general mass, $m$, in de Sitter space therefore takes the form~\cite{Anninos:2014lwa}
\beq
{\cal K}_\nu(k,\eta) = \left(\frac{\eta}{\eta_*}\right)^{3/2} \frac{H_\nu^{(2)}(-k\eta)}{H_\nu^{(2)}(-k\eta_*)}\,,
\label{eq:massivescalarbulkboundary}
\eeq
where $H_\nu^{(2)}(-k\eta)$ is a Hankel function of the second kind, whose order is set by the mass through the relation $\nu \equiv \sqrt{9/4-m^2/H^2}$. 
 Similarly, the bulk-to-bulk propagator for a massive scalar is 
\beq
\begin{aligned}
    \mathcal{G}_\nu(k;\eta,\eta')\equiv\frac{\pi}{4}(\eta\eta')^{3/2}\Bigg[ &H_{\nu}^{(2)}(-k\eta')H_{\nu}^{(1)}(-k\eta)\theta(\eta-\eta')+H_{\nu}^{(2)}(-k\eta)H_{\nu}^{(1)}(-k\eta')\theta(\eta'-\eta)\\
    &-\frac{H_{\nu}^{(1)}(-k\eta_*)}{H_{\nu}^{(2)}(-k\eta_*)}H_{\nu}^{(2)}(-k\eta)H_{\nu}^{(2)}(-k\eta')\Bigg]\,.
\end{aligned}
\label{eq:massivebulkbulkprop}
\eeq
Using these expressions, along with the relevant Feynman rules for the vertices, we can compute the wavefunction in the same way as in the flat-space examples. Later, we will also require expressions for spinning fields in de Sitter space, but we will introduce these as the need arises.

\subsection{Singularities of Correlators}
\label{sec:singularities}
The singularities of cosmological correlators will play an important role in the following, so we quickly review the singularity structure of wavefunction coefficients. Further details can be found in~\cite{Arkani-Hamed:2017fdk,Arkani-Hamed:2018bjr,Benincasa:2018ssx,Baumann:2020dch}. In the simple flat-space examples described above, we saw that correlators have singularities when the energies flowing into any subgraph add up to zero. This is, in fact,
a general feature of all correlators.

\vskip4pt
Nearly all correlators are singular when the energies of all the external particles add up to zero, and the residue of this ``total energy singularity" is the flat-space scattering amplitude associated with the process~\cite{Maldacena:2011nz,Raju:2012zr}.\footnote{Notable exceptions are correlation functions of spinning fields with equal helicities. In this case, the corresponding flat-space amplitude vanishes identically, so these correlators do not have total energy singularities.} 
 In addition, correlators arising from the bulk exchange of particles have ``partial energy singularities," when the energies flowing into a subgraph add up to zero. At this kinematic locus, the wavefunction factorizes into a product of a lower-point wavefunction and a lower-point scattering amplitude~\cite{Arkani-Hamed:2017fdk,Arkani-Hamed:2018bjr,Benincasa:2018ssx,Baumann:2020dch}.

\vskip4pt
The origin of the total energy singularity is easy to understand from the perspective of the time integrals that arise in the computation of the wavefunction.  In the limit $E \equiv k_1 + \cdots+ k_n \to 0$, 
the early-time  ($\eta\to 0$) part of the integration is unsuppressed and so leads to a divergence.\footnote{In the case of scattering amplitudes, these time integrals run from $t=-\infty$ to $+\infty$ and because of time-translation invariance give rise to an energy-conserving delta function, $\delta(E)$. In the cosmological context, the integrals range from $-\infty$ to $0$, time-translation symmetry is broken and hence energy is {\it not} conserved. Instead of an energy-conserving delta function, we get an energy singularity.}
Because the Feynman rules for the wavefunction are closely related to the corresponding S-matrix rules, the coefficient of this singularity is the flat-space scattering amplitude associated with the same process~\cite{Maldacena:2011nz,Raju:2012zr}
\beq
\lim_{E\to 0}\psi_n = \frac{A_n}{E^p}\,,
\eeq
where we are not being careful about the overall phase.
Note that in cosmological spacetimes, the precise order of the singularity,  $p$, depends on the details of the interactions in the theory, but the existence of some singularity is generic because the de Sitter mode functions  in the early time limit asymptote to a power of $\eta$ times a plane wave. For example, the bulk-to-boundary propagator becomes
\beq
{\cal K}_\Delta(k,\eta) \underset{\eta\to-\infty}{\sim} e^{\frac{i\pi}{2}(\Delta+1)}\frac{\pi^\frac{1}{2}}{2^{\Delta-2}} \frac{(-k\eta_*)^{\Delta-3}}{\Gamma[\Delta-\frac{3}{2}]}\left(-ik\eta+\frac{(\Delta-1)(\Delta-2)}{2}+\cdots\right)e^{ik\eta}\,,
\label{eq:bulkboundarypropnearinf}
\eeq
which makes it clear that correlators will have a  singularity at $E \to 0$.

\vskip4pt
The presence of the partial energy singularities can be understood in a similar way---they come from the region of integration where one of the bulk vertices is in the far past. 
In this limit, the bulk-to-bulk propagator~\eqref{eq:bbulkflat} (in flat space) takes the following factorized form
\beq
{\cal G}(k; t\to-\infty, t') = \frac{e^{ikt}}{2k}\Big(e^{-ikt'}-e^{ikt'}\Big)\,.
\label{eq:factorizedflatspacebbprop}
\eeq
We see  that the factor associated to the $t$ vertex (the one taken to the far past) behaves like a bulk-to-boundary propagator, generating the scattering amplitude associated to this vertex.
The other vertex produces a shifted wavefunction. The situation in de Sitter space is essentially the same because the bulk-to-bulk propagator factorizes similarly to~\eqref{eq:factorizedflatspacebbprop} when one of the vertex times is taken to the infinite past. Note also that this causes the time integrations to factorize, because only a single time-ordering contributes.

\vskip4pt
As a concrete example, consider an exchange four-point function in the $s$-channel. When the energies entering the left vertex add up to zero, $E_L = k_{12}+s\to0$, we get~\cite{Baumann:2020dch}
\beq
\lim_{E_L\to 0}\psi_4 = \frac{\tl A^{(L)}_3 \times \tl \psi^{(R)}_3}{E_L^q}\,,
\label{eq:partialEformula}
\eeq
where $q$ is the order of the singularity, and we have defined
\begin{align}
\label{eq:dressed3pt}
\tl A_3^{(L)} (k_1,k_2,s) &\equiv  k_1^{\Delta_1-2}k_2^{\Delta_2-2}s^{\Delta-2}A_3 (k_1,k_2,s) \,,\\
\tl \psi_3^{(R)}(k_3,k_4,s) &\equiv \frac{(-1)^\Delta}{2s^{2\Delta-3}}\Big(\psi_3 (k_3,k_4,-s)-\psi_3(k_3,k_4,s)\Big)\,.
\label{eq:shifted3pt}
\end{align}
The scattering amplitude is dressed with some energy factors required by dimensional analysis
and the shifted wavefunction reduces to~\eqref{eq:flatspaceshiftedwf1} in flat space.
In de Sitter space, the weight $\Delta$ is set by the masses of the corresponding bulk fields through the usual relation\footnote{This scaling corresponds to the subleading fall-off in de Sitter space, which is the relevant scaling for the wavefunction coefficients of light fields. See~\cite{Baumann:2020dch} for more details.}
\beq
\Delta =
 \begin{cases}
   \displaystyle \  \frac{3}{2}+\sqrt{\frac{9}{4}-\frac{m^2}{H^2}} & \text{(scalars)}\, , \\[16pt]
   \displaystyle \   \frac{3}{2}+\sqrt{\left(\ell-\frac{1}{2}\right)^2-\frac{m^2}{H^2}} & \text{(spinning fields)}\, ,\\
  \end{cases}
  \label{eq:masstodeltarelation}
\eeq
where $\ell$ is the spin of the bulk field. Note that we can recover our flat-space formulas by setting $\Delta = 2$ for all fields. 
Of course, there is an analogous $E_R$ singularity, whose residue is given in terms of $\tl\psi^{(L)}_3$ and $\tl A^{(R)}_3$. Though we have only shown formulas involving scalars, the obvious generalizations involving spinning fields also hold. See~\cite{Baumann:2020dch} for some examples.

\vskip4pt 
The generalization to more complicated graphs is straightforward. Anytime the total energy flowing into a subgraph vanishes, a correlator has a singularity, whose coefficient is related to the amplitude corresponding to the subgraph whose energy is conserved, dressed by factors of the energies as in~\eqref{eq:dressed3pt} in de Sitter space. We can understand this singularity as coming from the part of the time integrals where all vertex insertions comprising a subgraph are in the far past. Any vertices for which the incoming energies are not taken to zero will lead to shifted wavefunctions, where the internal line is shifted in the same way as in~\eqref{eq:shifted3pt}.

\vskip4pt
Although the total and partial energy singularities cannot be probed by physical processes (some of the energies have to be negative), they can be accessed by analytic continuation. Remarkably, the presence of these singularities nevertheless governs much of the structure of correlators, so that the central challenge is to understand how to extend the correlator away from these singular points.
In this paper, we will study several different ways to accomplish this goal for correlators that are rational functions of the energies.

\vskip4pt
An important additional constraint will be that the correlators do {\it not} have any singularities for physical momentum configurations. In particular, we will see that demanding the absence of spurious singularities in {\it folded} (or collinear) configurations (e.g.~at $k_{12}= s$) is a powerful constraint on the structure of correlation functions in de Sitter space. 
Such folded singularities cannot arise for the Bunch--Davies vacuum~\cite{Arkani-Hamed:2018kmz} and their absence is a litmus test for the quantum origin of cosmic structure~\cite{Green:2020whw}.

\newpage
\section{Cutting: Constraints From Unitarity}
\label{sec:cuttingsec}
Unitarity---the conservation of probability---is a fundamental feature of quantum mechanics.
It therefore plays an important role in defining consistent observables in any quantum-mechanical theory.
Until recently, however, it wasn't known how the constraints of bulk unitary are encoded in cosmological boundary correlators. In~\cite{Goodhew:2020hob,Cespedes:2020xqq}, it was pointed out that unitarity implies specific relations that cosmological correlators have to satisfy in perturbation theory, quite similar to the optical theorem in flat space. Moreover, these relations can be systematized into a set of cutting rules for cosmological correlators~\cite{Meltzer:2020qbr,Goodhew:2021oqg,Melville:2021lst} that are spiritually similar to the Cutkosky rules for the S-matrix~\cite{Cutkosky:1960sp,tHooft:1973wag,Sterman:1994ce,Veltman:1994wz}.

\vskip4pt
In this section, we will  derive these cutting rules and explain how they can be used as a tool for bootstrapping correlators.  We first describe the cutting rules in flat space, before generalizing to de Sitter space, which is conceptually very similar. 

\subsection{Cutting Rules in Flat Space}

Part of the difficulty involved in bulk perturbation theory is the presence of nested time integrals, which are challenging to compute.  We therefore would like to ascertain some features 
without having to perform these integrals explicitly. We can do this by considering a suitable combination of the wavefunction coefficients $\psi$ and conjugate wavefunction coefficients $\bar\psi$,
engineered so that the bulk-to-bulk propagator~\eqref{eq:bbulkflat} and the anti-bulk-to-bulk propagator
\beq
\bar {\cal G}(k;t,t') = - {\cal G}(-k;t,t')\,,
\label{eq:antiandregularbbprop}
\eeq
appear as the sum
\begin{tcolorbox}[colframe=white,arc=0pt,colback=greyish2]
%\vspace{-7pt}
\beq
\widetilde {\cal G}(k;t,t') \equiv
 {\cal G}(k;t,t')+\bar {\cal G}(k;t,t') = -\frac{1}{2k}\big(e^{-ikt}-e^{ikt}\big)\big(e^{-ikt'}-e^{ik t'}\big)\,.
 \label{eq:flatcutprop}
\eeq
\end{tcolorbox}
\noindent
We will refer to $\tl{\cal G}$ as the {\it cut propagator}. This combination trivializes the time-ordering involved in the bulk integrations and therefore simplifies the computation of particular sums of $\psi$ and $\bar\psi$. Notice that~\eqref{eq:flatcutprop} can be written in terms of the bulk-to-boundary propagator as
\beq
\tl{\cal G}(k;t,t') = -\frac{1}{2k}\Big({\cal K}(-k,t)-{\cal K}(k,t)\Big)\Big({\cal K}(-k,t')-{\cal K}(k,t')\Big)\,.
\label{eq:flatcutasbbprop}
\eeq
We will see that introducing this combination of propagators has the interpretation of cutting the internal line of a graph, as in the 
S-matrix cutting rules. The cutting rules essentially systematize finding combinations that simplify time integrals, and 
naturally produce identities involving both $\psi$ and~$\bar\psi$. In a unitary theory, $\bar\psi$ can be related back to $\psi$ which produces identities satisfied by the wavefunction $\psi$ alone.

\vskip4pt
In this section, we describe the cutting rules for wavefunction coefficients in their simplest manifestation by considering the theory~\eqref{equ:S}, which involves a single scalar field in flat spacetime with polynomial self-interactions. 
The vertex factors are just constants, $ig_n$, and we will typically set $g_n \equiv 1$ to avoid clutter.

\subsubsection{A Simple Example}

To illustrate the cutting rule in a simple example, we again consider the
four-point function arising from tree-level exchange in $\phi^3$ theory. The wavefunction coefficient corresponding to the $s$-channel contribution was given in~\eqref{eq:2sitegraph}:
\beq
 \psi_4(k_{12},k_{34}) 
= \  -\int_{-\infty}^0\rd t \hskip 1pt \rd t'\,e^{ik_{12}t}\,{\cal G}(s; t, t')\,e^{ik_{34}t'}\,.
\eeq
Note that the bulk-to-bulk propagator in~\eqref{eq:bbulkflat} leads to nested time integrals, which can be done explicitly in this simple example, but are harder to perform in more general situations. 
However, we can simplify the integrals involved
by considering the sum of $\psi_4$ and its conjugate $\bar\psi_4$:
\beq
\psi_4(k_{12},k_{34})+\bar\psi_4(k_{12},k_{34}) = - \int_{-\infty}^0\rd t \hskip 1pt \rd t'\,e^{ik_{12}t}\,\tl{\cal G}(s; t, t')\,e^{ik_{34}t'}\,,
\label{eq:2sitecut}
\eeq
where $\tl{\cal G}$ is the cut propagator~\eqref{eq:flatcutprop}.
We can perform the two time integrals over $t$ and $t'$ separately. Each integral leads to a {\it shifted three-point function}
\beq
\tl\psi_3(k_{12}\mp s) \equiv \frac{i}{2s}\int_{-\infty}^0\rd t\, e^{ik_{12} t}\left(e^{-is t}-e^{is t}\right) = \frac{1}{2s}\Big(\psi_3(k_{12}-s)-\psi_3(k_{12}+s)\Big)\,,
\label{eq:shiftedWF}
\eeq
so that~\eqref{eq:2sitecut} becomes 
\beq
\psi_4(k_{12},k_{34})+\bar\psi_4(k_{12},k_{34}) = -2s\,\tl\psi_3(k_{12}\mp s)\,\tl{\bar\psi}_3(k_{34} \mp s)\,.
\label{eq:2sitepbarp}
\eeq
We can interpret the right-hand side as cutting the internal line connecting the two vertices, and then shifting the energies of these nodes by the energy of the cut line. At this point, the relation~\eqref{eq:2sitepbarp} is just an algebraic identity between $\psi_4$ and the auxiliary object $\bar\psi_4$. 
However, using~\eqref{eq:unitarityflat}---which holds in a unitary theory---we can write the left-hand side in terms of  $\psi_4$ alone\footnote{In the simple flat-space examples that we are considering, we will never encounter branch cuts, so we do not have to specify the precise analytic continuation to negative energies, but this will be important in the generalization to de Sitter space, as we discuss in Section~\ref{sec:dscutting}.}
\begin{tcolorbox}[colframe=white,arc=0pt,colback=greyish2]
\vspace{-3pt}
\beq
\psi_4(k_{12},k_{34})+\psi^*_4(-k_{12},-k_{34}) = -2s\,\tl\psi_3(k_{12}\mp s)\,\tl{\psi}_3(-k_{34} \mp s)\,.
\label{eq:2siteformula}
\eeq
\end{tcolorbox}
\noindent
For later convenience, we will use the following shorthand for the left-hand side of the cutting rule
\beq
{\rm Disc}[\psi_4] \equiv \psi_4(k_{12},k_{34})+\psi_4^*(-k_{12},-k_{34})\,,
\label{eq:discdef}
\eeq
although we must emphasize that this operation is not precisely the same as extracting the discontinuity of a complexified function.\footnote{The notation follows that of~\cite{Cespedes:2020xqq,Jazayeri:2021fvk,Melville:2021lst,Goodhew:2021oqg}, and is meant to recall the discontinuity operation involved in the cuts of scattering amplitudes. It is also important to note that our definition of this operation differs slightly from that of~\cite{Cespedes:2020xqq,Jazayeri:2021fvk,Melville:2021lst,Goodhew:2021oqg}, because of different conventions for factors of $i$. See Appendix~\ref{app:cutproof} for more details and applications.}
Note that $\tl\psi_3(k_{34} \mp s) = \tl\psi_3(-k_{34} \mp s)$, but it is convenient to write things as in~\eqref{eq:2siteformula} in order to give the final formula a diagrammatic interpretation. The formula~\eqref{eq:2siteformula} also provides an additional perspective on the partial energy singularities discussed in Section~\ref{sec:preliminaries}. By taking partial energy limits of this formula we can reproduce the partial energy singularity formulas~\eqref{eq:partialEformula}.

\vskip4pt
It is straightforward to check that the cutting rule~\eqref{eq:2siteformula} is satisfied by the wavefunction coefficients in this example. Given the three-point function~\eqref{eq:flat3pt} and the four-point function~\eqref{eq:flatspacescalar4pt}, it is a matter of simple algebra to compute the two sides of~\eqref{eq:2siteformula} and verify that the relation holds.

\vskip4pt
In order to extract a more general lesson from this simple example, it is useful to 
move the shifted wavefunctions in~\eqref{eq:2siteformula} to the left-hand side. The resulting formula can then be given a simple diagrammatic interpretation: 
\vspace{1pt}
\beq
\raisebox{-33pt}{
\begin{tikzpicture}[line width=1. pt, scale=2]

\draw[fill=black] (0,0) -- (1,0);
\draw[lightgray, line width=1.pt] (0,0) -- (-0.3,0.75);
\draw[lightgray, line width=1.pt] (0,0) -- (0.3,0.75);
%\draw[lightgray, line width=1.pt] (0,0) -- (-0.1,0.75);
%\draw[lightgray, line width=1.pt] (0,0) -- (0.1,0.75);

\draw[lightgray, line width=1.pt] (1,0) -- (0.7,0.75);
\draw[lightgray, line width=1.pt] (1,0) -- (1.3,0.75);
%\draw[lightgray, line width=1.pt] (1,0) -- (0.9,0.75);
%\draw[lightgray, line width=1.pt] (1,0) -- (1.1,0.75);

\draw[lightgray, line width=2.pt] (-0.5,0.75) -- (1.5,0.75);
\draw[fill=black] (0,0) circle (.03cm);
%\draw[fill=lightgray] (1,0.21) circle (.2cm);
\draw[fill=black] (1,0) circle (.03cm);
\node[scale=1] at (0,-.15) {$k_{12}$};
\node[scale=1] at (1,-.15) {$k_{34}$};
\node[scale=1] at (0.5,-.12) {$s$};

\draw[red3, line width=1.5pt,opacity=.85] (1.25,-.3) -- (1.25,0.3);
\end{tikzpicture}
}+
\raisebox{-33pt}{
\begin{tikzpicture}[line width=1. pt, scale=2]
\draw[fill=black] (0,0) -- (1,0);
\draw[lightgray, line width=1.pt] (0,0) -- (-0.3,0.75);
\draw[lightgray, line width=1.pt] (0,0) -- (0.3,0.75);
%\draw[lightgray, line width=1.pt] (0,0) -- (-0.1,0.75);
%\draw[lightgray, line width=1.pt] (0,0) -- (0.1,0.75);

\draw[lightgray, line width=1.pt] (1,0) -- (0.7,0.75);
\draw[lightgray, line width=1.pt] (1,0) -- (1.3,0.75);
%\draw[lightgray, line width=1.pt] (1,0) -- (0.9,0.75);
%\draw[lightgray, line width=1.pt] (1,0) -- (1.1,0.75);

\draw[lightgray, line width=2.pt] (-0.5,0.75) -- (1.5,0.75);
\draw[fill=black] (0,0) circle (.03cm);
\draw[fill=white] (0,0) circle (.03cm);
%\draw[fill=lightgray] (1,0.21) circle (.2cm);
\draw[fill=black] (1,0) circle (.03cm);
\draw[fill=white] (1,0) circle (.03cm);
\node[scale=1] at (0,-.15) {$-k_{12}$};
\node[scale=1] at (1,-.15) {$-k_{34}$};
\node[scale=1] at (0.5,-.12) {$s$};

\draw[red3, line width=1.5pt,opacity=.85] (-.25,-.3) -- (-0.25,0.3);
\end{tikzpicture}
}+
\raisebox{-33pt}{
\begin{tikzpicture}[line width=1. pt, scale=2]

\draw[fill=black,dashed] (0,0) -- (1,0);
\draw[lightgray, line width=1.pt] (0,0) -- (-0.3,0.75);
\draw[lightgray, line width=1.pt] (0,0) -- (0.15,0.75);
%\draw[lightgray, line width=1.pt] (0,0) -- (-0.1,0.75);
%\draw[lightgray, line width=1.pt] (0,0) -- (0.1,0.75);

\draw[red3, line width=1.pt,opacity=.65] (0,0) -- (0.45,0.75);
\draw[red3, line width=1.pt,opacity=.65] (1,0) -- (0.55,0.75);

\draw[lightgray, line width=1.pt] (1,0) -- (0.85,0.75);
\draw[lightgray, line width=1.pt] (1,0) -- (1.3,0.75);
%\draw[lightgray, line width=1.pt] (1,0) -- (0.9,0.75);
%\draw[lightgray, line width=1.pt] (1,0) -- (1.1,0.75);

\draw[lightgray, line width=2.pt] (-0.5,0.75) -- (1.5,0.75);
\draw[fill=black] (0,0) circle (.03cm);
%\draw[fill=lightgray] (1,0.21) circle (.2cm);
\draw[fill=black] (1,0) circle (.03cm);
\draw[fill=white] (1,0) circle (.03cm);
\node[scale=1] at (0,-.15) {$k_{12}$};
\node[scale=1] at (1,-.15) {$-k_{34}$};

\draw[red3, line width=1.5pt,opacity=.85] (.5,-.3) -- (.5,0.3);
%\node[scale=1] at (0.53,-.12) {$s_{12}$};
\node[scale=1] at (0.7,.3) {${\color{red3} s}$};
\node[scale=1] at (0.28,.3) {${\color{red3} s}$};

\end{tikzpicture}
} = 0\,.
%\label{eq:2sitepicture}
\nonumber
\eeq
This corresponds to partitioning the graph 
into two sets of vertices, which we color black and white. The $\bullet$ vertices are treated in the usual way, while the energies of $\circ$ vertices must be flipped. If a cut through the graph intersects an internal line, that line is replaced with two lines to the boundary each representing a sum of two terms, 
carrying $\mp$ the energy of the internal line.
Note that we have also included two trivial ``cuts" where the cut runs either to the left or right of the entire graph. These cuts correspond to the left-hand side of the formula~\eqref{eq:2siteformula}.  The diagrammatic version of the cutting rules is then the statement that the sum of all cuts, including the trivial cuts, vanishes.

\vskip4pt
In the remainder of the paper, we will only require expressions for the cuts of graphs involving single exchanges at tree level, as described by the cutting rule in~\eqref{eq:2siteformula}.
It is possible, however, to produce identities for arbitrary graphs following the same steps. We elaborate on these more general situations in Section~\ref{sec:cutsummary} and~Appendix~\ref{app:cutproof}.

\subsubsection{Including Spin}
\label{sec:flatspacespin}
So far, we have described the cutting rule in the simplest possible example involving only a single scalar field. However, many cases of practical interest involve fields with spin on both internal and external lines. Fortunately, the generalization to this situation is relatively straightforward. In cases with spinning external lines, everything proceeds as before, the only subtlety being that the lower-point shifted wavefunctions will now involve spinning external lines.

\vskip4pt
The generalization to the cutting of spinning internal lines is also simple, but requires a small amount of work. Essentially, the only new ingredient  is the bulk-to-bulk propagator for fields with spin. In the following, we will specialize to the case of massless fields.
This is partly for simplicity, but also because we expect that in the extension to de Sitter space, cutting will provide the most useful information for rational correlators, like those involving massless spinning particles in four dimensions. 

\vskip4pt 
For a massless spin-1 particle, the bulk-to-bulk propagator is most naturally expressed as a frequency integral in axial gauge (where $A_0=0$)~\cite{Raju:2011mp} 
\beq
{\cal G}_{ij}(\vec k;t,t') = -i\int_0^\infty\frac{\rd\omega}{2\pi}(\Pi_{1,1}^{\hskip 1pt\omega})_{ij}\frac{\big(e^{-i\omega t}-e^{i\omega t}\big)\big(e^{-i\omega t'}-e^{i\omega t'}\big)}{\omega^2-k^2+i\epsilon}\,,
\label{eq:spin1bulkbulkprop}
\eeq
where the tensor $(\Pi_{1,1}^{\hskip 1pt\omega})_{ij}$ is defined as
\beq
(\Pi_{1,1}^{\hskip 1pt\omega})_{ij} \equiv \delta_{ij}-\frac{k_ik_j}{\omega^2}.
\eeq
On shell (i.e.~when $\omega = k$), this tensor is transverse and traceless, but away from this point it is not, which accounts for the exchange of lower-helicity potential modes. In analogy to the scalar case, we can define the cut propagator as the sum of ${\cal G}_{ij}$ and its complex conjugate, which takes the form\footnote{In deriving this, we use the standard distributional identity
\be
\frac{1}{x+i\epsilon} - \frac{1}{x-i\epsilon} = -2\pi i \delta(x)\,,
\ee
which also lies at the heart of the S-matrix cutting formulae.
} 
%
%\begin{tcolorbox}[colframe=white,arc=0pt,colback=greyish2]
%\vspace{-4pt}
\beq
\widetilde{\cal G}_{ij}(\vec k;t,t') = -(\Pi_{1,1})_{ij}\frac{1}{2k}\big(e^{-ik t}-e^{ikt}\big)\big(e^{-ik t'}-e^{ikt'}\big)\,,
\label{eq:flatspin1cut}
\eeq
%\end{tcolorbox}
\noindent
where $(\Pi_{1,1})_{ij}$ is the transverse-traceless projector
\beq
(\Pi_{1,1})_{ij} \equiv \delta_{ij}-\frac{k_ik_j}{k^2}.
\label{eq:spin1ttproj}
\eeq
Note that~\eqref{eq:flatspin1cut} is identical to~\eqref{eq:flatcutprop}, except for the additional tensor factor carrying the polarization information. Consequently, the cutting rules involving massless spin-1 exchange will be essentially the same as those for scalar internal lines. We only have to contract the indices of the lower-point wavefunctons that the graph splits into with $ (\Pi_{1,1})_{ij} $. An interesting corollary to this is that only the highest-helicity propagating components of exchanges contribute to cuts, as we will see in some examples below.

\vskip4pt 
The analysis for a massless spin-2 particle proceeds similarly. Notice that in the spin-1 cut propagator~\eqref{eq:flatspin1cut} we are effectively putting the exchanged line on-shell, so that only the helicity-1 modes propagate. The same thing happens in the spin-2 case: only the helicity-2 part of the bulk-to-bulk propagator survives, so that we get (in axial gauge, where $h_{0\mu}=0$)
%
%\begin{tcolorbox}[colframe=white,arc=0pt,colback=greyish2]
%\vspace{-4pt}
\beq
\widetilde{\cal G}^{i_1i_2}_{j_1j_2}(\vec k;t,t') =-(\Pi_{2,2})^{i_1i_2}_{j_1j_2}\frac{1}{2k}\big(e^{-ik t}-e^{ikt}\big)\big(e^{-ikt'}-e^{ikt'}\big)\, ,
\label{eq:flatspin2cut}
\eeq
%\end{tcolorbox}
\noindent
which again differs from the scalar cut propagator only by the transverse-traceless spin-2 projector
\beq
(\Pi_{2,2})^{i_1i_2}_{j_1j_2} \equiv \pi^{(i_1}_{(j_1} \pi^{i_2)}_{j_2)}-\frac{1}{2}\pi^{i_1i_2}\pi_{j_1j_2}\,, \quad {\rm with}\quad \pi_{ij} \equiv  \delta_{ij}-\frac{k_ik_j}{k^2}.
\label{eq:spin2ttproj}
\eeq
When we cut diagrams involving internal spin-2 fields, we therefore just have to sum over the exchanged helicity-2 polarizations using the $\Pi_{2,2}$ projector. The generalization to higher-spin particles is straightforward, the only substantive difference is that higher-spin transverse-traceless projectors $\Pi_{\ell,\ell}$ appear multiplying the cut propagator. (See Appendix C of~\cite{Baumann:2019oyu} for explicit expressions for these projectors.)

\subsection*{Examples}
The simplest example where cutting a spinning internal line is important is the four-point function of external scalars exchanging an internal particle with spin. In this case, the cutting rule has the following diagrammatic expression
\begin{equation*}
\raisebox{-33pt}{
\begin{tikzpicture}[line width=1. pt, scale=2,
sines/.style={
        line width=0.75pt,
        line join=round, 
        draw=black, 
        decorate, 
        decoration={complete sines, number of sines=4, amplitude=4pt}
    }
]
\draw[lightgray, line width=1.pt] (0,0) -- (-0.3,0.75);
\draw[lightgray, line width=1.pt] (0,0) -- (0.3,0.75);

\draw[lightgray, line width=1.pt] (1,0) -- (0.7,0.75);
\draw[lightgray, line width=1.pt] (1,0) -- (1.3,0.75);

\draw[lightgray, line width=2.pt] (-0.5,0.75) -- (1.5,0.75);
\draw[white,postaction={sines}] (0,0) -- (1,0);
%\draw[fill=black] (0,0) -- (1,0);
\draw[fill=black] (0,0) circle (.03cm);
\draw[fill=black] (1,0) circle (.03cm);
%\node[scale=1] at (0,-.15) {$x_1$};
%\node[scale=1] at (1,-.15) {$x_2$};
%\node[scale=1] at (0.5,-.12) {$s_{12}$};
\draw[red3, line width=1.5pt,opacity=.85] (1.25,-.3) -- (1.25,0.3);
\end{tikzpicture}
}
\,+\,
\raisebox{-33pt}{
\begin{tikzpicture}[line width=1. pt, scale=2,
sines/.style={
        line width=0.75pt,
        line join=round, 
        draw=black, 
        decorate, 
        decoration={complete sines, number of sines=4, amplitude=4pt}
    }
]
\draw[lightgray, line width=1.pt] (0,0) -- (-0.3,0.75);
\draw[lightgray, line width=1.pt] (0,0) -- (0.3,0.75);

\draw[lightgray, line width=1.pt] (1,0) -- (0.7,0.75);
\draw[lightgray, line width=1.pt] (1,0) -- (1.3,0.75);

\draw[lightgray, line width=2.pt] (-0.5,0.75) -- (1.5,0.75);
\draw[white,postaction={sines}] (0,0) -- (1,0);
%\draw[fill=black] (0,0) -- (1,0);
\draw[fill=white] (0,0) circle (.03cm);
\draw[fill=white] (1,0) circle (.03cm);
%\node[scale=1] at (0,-.15) {$x_1$};
%\node[scale=1] at (1,-.15) {$x_2$};
%\node[scale=1] at (0.5,-.12) {$s_{12}$};
\draw[red3, line width=1.5pt,opacity=.85] (-.25,-.3) -- (-0.25,0.3);
\end{tikzpicture}
}
\,+\,
\raisebox{-33pt}{
\begin{tikzpicture}[line width=1. pt, scale=2,
sines/.style={
        line width=0.75pt,
        line join=round, 
        draw=black, 
        decorate, 
        decoration={complete sines, number of sines=4, amplitude=4pt}
            },
sines2/.style={
        line width=0.75pt,
        line join=round, 
        draw=red3, 
        decorate, 
        decoration={complete sines, number of sines=4, amplitude=4pt}
            }
]
\draw[lightgray, line width=1.pt] (0,0) -- (-0.3,0.75);
\draw[lightgray, line width=1.pt] (0,0) -- (0.15,0.75);

\draw[lightgray, line width=1.pt] (1,0) -- (0.85,0.75);
\draw[lightgray, line width=1.pt] (1,0) -- (1.3,0.75);

\draw[white,postaction={sines},dashed] (0,0) -- (1,0);
%\draw[fill=black] (0,0) -- (1,0);
%\node[scale=1] at (0,-.15) {$x_1$};
%\node[scale=1] at (1,-.15) {$x_2$};
%\node[scale=1] at (0.5,-.12) {$s_{12}$};
\draw[red3, line width=1.5pt,opacity=.85] (.5,-.3) -- (.5,0.3);
\draw[white, line width=1.pt,opacity=.65,postaction={sines2}] (0,0) -- (0.45,0.75);
\draw[white, line width=1.pt,opacity=.65,postaction={sines2}] (1,0) -- (0.55,0.75);
\draw[lightgray, line width=2.pt] (-0.5,0.75) -- (1.5,0.75);
\draw[fill=black] (0,0) circle (.03cm);
\draw[fill=white] (1,0) circle (.03cm);
\end{tikzpicture}
} = 0\,,
\end{equation*}
where we note the appearance of shifted three-point functions involving two scalars and one spinning field in the cut diagram.

\begin{itemize}
\item {\it Spin-1 exchange:} We first consider the wavefunction coefficient arising from the exchange of a massless spin-1 field, $\psi^{(J)}$. Since the interactions are Weyl invariant, this correlator is the same in flat space and in de Sitter. The cutting rule for the $s$-channel correlator is 
\beq
{\rm Disc}[\psi_4^{(J)}] = -2s\, \tl\psi_{\vp\vp J}^i (\Pi_{1,1})_{ij} \tl\psi_{J\vp\vp }^j  \equiv -2s\, \tl\psi_{\vp\vp J} \otimes \tl\psi_{J\vp\vp }\,,
\label{eq:flatspin1cuttingrule}
\eeq
where we have introduced the symbol $\otimes$ to denote the contraction with the polarization tensor.
From a direct calculation, one finds~\cite{Arkani-Hamed:2018kmz,Baumann:2019oyu}
\beq
\psi_4^{(J)} = -\frac{\tl\Pi_{1,1}}{EE_LE_R} +\frac{\tl\Pi_{1,0}}{E}\,,
\label{eq:flatspin1exc}
\eeq
where $\tl\Pi_{1,1}$ and $\tl\Pi_{1,0}$ arise from the sums over exchanged polarizations and are given by
\be
\label{eq:p11def}
\tl\Pi_{1,1} &\equiv   \alpha^i \pi_{ij} \beta^j = k_{12}k_{34}\frac{\alpha \beta}{s^2}- t^2+u^2\,, \\
\tl\Pi_{1,0} &\equiv \frac{\alpha \beta}{s^2}, \label{eq:p10def}
\ee
with 
\beq
\begin{aligned}
\vec\alpha &\equiv \vec k_1-\vec k_2\,,\quad \vec\beta \equiv \vec k_3-\vec k_4\,,\\ 
\alpha &\equiv k_1-k_2\,, \quad \beta \equiv k_3-k_4\, .
\end{aligned}
\label{equ:AlphaBetaDef}
\eeq
 (Note that $\alpha$ and $\beta$ are {\it not} the magnitudes of $\vec\alpha$ and $\vec\beta$.)

From the explicit expression~\eqref{eq:flatspin1exc}, it is straightforward to compute the cut: 
\beq
{\rm Disc}[\psi_4^{(J)}] = \psi_4^{(J)}(k_{12},k_{34})+\psi_4^{(J)}(-k_{12},-k_{34})= \frac{2 s\, \tl\Pi_{1,1}}{(k_{12}^2-s^2)(k_{34}^2-s^2)}\,.
\label{eq:spin1cut}
\eeq
Notice that only the highest-helicity part of the exchange contributes to the cut, the helicity-0 potential part of the interaction has dropped out due to its singularity structure. This is of course how it must be, because on the right-hand side of the cutting rule~\eqref{eq:flatspin1cuttingrule} we only sum over the highest-helicity polarizations.

To independently compute the right-hand side of~\eqref{eq:flatspin1cuttingrule}, we need the three-point function involving two scalars and a massless spin-1 field:
\beq
\psi_{\varphi\varphi J}(k_1,k_2,s) = \frac{i\vec \alpha\cdot\vec\xi}{k_{12}+s}\,.
\eeq
We then calculate the shifted wavefunction as in~\eqref{eq:shiftedWF}, shifting the spin-1 line,
and contract it with the same object permuted as $\{1,2\}\mapsto\{3,4\}$ using the projector~\eqref{eq:spin1ttproj}.
From the definition~\eqref{eq:p11def}, we find 
\beq
-2s\, \tl\psi_{\vp\vp J} \otimes \tl\psi_{J\vp\vp } =\frac{2 s\,\tl\Pi_{1,1}}{(k_{12}^2-s^2)(k_{34}^2-s^2)}\,,
\eeq
which is identical to~\eqref{eq:spin1cut}, as expected.

\item {\it Spin-2 exchange:} The case of spin-2 exchange is very similar. The relevant cut is now 
\beq
{\rm Disc}[\psi_4^{(T)}] =
-2s\,   \tl\psi_{\vp\vp T}\otimes \tl\psi_{T\vp\vp } \,,
\label{eq:flatspin2cuttingrule}
\eeq
where the three-point function of one graviton and two scalars is given by
\beq
\psi_{\vp\vp T}(k_1,k_2,s)= \frac{\big(\vec \alpha \cdot\vec\xi \hskip 3pt\big)^2}{2(k_{12}+s)}\,.
\eeq
Computing the shifted wavefunction and contracting with the spin-2 transverse-traceless projector~\eqref{eq:spin2ttproj}, we obtain
\beq
-2s\,   \tl\psi_{\vp\vp T}\otimes \tl\psi_{T\vp\vp }  = - \frac{ s\,\tl\Pi_{2,2}}{3(k_{12}^2-s^2)(k_{34}^2-s^2)}\,,
\label{eq:spin2rhscut}
\eeq
where we have defined
\beq
\tl\Pi_{2,2} \equiv \frac{3}{2}\alpha^i\alpha^j(\Pi_{2,2})_{ij}^{kl}\beta_k\beta_l \,.
 \label{eq:pi22polsum}
\eeq
We would like to check this expression for the cut against the full answer. The correlator in flat space can be obtained in various ways, where the explicit expression is\footnote{This correlator is actually an example where the full answer can be bootstrapped from physical principles. The entire first line of~\eqref{eq:flatspacespin2exc} is fixed by the requirements that it has the correct residues on its partial energy and total energy singularities (as we will explore more fully in the following). In particular, the polarization sums appear in the appropriate combination to reduce to a Legendre polynomial as $E\to0$ (see Appendix~\ref{app:pols}), which is the signature of massless particle exchange in the scattering amplitude. Aside from reproducing these limits correctly, the second line of~\eqref{eq:flatspacespin2exc} is fixed by demanding that the correlator vanishes when any of its external momenta are taken to be soft, which is a consequence of the shift symmetry of the interactions. 
Putting these things together, and demanding the absence of any other singularities in the physical region, leads to~\eqref{eq:flatspacespin2exc}. The resulting expression can also be verified by a direct bulk calculation.} 
\begin{tcolorbox}[colframe=white,arc=0pt,colback=greyish2]
\beq
\begin{aligned}
\label{eq:flatspacespin2exc}
\psi^{(T)}_4	=~&\frac{1}{6EE_LE_R}\tl\Pi_{2,2}-\frac{1}{6E}\tl\Pi_{2,1}+\frac{1}{6E}\tl\Pi_{2,0}\\
	& -\frac{k_{12}k_{34}+s^2}{6E} -\frac{1}{4}\left(\frac{k_{12}\beta^2}{s^2}+\frac{k_{34} \alpha^2}{s^2}\right)+ \frac{1}{4}E\,.
\end{aligned}
\eeq 
\end{tcolorbox}
In addition to~\eqref{eq:pi22polsum}, we have defined the polarization sums $\tl\Pi_{2,1}$ and $\tl\Pi_{2,0}$ by (the motivations for these definitions can be found in Appendix~\ref{app:pols})
\begin{align}
\tl\Pi_{2,1} &\equiv -\frac{s^2}{k_{12}k_{34}} \frac{3}{2 s^2}\alpha_i \alpha_j (\Pi_{2,1})^{ij}_{lm}  \beta^l \beta^m=3  \alpha \beta \frac{\alpha^i \pi_{ij}  \beta^j}{s^2}\, , \\[4pt]
\tl\Pi_{2,0} &\equiv \frac{E_LE_R-sE}{4}\left(1-3\frac{\alpha^2}{s^2} \right)\left(1-3 \frac{ \beta^2}{s^2}\right) .
\end{align}
Given the expression~\eqref{eq:flatspacespin2exc}, it is straightforward to check that it has the expected cut
\beq
{\rm Disc}[\psi_4^{(T)}] = \psi_4^{(T)}(k_{12},k_{34})+\psi_4^{(T)}(-k_{12},-k_{34}) = -\frac{ s\,\tl\Pi_{2,2}}{3(k_{12}^2-s^2)(k_{34}^2-s^2)}\,,
\eeq
which indeed agrees with~\eqref{eq:spin2rhscut}. Note that, like in the spin-1 example, only the highest-helicity part of the exchange (proportional to $\tl\Pi_{2,2}$) contributes to the cut of the diagram, as expected from the form of the cut propagator~\eqref{eq:flatspin2cut}.

\end{itemize}

\subsection{Generalization to de Sitter Space}
\label{sec:dscutting}

Our discussion of cutting correlators in flat space has only relied on very general properties of the Green's functions used to compute the wavefunction, so the essential features generalize almost immediately to other spacetimes. Our particular interest will be in de Sitter backgrounds because the relevant propagators take a simplified form, but the extension to even more general situations is straightforward.

\vskip4pt
Recall that the line element of de Sitter space in the flat slicing is given by~\eqref{eq:dsmetric}.
The bulk dynamics of fields in de Sitter space give rise to correlations between fluctuations on the late-time ($\eta = 0$) surface. In the conventional inflationary picture, this is the reheating surface where these perturbations serve as the initial conditions for the subsequent evolution of the universe. The goal is to understand the structure of these fluctuations without tracking their detailed time evolution through the de Sitter spacetime. Cuts of correlators give us some partial information about this time evolution.
An interesting feature of the cutting rules in de Sitter space is that they apply equally well in cases where the
symmetries of the background spacetime are broken by the interaction vertices. These symmetry-breaking interactions are phenomenologically relevant in inflationary models with large non-Gaussianities.

\vskip4pt
 Only two generalizations of the previous discussion are needed to import our results to de Sitter space. First, the bulk-to-bulk propagators are slightly different, so we have to re-derive the cut propagators. Second, the bulk-to-boundary propagators are no longer just exponentials, so we have to be careful about the precise analytic continuation that relates $\bar\psi$ to $\psi$.

\subsubsection*{Scalar propagators}

We begin by collecting the necessary ingredients for cutting scalar lines in de Sitter. The bulk-to-boundary propagator for a general massive scalar is~\eqref{eq:massivescalarbulkboundary}. 
In flat space, the conjugate wavefunction coefficients $\bar\psi$ are related to the complex conjugate of the ordinary wavefunction coefficients, but with the external energies analytically continued to negative values.  To apply a similar analytic continuation in de Sitter space, we note
that the complex conjugate of the bulk-to-boundary propagator satisfies~\cite{Goodhew:2020hob}
\beq
{\cal K}_\nu^*(k,\eta) = {\cal K}_\nu(e^{-i\pi}k,\eta)\,.
\label{eq:analticcontin}
\eeq
This relation instructs us to analytically continue by rotating the energies clockwise
 in the complex plane to negative arguments. 
 In flat space, the details of this analytic continuation did not matter because none of the answers had branch cuts, but some correlators in de Sitter space do, so we should navigate the branch cuts according to this prescription. In the following, we will typically leave this implicit, but negative energies should be understood in this sense.

\vskip 4pt
The bulk-to-bulk propagator for a massive scalar is given by~\eqref{eq:massivebulkbulkprop}.
Using this expression, it is easy to show that the cut propagator $\widetilde{\cal G}_\nu \equiv {\cal G}_\nu +{\cal G}^*_\nu$ can be written in terms of the bulk-to-boundary propagator as\footnote{In this expression, we have pulled out an overall factor of $-1$ on the right-hand side for uniformity with the other cut propagator expressions that we have presented. When evaluating this expression for explicit mass values, for example for $\nu = 3/2$ (appropriate for massless fields), this sign can be compensated by choosing the branch of the square root that appears appropriately. The result then recovers the previously shown expressions.} 
%
%\begin{tcolorbox}[colframe=white,arc=0pt,colback=greyish2]
\beq
\widetilde {\cal G}_\nu(k;\eta, \eta') =  -P_{\nu}(k)\Big({\cal K}_\nu(-k,\eta)-{\cal K}_\nu(k,\eta)\Big)\Big({\cal K}_\nu(-k,\eta')-
    {\cal K}_\nu(k,\eta')\Big)\,,
    \label{eq:dScutprop}
\eeq
%\end{tcolorbox}
\noindent
where  we have introduced the power spectrum of the exchanged field
\beq
P_{\nu}(k) \equiv \frac{\pi}{4}\eta_*^3H^{(1)}_{\nu}(-k\eta_*)H^{(2)}_{\nu}(-k\eta_*)\, .
\eeq
The cut propagator~\eqref{eq:dScutprop} is structurally the same as the formula~\eqref{eq:flatcutasbbprop} for the flat-space cut propagator. In particular, it also factorizes. As a result, the cutting formulas in de Sitter are essentially identical---we just have to replace all flat-space objects with their de Sitter counterparts, to obtain
\begin{tcolorbox}[colframe=white,arc=0pt,colback=greyish2]
\vspace{-4pt}
\beq
\psi_4(k_{12},k_{34})+\psi^*_4(-k_{12},-k_{34}) = -P_{\nu}(s)\,\tl\psi_3(k_{12}\mp s)\,\tl{\psi}_3(-k_{34} \mp s)\,,
\label{eq:2siteformuladS}
\eeq
\end{tcolorbox}
\noindent
where the shifted wavefunctions are defined as in~\eqref{eq:shifted3pt} and the analytic continuation to negative arguments is clockwise in the complex plane as in~\eqref{eq:analticcontin}.

\subsubsection*{Spinning propagators}
It is also straightforward to update our discussion of spinning internal lines to cover the de Sitter case. First, notice that the bulk-to-boundary propagators of the (transverse parts of) spinning fields are just given by transverse-traceless projectors times~\eqref{eq:massivescalarbulkboundary} at special values of $\nu$, so we can do the same analytic continuation of the external energies as for the scalar case.

\vskip4pt
A massless spin-1 field has a quadratic action that is Weyl invariant (in $D=4$), so its propagators in de Sitter are identical to those in flat space. This implies that we can use the cut propagator~\eqref{eq:flatspin1cut} also in de Sitter space.

\vskip4pt
The massless spin-2 propagator does require a new calculation. We can write the bulk-to-bulk propagator in axial gauge as~\cite{Raju:2011mp}
\beq
    {\cal G}^{i_1i_2}_{j_1j_2}(\vec k;\eta,\eta')=-i\int_0^{\infty}\frac{\rd\omega^2}{2}({\Pi}_{2,2}^{\hskip 1pt\omega})^{i_1i_2}_{j_1j_2}\frac{(\eta\eta')^{3/2}J_{3/2}(\omega\eta)J_{3/2}(\omega\eta')}{\omega^2-k^2+i\epsilon}\,,
\eeq
where now Bessel functions instead of plane waves appear in the numerator and $({\Pi}_{2,2}^{\hskip 1pt\omega})^{i_1i_2}_{j_1j_2}$ is the tensor~\eqref{eq:spin2ttproj} with $k^2\mapsto \omega^2$. By adding this propagator with its complex conjugate, we can extract the cut propagator as before
%
%\begin{tcolorbox}[colframe=white,arc=0pt,colback=greyish2]
%\vspace{-0.5cm}
\beq
\widetilde{\cal G}_{j_1j_2}^{i_1i_2}(\vec k;\eta,\eta')=-\frac{({\Pi}_{2,2})^{i_1i_2}_{j_1j_2}}{2k^3}\Big({\cal K}_{3/2}(-k,\eta)-{\cal K}_{3/2}(k,\eta)\Big)\Big({\cal K}_{3/2}(-k,\eta')-{\cal K}_{3/2}(k,\eta')\Big)\,.
\label{eq:dsmasslessspin2cutprop}
\eeq
%\end{tcolorbox}
%
Note that the result could be written in terms of the massless scalar bulk-to-boundary propagator,
\beq
{\cal K}_{3/2}(k,\eta) = (1-ik\eta)e^{ik\eta}\,,
\eeq
because the mode function of the massless graviton is the same as that of a massless scalar field. The generalization to higher spin is straightforward: the cut propagator is the transverse-traceless projector for spin $\ell$ multiplied by the scalar cut propagator~\eqref{eq:dScutprop}, with $\nu =  \ell-1/2$.

\vskip4pt
We have now assembled all the ingredients necessary to compute cuts of correlators in de Sitter space. All of the flat-space combinatorics go through unchanged because they only relied on the relation between the bulk-to-bulk, anti-bulk-to-bulk and cut propagators: ${\cal G}+\bar{\cal G} = \tl{\cal G}$.

\subsubsection*{Examples}

Our ultimate goal is to use the cuts of correlators as additional constraints to aid in bootstrapping them from the boundary, but it is useful to first verify that a couple examples in de Sitter space have the expected cuts.

\begin{itemize}

\item {\it Conformally coupled scalar:} Perhaps the simplest example is the four-point function arising from a $\phi^3$ interaction involving conformally coupled scalars. In this case, the cut of the four-point wavefunction is given by
\beq
{\rm Disc}[\psi_{\varphi^4}] = -2s\,\widetilde\psi_{\varphi^3}(k_{12}\mp s)\,\widetilde\psi_{\varphi^3}(-k_{34}\mp s)\,.
\label{equ:SIL}
\eeq
We can verify this formula using the explicit expressions for the three and four-point functions computed in de Sitter\footnote{The expression for $\psi_{\vp^4}$ differs slightly from the one in~\cite{Arkani-Hamed:2018kmz}, which is the corresponding correlation function. The wavefunction coefficient and correlator differ by a homogeneous term which affects the sign and coefficient of the $\pi^2$ term. See~\cite{Arkani-Hamed:2015bza,Hillman:2019wgh} for a derivation of the wavefunction expression.} 
\begin{align}
\psi_{\varphi^4}&= \frac{1}{2s}\left[{\rm Li}_2\left(\frac{k_{12}-s}{E}\right)+{\rm Li}_2\left(\frac{k_{34}-s}{E}\right)+\log\left(\frac{k_{12}+s}{E}\right)\log\left(\frac{k_{34}+s}{E}\right)-\frac{\pi^2}{6}\right], \label{equ:psi4}\\
\psi_{\varphi^3}&= i\log(K/\mu)-\frac{\pi}{2}\,.
\label{equ:psi3ds}
\end{align}
Notice that there is already some nontrivial structure to the $\psi_{\vp^3}$ wavefunction coefficient in de Sitter space. Aside from the non-local part being purely imaginary (as can be verified by a direct calculation~\cite{Arkani-Hamed:2015bza}), there is a precise (real) local term required in order for the answer to have a vanishing cut:
\beq
\psi_{\varphi^3}(K)+\Big[\psi_{\varphi^3}(e^{-i\pi}K)\Big]^* =  i\log(K/\mu) -  i\log(e^{i\pi}K/\mu) - \pi = 0\,.
\label{eq:cphi33pt}
\eeq
A similar example involving the three-point function of massless scalars was discussed in~\cite{Goodhew:2020hob}.

In order to verify the cutting rule in~\eqref{equ:SIL},
we compute its left-hand side using Euler's identity for the dilog
\beq
{\rm Li}_2(z) = -{\rm Li}_2(1-z)-\log(1-z)\log(z)+\frac{\pi^2}{6}\,,
\eeq
which, after some algebra, leads to
\beq
{\rm Disc}[\psi_{\varphi^4}]  = \psi_{\varphi^4}(k_{12},k_{34})+\psi^*_{\varphi^4}(-k_{12},-k_{34}) = \frac{1}{2s}\log\left(\frac{k_{12}-s}{k_{12}+s}\right)\log\left(\frac{k_{34}-s}{k_{34}+s}\right) .
\label{eq:rhsconfcoupled4pt}
\eeq
This is exactly minus the product of shifted three-point functions, accounting for the fact that we have to flip the energies in one of them.

There is a parallel path to checking the cut of the conformally coupled four-point function, which will be conceptually useful later on. The idea is to express the de Sitter answer by operating on the flat-space correlator~\eqref{eq:flatspacescalar4pt} in some way---in this case the relevant operation is integration with respect to energies~\cite{Arkani-Hamed:2015bza}
\beq
\psi_{\varphi^4} = -\int_{k_{12}}^\infty\rd \tilde k_{12}\int_{k_{34}}^\infty\rd \tilde k_{34}\,\frac{1}{(\tilde k_{12}+\tilde k_{34})(\tilde k_{12}+s)(\tilde k_{34}+s)}\,.
\label{eq:cc4ptint}
\eeq
It is most convenient to compute the cut in this case using the formula~\eqref{eq:antiandregularbbprop} which implies that the cut is given by subtracting~\eqref{eq:cc4ptint} with the sign of $s$ flipped. This leads to 
\beq
{\rm Disc}[\psi_{\varphi^4}] =  2s\int_{k_{12}}^\infty\rd \tilde k_{12}\,\frac{1}{(\tilde k_{12}^2-s^2)}\int_{k_{34}}^\infty\rd \tilde k_{34}\,\frac{1}{(\tilde k_{34}^2-s^2)}\,,
\label{eq:integratecutflat4pt}
\eeq
which integrates to~\eqref{eq:rhsconfcoupled4pt}, as expected. Notice that in this case the de Sitter space cut follows straightforwardly from the cut of the integrand, which is just the flat-space wavefunction. We explore this theme more fully in Section~\ref{sec:lifting}.

\item {\it Massless spin-2 exchange:} As a simple example involving a cut of an internal line with spin, we consider the four-point correlator arising from the exchange of a massless spin-2 field between conformally coupled scalar fields. In this case, the relevant three-point data is the wavefunction coefficient for the interaction between two scalars and a massless spin-2
\begin{align}
\psi_{\varphi^2 T} &=  \frac{k_{12}+2s}{2(k_{12}+s)^2} \big(\vec \alpha \cdot\vec\xi \hskip 3pt\big)^2\,,
\end{align}
so that the wavefunction with a shifted graviton leg is
\be
\widetilde \psi_{\varphi^2 T} =  - \frac{1}{(k_{12}^2-s^2)^2}  \big(\vec \alpha \cdot\vec\xi \hskip 3pt\big)^2\,.
\ee
This allows us to construct the cut 
\be
\begin{aligned}
{\rm Disc}[\psi_4^{(T)}] = \psi_4^{(T)}(k_{12},k_{34})+\psi_4^{(T)}(- k_{12},-k_{34}) &= -2s^3 \widetilde\psi_{\varphi^2 T}\otimes \widetilde\psi_{T\varphi^2}\\
&= -2 s^3 \frac{2\tl\Pi_{2,2}}{3(k_{12}^2-s^2)^2(k_{34}^2-s^2)^2}\,.
\end{aligned}
\label{eq:spin2exccut}
\ee
On the other hand, the full answer is~\cite{Arkani-Hamed:2018kmz, Baumann:2019oyu} 
\beq
\psi^{(T)}_4 = -\frac{sE+E_LE_R}{3E^3E_L^2E_R^2} \tl\Pi_{2,2} + \frac{1}{3E^3}\tl\Pi_{2,1} - \frac{1}{3E^3 } \tl \Pi_{2,0}\,.
\label{eq:dscc4ptspin2}
\eeq
Note that the result of an explicit bulk calculation contains additional contact solution pieces, as in the second line of~\eqref{eq:flatspacespin2exc}, which are de Sitter invariant by themselves. This freedom to add invariant contact solutions reflects an inherent ambiguity in exchange correlators. However, the helicity-2 part of~\eqref{eq:dscc4ptspin2}, involving partial energy singularities, is an unambiguous signature of a massless spin-2 exchange. The only ambiguity is in its extension to a fully de Sitter-invariant correlator, which is reflecting the freedom to add contact interactions in the bulk. Taking the cut of~\eqref{eq:dscc4ptspin2}, it is straightforward to see that we reproduce~\eqref{eq:spin2exccut}. 

\end{itemize}

\subsection{Summary: Cutting Rules}
\label{sec:cutsummary}
In the remainder of the text, we will only require the simplest cutting formulas that have already been introduced, involving only a single internal line. However, it is possible to generalize the cutting rules to arbitrary graphs. Here, we just state the procedure without justification, and refer the interested reader to Appendix~\ref{app:cutproof} for the details.

\vskip4pt
We have seen that the various cuts separate the graph into two pieces, where on one side we compute the ordinary wavefunction, and on the other side we compute the conjugate wavefunction (or the complex conjugated wavefunction at negative energies), both shifted by the internal energy of the cut line. This generalizes readily to a general graph. The procedure is the following: 
\begin{itemize}
\item Pick a direction to move through the graph. This is essentially a (partial) ordering of the vertices of the graph.
\vspace{-3pt}
\item Consider cuts that separate the graph into two parts, consistent with the partial order.
\vspace{-3pt}
\item  On one side of the cut, compute the wavefunction associated to the relevant graph, but use the cut propagator for the cut line.
\vspace{-3pt}
\item  On the other side of the cut, compute the wavefunction at negative energy, again using the  cut propagator
for internal lines and then take its complex conjugate. Alternatively, compute the conjugate wavefunction. 
%\vspace{-.3cm}
\end{itemize}
The sum over all cuts of this type vanishes, leading to the following schematic identity
\begin{tcolorbox}[colframe=white,arc=0pt,colback=greyish2]
%\vspace{-6pt}
\beq
\psi_n(X)+ \psi_n^*(-X) = -\sum_{\rm cuts} \psi_n\,,
\eeq
\end{tcolorbox}
\noindent
where we have separated out on the left-hand side the ``cuts" that pass either all the way to the left or to the right of the graph, and $X$ stands for all the external energies of the wavefunction.

\newpage
\section{Gluing: Singularities and Recursion} \label{sec:gluing}
The structure of cosmological correlation functions is in large part controlled by their singularities. In fact, in many cases the behavior near singularities is sufficient information to completely reconstruct the full correlator at tree level. In this section, we will describe how to construct higher-point rational wavefunction coefficients recursively by ``gluing" together lower-point building blocks. 
If all of a wavefunction's singularities are poles, the energy variables of the correlator can be deformed into the complex plane
and the physical correlator can be expressed as a sum over the residues of its singularities~\cite{Arkani-Hamed:2017fdk}. This is philosophically similar to the BCFW construction of scattering amplitudes~\cite{Britto:2005fq}. To write such a recursive formula, we need to know both the locations and residues of singularities of cosmological correlators. Fortunately, as we reviewed in Section~\ref{sec:singularities}, this information is well-understood at tree level: correlators have singularities when the energies of subgraphs add up to zero, and their residues involve scattering amplitudes and shifted correlators.

\subsection{A Recursion Relation}

The general strategy is to promote the correlator of interest to a complex function of its kinematic arguments, which can then be fixed by its singularities using the power of complex analysis.
At the kinematic level, an $n$-point wavefunction coefficient is a function of $n$ spatial momenta, $\vec k_a$, which are subject to the constraint of (spatial) momentum conservation. The lengths of these momenta (the ``energies") are importantly {\it not} required to add up to zero, in contrast to the situation for scattering amplitudes. Despite the fact that the wavefunction is fundamentally a function of {\it momenta}, the singularities of the wavefunction always appear at certain loci in {\it energy} space. Hence, it is most convenient to deform the energies into the complex plane. The simplest family of such deformations is to extend the internal and external energy variables by a single complex parameter, $z$, as~\cite{Arkani-Hamed:2017fdk}
\beq
\begin{aligned}
k_a& \mapsto k_a + c_a z\,,\\
|\vec k_{a} + \vec k_{b} | &\mapsto |\vec k_{a} + \vec k_{b} |  + d_{ab} z\,,
\end{aligned}
\label{eq:energydefs}
\eeq
where $c_a$ and $d_{ab}$ are arbitrary numerical coefficients parameterizing the fact that we can do a different deformation for each energy variable. Since the correlators do not conserve energy, there is no constraint on the deformation, in stark contrast to the situation for amplitudes, where it is important for the complex deformation to be consistent with momentum conservation. Despite the simplicity of thinking of the correlator as a function of its energy arguments and deforming them into the complex plane, this type of energy deformation is quite subtle.  Fundamentally, the energies are not  independent kinematic variables, but are the norms of spatial momenta. However, a generic deformation~\eqref{eq:energydefs} cannot be induced by such a deformation of the momenta. Instead, these complex deformations should be thought of as a formal device to extract the singularity structure of the wavefunction coefficients.\footnote{This requires us to separate the part of the correlator that depends on momenta---which will enter through the residues---from the singularity structure that depends only on the energies. This split is slightly ambiguous, but the problem is sufficiently constrained that the ambiguity will be restricted to terms that are regular in all kinematic limits. These leftover terms are somewhat analogous to undetermined rational terms that can appear in constructions based on generalized unitarity. }

\begin{figure}[t!]
   \centering
            \includegraphics[width=.6\textwidth]{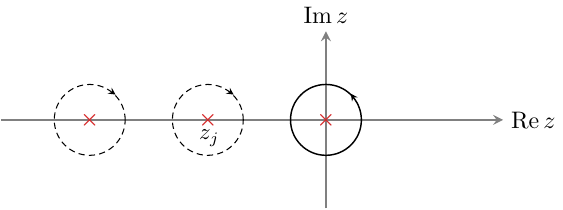}
   \caption{Schematic of the contour deformation used to derive~\eqref{eq:residue}. The solid contour centered around $z=0$, whose residue is the original wavefunction coefficient, $\psi(0)$, can be deformed into the dashed contour to write the wavefunction as a sum over the residues of the other poles of $\psi(z)$, located at $z_j$. }
  \label{fig:residue}
\end{figure}

\vskip4pt
Deformations of the form~\eqref{eq:energydefs} promote the wavefunction coefficient to a complex function $\psi(z)$. Our interest is in the physical wavefunction coefficient, which is $\psi(z=0)$. Using Cauchy's integral formula, we can write this as
\beq
\psi(0) = \frac{1}{2\pi i}\underset{{\cal C}}{\oint}\rd z\, \frac{\psi(z)}{z}\,,
\eeq
where the integration is along a small circle ${\cal C}$ centered around $z=0$ (see Fig.~\ref{fig:residue}). {\it If all the singularities of $\psi(z)$ are poles}, then we can deform the integration contour and write
 the integral as a sum of residues 
\beq
\psi(0) = -\sum_{j} \underset{z=z_j}{\rm Res} \left(\frac{\psi(z)}{z}\right) + B_\infty\,,
\label{eq:residue}
\eeq
where $z_j$ are the poles of the function $\psi(z)$---which are related to the total energy and partial energy singularities of the wavefunction---and $B_\infty$ is a possible  boundary contribution from $z\to \infty$. In cases where the pole at infinity is absent, 
we can reconstruct correlators completely from knowledge of their residues. Typically, the vanishing of the pole at infinity follows from $\psi(z)$ going to zero as $z\to \infty$. Theories for which this happens are the analogues of BCFW-constructible theories.

\vskip4pt
In flat space, all of the singularities are simple poles, so that we can apply the formula~\eqref{eq:residue} directly.
In cosmological spacetimes, the situation is slightly more complicated. For our purposes, the primary subtlety is that correlators in de Sitter space often have a series of higher-order poles, so that we need to know the Laurent expansion of a correlator in the vicinity of its singularities
in order to apply the formula~\eqref{eq:residue}. As described in Appendix~\ref{app:Residues}, the residue at a pole of order $n$ is related to the Laurent expansion of $\psi$ through
\beq 
\label{eq:GenResidue}
\underset{z=z_j}{\rm Res} \left(\frac{\psi(z)}{z}\right)= - \sum_{l=1}^{n} \frac{R^{(l)}(K_a)}{(-z_j)^l} \, ,
\eeq
where $K_a \equiv \{k_a,|\vec{k}_a+\vec{k}_b|\}$ collectively denotes the energy variables that the correlator depends on and the functions $R^{(l)}$ are the coefficients of the singular terms appearing in the Laurent expansion of $\psi(z)$ around the pole, i.e.
\beq
\lim_{z\to z_j} \psi(z) \vphantom{\frac{1}{x}} = \sum_{l=1}^{n} \frac{R^{(l)}(K_a)}{(z-z_j)^l} ~+~{\rm finite~terms}\, .
\eeq
This implies that we need to know all singular terms of the Laurent series of $\psi(z)$ around the singularities. 
As we will see, the subleading partial energy coefficients in de Sitter space can be inferred from the cutting rule~\cite{Jazayeri:2021fvk}.

\subsection{Energy Deformations in Flat Space}

The recursive procedure is simplest to illustrate through examples. We begin by considering some examples in flat space, 
which are simpler to construct via recursion than de Sitter correlators because they only have simple poles.
Moreover, as we will see in Section~\ref{sec:lifting}, in many cases it is possible to lift a correlation function computed in flat space to its de Sitter version by acting with a set of transmutation operations. Therefore, we can often first construct a correlator in flat space---taking advantage of the simpler singularity structure---and then transform it to get the correlator of interest in cosmology.

\paragraph{Four-point $\boldsymbol{\phi^3}$ correlator}
As our first example, we consider the four-point correlator in $\phi^3$ theory in flat space---the $s$-channel part of the correlator was given by~\eqref{eq:flatspacescalar4pt}. Famously, the corresponding scattering amplitude is {\it not} BCFW-constructible, so it might seem somewhat surprising that the correlator can be constructed via recursion. The important difference is that we are using information about the total energy singularity of the correlator, for which the scattering amplitude is an {\it input}. In effect, this is fixing the (absence of a) contact interaction in the theory, and so the corresponding correlator can be generated recursively.

\vskip4pt
In this case, it is sufficient to deform only one of the external energies. For example, we can take
\beq
k_1 \mapsto k_1+z\,.
\label{eq:phi3deformation}
\eeq
Since the correlator does not conserve energy, we do not have to compensate this shift by deforming any of the other variables.
The singularities of the deformed tree-level wavefunction, $\psi_4(z)$ can be inferred from the possible singularities of the undeformed wavefunction: it can have a singularity when the total energy adds up to zero or when a partial energy involving $k_1$ adds up to zero. The deformed wavefunction $\psi_4(z)$ therefore has (simple) poles at the locations
\beq
\begin{aligned}
z_E &\equiv -(k_1+k_2+k_3+k_4)= - E\,,\\
z_L^{(s)} &\equiv \ \, -(k_{12}+s)\ = - E_L^{(s)}\,,\\
z_L^{(t)} &\equiv \ \,-(k_{14}+t)\ \, =- E_L^{(t)}\,,\\
z_L^{(u)} &\equiv \ \, -(k_{13}+u)\ =- E_L^{(u)}\,.
\end{aligned}
\eeq
Since we have shifted $k_1$, the complex shift only sees the left partial energy singularities of the correlator (along with the total energy singularity). However, this is enough information to reconstruct the correlator because the residues of the right partial energy singularities are not independent from these~\cite{Benincasa:2018ssx}.
We further know that the residues are\footnote{The relative normalizations of the residues are fixed by noting that both the three- and four-point objects used to build the residues are related to flat-space scattering amplitudes. Demanding the correct relative normalization between the scattering amplitudes fixes the relative normalization between the residues. We have also set the $\phi^3$ coupling to $1$ for simplicity.} 
\begin{align}
\underset{z=z_E}{\rm Res}\, \psi_4(z) &= A_4 = -\frac{1}{k_{34}^2-s^2}-\frac{1}{k_{23}^2-t^2}-\frac{1}{k_{24}^2-u^2}\,,\\
\underset{z=z_L^{(s)}}{\rm Res}\,  \psi_4(z) &= A_3 \times \widetilde\psi_3  = \frac{1}{k_{34}^2-s^2}\,,\\
\underset{z=z_L^{(t)}}{\rm Res}\,  \psi_4(z) &= A_3 \times \widetilde\psi_3  = \frac{1}{k_{23}^2-t^2}\,,\\
\underset{z=z_L^{(u)}}{\rm Res}\,  \psi_4(z) &= A_3 \times \widetilde\psi_3  = \frac{1}{k_{24}^2-u^2}\, ,
\end{align}
where $A_4 = A_4(z_E)$, $A_3=A_3(z_L)$ and $\widetilde \psi_3= \widetilde \psi_3(z_L)$.
Note that in the final equality, the residues do not depend on $k_1$, because this is the energy variable that we deformed (see Appendix~\ref{app:Residues}). Summing up the residues according to~\eqref{eq:residue}, we get
\beq
\begin{aligned}
\psi_4(0)= & -\frac{1}{E}\left(\frac{1}{k_{34}^2-s^2}+\frac{1}{k_{23}^2-t^2}+\frac{1}{k_{24}^2-u^2}\right) \\
&+ \frac{1}{E_L^{(s)}} \frac{1}{k_{34}^2-s^2}+\frac{1}{E_L^{(t)}} \frac{1}{k_{23}^2-t^2}+\frac{1}{E_L^{(u)}} \frac{1}{k_{24}^2-u^2}\,.
\end{aligned}
\eeq
Interestingly, each residue separately has an unphysical folded singularity (e.g.~at $k_{34} = s$), but these all cancel in the sum of terms to give 
\begin{tcolorbox}[colframe=white,arc=0pt,colback=greyish2]
\beq
\psi_4 = \frac{1}{EE_L^{(s)}E_R^{(s)}}+\frac{1}{EE^{(t)}_LE^{(t)}_R}+\frac{1}{EE^{(u)}_LE^{(u)}_R}\,,
\eeq
\end{tcolorbox}
\noindent
which is exactly the right answer; cf.~\eqref{eq:flatspacescalar4pt}. It is easy check that this correlator has the correct cuts, which as expected are reproduced solely by the partial energy singularities, with the contribution from the total energy singularity dropping out.\footnote{We can also understand the lack of a boundary term by noting that the shift~\eqref{eq:phi3deformation} causes the entire correlator to go to zero as $z\to\infty$. This is in contrast to the usual  S-matrix BCFW shifts, for which one of the channels survives at infinity and must be reproduced by a boundary term.}

\begin{figure}[t!]
   \centering
            \includegraphics[width=.8\textwidth]{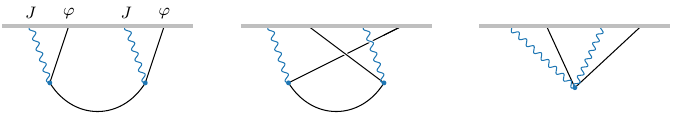}
   \caption{Illustration of the different contributions to the Abelian Compton correlator. Consistency requires the exchange of particles in both the $s$ and $t$-channels, along with a particular contact contribution.  In the recursive construction of the correlator, these contributions are linked together by the total energy singularity.}
  \label{fig:JOJOcorr}
\end{figure}

\paragraph{Abelian Compton scattering} 
A second instructive example is provided by the correlator arising from (Abelian) Compton scattering, where a photon scatters off a scalar particle (see Fig.~\ref{fig:JOJOcorr}). Since some of the  external particles now carry spin, there must be exchanges in more than one channel in order for the final correlator to be consistent.

\vskip4pt
We consider the same deformation as in the previous example 
\beq
k_1\mapsto k_1+z\,.
\eeq
To catalog the singularities of the deformed correlator, $\psi_{J\varphi J\varphi}(z)$, we note that the  corresponding scattering amplitude  only has exchanges in the $s$ and $t$-channels. This implies that the only poles of the deformed correlator are at
\beq
\begin{aligned}
z_E &= -E\,,\\
z_L^{(s)} &= -E_L^{(s)}\,,\\
z_L^{(t)} &= -E_L^{(t)}\,,
\end{aligned}
\eeq
and that the residues associated to these singularities are
\begin{align}
\underset{z=z_E}{\rm Res} \, \psi_{J\varphi J\varphi}(z) &= A_{J\varphi J\varphi}
= -4\left(\frac{(\vec \xi_1\cdot \vec k_2)(\vec \xi_3\cdot \vec k_4) }{k_{34}^2-s^2}+\frac{(\vec \xi_1\cdot \vec k_4)(\vec \xi_3\cdot \vec k_2) }{k^2_{23}-t^2}-\frac{ \vec\xi_1\cdot \vec\xi_3 }{2}\right),
\label{equ:zEres}
\\
\underset{z=z_L^{(s)}}{\rm Res}\, \psi_{J\varphi J\varphi}(z) &=  A_{J\varphi^2}\times \widetilde\psi_{J\varphi^2}= 2(\vec\xi_1\cdot\vec k_2) \,\frac{2(\vec\xi_3\cdot\vec k_4)}{k_{34}^2-s^2}\,,\\
\underset{z=z_L^{(t)}}{\rm Res}\,\psi_{J\varphi J\varphi}(z) &= A_{J\varphi^2}\times \widetilde\psi_{J\varphi^2}= 2(\vec\xi_1\cdot\vec k_4) \,\frac{2(\vec\xi_3\cdot\vec k_2)}{k_{23}^2-t^2}\,.
\end{align}
Using~\eqref{eq:residue} to sum up these residues, we get
\beq
\begin{aligned}
\psi_{J\varphi J\varphi} &= -\frac{4}{E}\left(\frac{(\vec \xi_1\cdot \vec k_2)(\vec \xi_3\cdot \vec k_4) }{k_{34}^2-s^2}+\frac{(\vec \xi_1\cdot \vec k_4)(\vec \xi_3\cdot \vec k_2) }{k_{23}^2-t^2}-\frac{ \vec\xi_1\cdot \vec\xi_3 }{2}\right)\\
&~~~~
+ \frac{1}{E_L^{(s)}} \frac{4(\vec\xi_1\cdot\vec k_2)(\vec\xi_3\cdot\vec k_4)}{k_{34}^2-s^2}+ \frac{1}{E_L^{(t)}}\frac{4(\vec\xi_1\cdot\vec k_4)(\vec\xi_3\cdot\vec k_2)}{k_{23}^2-t^2}\,.
\end{aligned}
\eeq
Despite all terms having unphysical folded singularities at $k_{34}=s$ and $k_{23}=t$, they cancel in the sum to generate\footnote{Alternatively, we can think of the cancellation of these unwanted singularities as an {\it input} that will fix the relative normalization of the various contributions to be consistent with charge conservation~\cite{Baumann:2020dch,1866373}.}
\begin{tcolorbox}[colframe=white,arc=0pt,colback=greyish2]
\beq
\psi_{J\varphi J\varphi}  = \frac{4}{E}\left(\frac{(\vec \xi_1\cdot \vec k_2)(\vec \xi_3\cdot \vec k_4) }{E_L^{(s)} E_R^{(s)}}+\frac{(\vec \xi_1\cdot \vec k_4)(\vec \xi_3\cdot \vec k_2) }{E_L^{(t)} E_R^{(t)}}+\frac{ \vec\xi_1\cdot \vec\xi_3 }{2}\right) ,
\label{eq:poljojoBCFW}
\eeq
\end{tcolorbox}
\noindent
which is indeed the correct answer~\cite{Baumann:2020dch}. The residue of the total energy singularity again drops out of the cuts, which are entirely reproduced by the partial energy singularities.

\vskip4pt
The various exchange channels of~\eqref{eq:poljojoBCFW} are linked together in an interesting way by the singularity structure. The residue of the total energy singularity---i.e.~the scattering amplitude~\eqref{equ:zEres}---has folded singularities in both the $s$ and $t$ variables, which requires the correlator to have partial energy singularities at both $E_L^{(s)}$ and $E_L^{(t)}$, with residues that cancel off these unwanted singularities. From this perspective, the rigid structure of the scattering amplitude
connects the channels of the correlator. This introduces a contact contribution to the correlator, which is also required by current conservation on the boundary~\cite{Baumann:2020dch}.
Despite these nice features, the final answer~\eqref{eq:poljojoBCFW} is still written in a form resembling a sum over channels. Its rich structure, however, suggests that there should be another presentation of the answer that makes its combined structure more manifest, and it would be extremely interesting to find such a representation of the correlator.

\vskip 4pt
Although we have focused on a simple correlator involving spinning particles in flat space, it is straightforward to generalize the procedure to more complicated examples like non-Abelian Compton scattering or pure Yang--Mills theory (see Appendix~\ref{app:YMapp} for the details). Due to the Weyl invariance of the interactions in these theories, the corresponding correlators are the same in de Sitter space. Furthermore, amplitudes in these theories are known to extremely high multiplicity, so we expect that the recursive approach will be particularly powerful in these cases.

\subsection{Generalization to de Sitter Space}
\label{sec:dsrecurs}
For applications to cosmology, we would like to extend the previous discussion to more general backgrounds, in particular to de Sitter space. An immediate complication is that the analytic structure of correlation functions in de Sitter can be markedly different from their flat-space counterparts. However, in some special circumstances de Sitter correlators are still rational functions of various combinations of energies, just with higher-order singularities, making it possible to extend the previous discussion to include these cases.

\vskip 4pt
The key insight that makes this possible is that the cut of the de Sitter four-point function relates the partial energy singularities---including its subleading divergences---to three-point functions~\cite{Jazayeri:2021fvk}. This determines the Laurent expansion around the partial energy singularities exactly. In order to completely fix the correlator, we need one additional constraint, which is that the residues of subleading total energy singularities are not independent from the residues of the subleading partial energy singularities~\cite{Benincasa:2019vqr}. Rather, they must combine in order to remove folded singularities, similar to the flat-space case.

\vskip 4pt
More concretely, the ($s$-channel) cut of a four-point exchange correlator in de Sitter can be written as
\beq
\psi_4(k_1,k_2,k_3,k_4)+\psi_4^*(-k_1,-k_2,-k_3,-k_4) = -\frac{1}{P_\sigma(s)} \,\widetilde\psi_3(k_1,k_2,s) \otimes \widetilde\psi_3(s,-k_3,-k_4)\,.
\label{eq:4ptcutdSsings}
\eeq
Crucially, the correlator $\psi_4^*(-k_1,-k_2,-k_3,-k_4)$ is {\it regular} as $E_L$ or $E_R$ go to zero, which follows from the assumption that the original correlator does not have any folded singularities (e.g.~when $k_{12}-s \to 0$). This implies that if we expand the right-hand side of~\eqref{eq:4ptcutdSsings} near $E_{L,R}=0$, then all of the divergences must be attributable to $\psi_4(k_1,k_2,k_3,k_4)$, which is precisely the object that we are interested in. 
In other words, in the limit $E_L\to 0$ the cutting rule implies that the Laurent series of $\psi_4$ can be written as~\cite{Jazayeri:2021fvk}  
\beq
\lim_{E_L\to0}\psi_4(E_L) = -\lim_{E_L\to0}\,\frac{1}{P_\sigma} \,\widetilde\psi_3 \otimes \widetilde\psi_3 = \frac{R^{(m)}_L}{E_L^m}+ \frac{R^{(m-1)}_L}{E_L^{m-1}}+\cdots+ \frac{R^{(1)}_L}{E_L}+\cdots\,,
\label{eq:ELlaurent}
\eeq
where $m$ is the order of the total energy singularity of the three-point function $\psi_3(k_1,k_2,s)$ and the coefficients of the expansion are 
\beq
R_L^{(l)} = \lim_{E_L\to 0}\frac{1}{(m-l)!}\frac{\rd^{m-l}}{\rd (E_L)^{m-l}}\left(-E_L^m\frac{1}{P_\sigma} \,\widetilde\psi_3 \otimes \widetilde\psi_3 \right) .
\eeq
A similar Laurent expansion holds around $E_R = 0$.

\vskip 4pt
A generic complex deformation of a correlator will also have a singularity when $z = -E$. The residue of the leading $E\to 0$ divergence is just the corresponding flat-space scattering amplitude, but the coefficients of the subleading terms are more subtle and are not obviously fixed by readily available information.\footnote{To avoid dealing with this subtlety, one can restrict to complex deformations that do not shift the total energy~\cite{Jazayeri:2021fvk}. The downside of this approach is that in many cases correlators then require boundary terms to be reconstructed. This can be understood from the presence of terms in the true correlator that have total energy singularities, but no $E_L$ or $E_R$ singularities. Under the complex deformation these terms do not shift and so survive as boundary terms as $z\to\infty$. In the way we are proceeding, demanding a particular residue for the total energy singularity is effectively fixing the freedom that would otherwise appear as a boundary term.} Fortunately, it is possible to deal with this complication by requiring that the subleading total energy singularities have the right coefficients to cancel off the would-be folded singularities in the Laurent coefficients in~\eqref{eq:ELlaurent}. This information turns out to be enough to completely fix the answer in many cases, so that we can apply~\eqref{eq:GenResidue} to reconstruct the correlator.

\vskip 4pt
It is simplest to illustrate the procedure through examples, so we now consider a few cases that demonstrate the general features.

\paragraph{Massless spin-2 exchange}
As a first example, we consider the ($s$-channel) correlator  arising from the exchange of a massless spin-2 field between conformally coupled scalars. 
 Because of the permutation symmetries of the correlator, it is convenient to do a slightly different complex deformation than before, namely 
\beq
k_{12}\mapsto k_{12}+z\,,
\eeq
which accesses the singularities of the correlator at $E_L=0$ and $E=0$. As we have just explained, the Laurent expansion about $E_L = 0$ is fixed by the cut of the correlator
\beq
\psi^{(T)}_4(k_{12},k_{34})+\psi^{(T)}_4(-k_{12},-k_{34}) = -2s^3 \tl\psi_{\vp\vp T}\otimes\tl\psi_{T\vp\vp}\,,
\label{eq:dsmasslessspin2exccut}
\eeq
where the right-hand side can be computed using the shifted wavefunction~\cite{Baumann:2020dch}
\beq
\widetilde\psi_{\vp\vp T} = \frac{[(\vec k_1 - \vec k_2)\cdot \vec \xi_s\,]^2}{(k_{12}^2-s^2)^2}\,,
\eeq
as well as an  analogous expression with $\vec k_{1,2}$ replaced by $\vec k_{3,4}$.
Contracting these three-point functions with the projector $(\Pi_{2,2})^{ij}_{kl}$  gives an explicit expression for the cut of the four-point function.
According to~\eqref{eq:GenResidue}, the expansion of this expression around $E_L\to 0$ is the residue of the deformed correlator
\beq
-\underset{z=z_L}{\rm Res} \bigg( \frac{\psi^{(T)}_4 (z)}{z} \bigg) = \left(-\frac{1}{E_L^2} \frac{4s}{E_R^2(k_{34}-s)^2} -\frac{1}{E_L} \frac{4}{E_R^2(k_{34}-s)^2} \right) \frac{ \tl\Pi_{2,2}}{12}
\,,
\label{equ:ELSing}
\eeq
where we have again utilized the polarization sum $\tl\Pi_{2,2}$ (see Appendix~\ref{app:pols}).
Note that the residue in \eqref{equ:ELSing} has singularities at $E_R=0$ and $k_{34}=s$. We know that the final answer must have $E_R$ singularities, but the residues of~\eqref{equ:ELSing} are not quite right. Further, we know that the folded singularities at $k_{34} = s$ must be absent in the final answer. Both of these problems are fixed by adding the residue at $E=0$, exactly as in the flat-space case.

\vskip 4pt
The coefficient of the leading total energy singularity is fixed by the corresponding scattering amplitude, 
\beq
\lim_{E \to 0} \psi_4^{(T)} =\frac{1}{E^3} \frac{1}{3S}\,S^2P_2\left(1+\frac{2U}{S}\right) ,
\label{equ:Elimit}
\eeq
where $P_2$ is a Legendre polynomial and $S$, $U$ are the flat-space Mandelstam variables. While this 
tells us the leading residue of the total energy pole evaluated on the kinematic locus $E=0$, we also need to know the subleading total energy coefficients, which we will fix below. 
The most convenient way to write the residue is
\beq
- \underset{z=z_E}{\rm Res} \bigg( \frac{\psi^{(T)}_4 (z)}{z} \bigg) =  \frac{1}{E^3}\frac{1}{3}\left(\frac{\tl\Pi_{2,2}}{k_{34}^2-s^2}+\tl\Pi_{2,1}- \widetilde \Pi_{2,0}\right) +\frac{R_E^{(2)}}{E^2} +\frac{R_E^{(1)}}{E} 
\, ,
\label{eq:s2excres1}
\eeq
where $R_E^{(2)}$ and  $R_E^{(1)}$ parameterize the subleading total energy poles that are not fixed by the limit~\eqref{equ:Elimit}. Note that any other way of writing the leading singularity in~\eqref{eq:s2excres1} can only differ by terms that can be absorbed into $R_E^{(2)}$ and  $R_E^{(1)}$, which we will see are fixed by other considerations. An important constraint that~\eqref{eq:s2excres1} satisfies is that it has vanishing cut, so that the contribution from the total energy singularity drops out of the cut of the reconstructed answer.  In the following, it will be convenient to combine the subleading total energy Laurent coefficients as ${\cal R}_E \equiv R_E^{(2)} + E R_E^{(1)}$.

\vskip4pt
Adding up~\eqref{equ:ELSing} and~\eqref{eq:s2excres1}, we obtain 
\beq
\begin{aligned}
\psi^{(T)}_4 =& \left(-\frac{1}{E_L^2} \frac{4s}{(k_{34}^2-s^2)^2} -\frac{1}{E_L} \frac{4}{(k_{34}^2-s^2)^2} \right) \frac{\tl\Pi_{2,2}}{12}\\
&\hspace{.3cm}+\frac{1}{E^3}\frac{1}{3}\left(\frac{\tl\Pi_{2,2}}{k_{34}^2-s^2}+\tl\Pi_{2,1}- \widetilde \Pi_{2,0}\right)+\frac{{\cal R}_E}{E^2}\,.
\end{aligned}
\label{eq:s2partialanswer}
\eeq
The subleading total energy pole coefficients, ${\cal R}_E$, are then fixed by demanding that~\eqref{eq:s2partialanswer} does not have any folded singularities, and that it has the correct residues on its $E_R$ singularities.
Expanding~\eqref{eq:s2partialanswer} in the folded limit, $k_{34} \to s$, and imposing the constraint that
the would-be folded singularities cancel, we obtain the following  Laurent series for the function ${\cal R}_E$:
\beq
\lim_{k_{34}\to s}{\cal R}_E =   \left(\frac{1}{(k_{34}-s)^2}\frac{k_{12}+2s}{s^2}-\frac{1}{k_{34}-s}\frac{k_{12}}{s^3} \right) \frac{\tl\Pi_{2,2}}{12}+\cdots .
\label{eq:erlaurentguess1}  
\eeq
Similarly, by expanding the cut~\eqref{eq:dsmasslessspin2exccut} as $E_R\to 0$, and comparing the result to~\eqref{eq:s2partialanswer}, we get the Laurent series for ${\cal R}_E$ in this limit
\beq
\lim_{E_R\to0}{\cal R}_E = \bigg(\frac{1}{E_R^2}\frac{k_{12}-2s}{s^2}+\frac{1}{E_R}\frac{k_{12}}{s^3} \bigg) \frac{\tl\Pi_{2,2}}{12}+\cdots. \hspace{1.4cm} \phantom{x} \label{eq:erlaurentguess2}
\eeq
From these two Laurent expansions, we can then reconstruct ${\cal R}_E$ as 
\begin{align}
\frac{{\cal R}_E }{E^2}  &= \frac{1}{E^2}\bigg[ \frac{1}{(k_{34}-s)^2}\frac{k_{12}+2s}{s^2}-\frac{1}{k_{34}-s}\frac{k_{12}}{s^3} 
+\frac{1}{E_R^2}\frac{k_{12}-2s}{s^2}+\frac{1}{E_R}\frac{k_{12}}{s^3} \bigg] \frac{\tl\Pi_{2,2}}{12}\nonumber \\[6pt]
&= \left( \frac{k_{34} }{E^2} + \frac{1}{E} \right)  \frac{1}{(k_{34}^2 -s^2)^2}\frac{\tl\Pi_{2,2}}{3}\,. 
\label{eq:spin2exccorrection}
\end{align}
Notice that this correction term is proportional to $\tl\Pi_{2,2}$, implying that only the highest-helicity part of the correlator needs to be corrected. This is physically sensible because it is only in this part that folded singularities appear in~\eqref{eq:s2partialanswer}.  Importantly, the expression~\eqref{eq:spin2exccorrection} does not have any $E_L$ singularities, so it does not ruin the partial energy singularities that were  fixed earlier in~\eqref{eq:s2partialanswer}. 
Substituting \eqref{eq:spin2exccorrection} into \eqref{eq:s2partialanswer}, we find
\begin{tcolorbox}[colframe=white,arc=0pt,colback=greyish2]
\beq
\psi^{(T)}_4 = -\frac{sE+E_LE_R}{3E^3E_L^2E_R^2} \tl\Pi_{2,2} + \frac{1}{3E^3}\tl\Pi_{2,1}- \frac{1 }{3E^3 }\tl\Pi_{2,0}\,,
\label{eq:spin2dsrecurse}
\eeq
\end{tcolorbox}
\noindent
which agrees with the result computed by other means in~\cite{Arkani-Hamed:2018kmz,Baumann:2020dch}.

\paragraph{Massless $\boldsymbol{\dot\phi^3}$ exchange}  
Another interesting example is the four-point function of a massless scalar with a $\dot \phi^3$ interaction, which breaks the boost symmetries of the de Sitter background.  Concretely, we want to reconstruct the four-point function arising in a model of the form
\beq
S = \int\rd t\hskip 1pt \rd^3x\,a^{3}(t)\left(\frac{1}{2}\dot\phi^2 - \frac{1}{a^2}(\vec\nabla\phi)^2+\frac{g_3}{3!}\dot\phi^3\right) .
\label{eq:dot3lagrangian}
\eeq
Note that this action is written in terms of cosmic (as opposed to conformal) time. De Sitter boost-breaking interactions of the form $\dot\phi^n$ are particularly simple because the corresponding bulk-to-boundary propagator is the same as in a theory of conformally coupled scalars, and the bulk-to-bulk propagator is simply related to the one in flat space. 
As a consequence, the final correlation functions are rational functions, and so are amenable to reconstruction via recursion.

\vskip4pt
For concreteness, we will focus on the correlator arising from $s$-channel exchange. The cut of the four-point function can then be written as
\beq
\psi_{4}(k_{12},k_{34})+\psi_{4}(-k_{12},-k_{34}) = -2s^3 \tl\psi_{\dot\phi^3}\times\tl\psi_{\dot\phi^3}\,,
\label{eq:dsdotphicut}
\eeq
where the shifted three-point functions can be computed from the three-point wavefunction coefficient
\beq
\psi_{\dot\phi^3}(k_1,k_2,s) = -\frac{2(k_1k_2s)^2}{(k_{12}+s)^3}\,,
\eeq
where we have set $g_3 = 1$.
This will allow us to compute the partial energy singularities of $\psi_4$. We again perform a complex deformation of the correlator
\beq
k_{12}\mapsto k_{12}+z\,,
\eeq
which will access the $E_L$ and $E$ singularities. Expanding the right-hand side of~\eqref{eq:dsdotphicut} around $E_L=0$ yields the Laurent series, and hence the residue at $z=z_L$:
\beq
-\underset{z=z_L}{\rm Res} \left( \frac{\psi_4 (z)}{z} \right) = \frac{1}{E_L^3} \frac{4 (k_1k_2k_3k_4s)^2(3k_{34}^2+s^2)}{E_R^3(k_{34}-s)^3}
\,.
\eeq
As before, this residue has singularities at $E_R= 0$ and $k_{34}=s$ that have to combine with the corresponding singularities of the residue at $E=0$.
The latter can be written as
\beq
-\underset{z=z_E}{\rm Res} \left( \frac{\psi_4 (z)}{z} \right) = -\frac{1}{E^5}\frac{24 (k_1k_2k_3k_4)^2k_{34}^2}{k_{34}^2-s^2}+\frac{{\cal R}_E}{E^4}
\,,
\eeq
where ${\cal R}_E \equiv R_E^{(4)} + \cdots +E^3 R_E^{(1)}$ again parameterizes the subleading singularities.
The full correlator can then be written as
\beq
\psi_4 = -\frac{1}{E^5}\frac{24 (k_1k_2k_3k_4)^2k_{34}^2}{k_{34}^2-s^2}+\frac{1}{E_L^3} \frac{4 (k_1k_2k_3k_4s)^2(3k_{34}^2+s^2)}{(k_{34}^2-s^2)^3}+\frac{{\cal R}_E}{E^4}\,.
\label{eq:dot3guess}
\eeq
Note that  there are now four unknown Laurent coefficients, but they will again be fixed by demanding the absence of folded singularities and the correct normalization of the $E_R$ singularities. Following the same steps as above, we find
\beq
\frac{{\cal R}_E}{E^4} = -\frac{4(k_1k_2k_3k_4s)^2}{(k_{34}^2-s^2)^2} \left[\frac{6 k_{34}}{E^4}  + \frac{1}{E^3}\frac{s^2+3 k_{34}^2}{k_{34}^2-s^2}\right] .
\label{eq:subleadingEdot3}
\eeq
This expression has a string of subleading total energy singularities, but is  regular in the limit $E_L \to 0$, as it must be in order to not spoil the limit~\eqref{eq:dot3guess}. Substituting ~\eqref{eq:subleadingEdot3} into~\eqref{eq:dot3guess}, we obtain the full correlator 
\begin{tcolorbox}[colframe=white,arc=0pt,colback=greyish2]
\vskip -10pt
\beq
\begin{aligned}
\psi_4 =(k_1k_2k_3k_4)^2\bigg[&2s\bigg(\frac{6}{E^5}\left[\frac{1}{E_L}+\frac{1}{E_R}\right]+\frac{3}{E^4}\left[\frac{1}{E_L^2}+\frac{1}{E_R^2}\right] \\
&\hspace{3.3cm}\,+\frac{1}{E^3}\left[\frac{1}{E_L^3}+\frac{1}{E_R^3}\right]-\frac{1}{E_L^3E_R^3}\bigg)-\frac{24}{E^5}\bigg]\,.
\end{aligned}
\label{eq:dotphi34ptfunction}
\eeq
\end{tcolorbox}
\noindent
This final expression has only physical singularities, as expected, and agrees with a direct bulk calculation~\cite{Goodhew:2020hob,Jazayeri:2021fvk}.

\vskip4pt
It is conceptually very satisfying that rational correlation functions can be completely fixed by their behavior in the vicinity of their singularities. However, as we have seen in these examples, considerably more labor is involved in reconstructing the de Sitter answers from this singularity information than for their flat-space counterparts. As we will show in the next section, in many cases we can bypass the construction in de Sitter space and instead transmute the corresponding flat-space correlation functions into their de Sitter counterparts. This makes manifest that many of the subleading singularities must have related coefficients, being derivable from simple structures in flat space.

\newpage
\section{Lifting: From Flat Space to de Sitter}
\label{sec:lifting}

The singularity structure of wavefunction coefficients is substantially simpler in flat space than in de Sitter space. The poles of flat-space correlators are simple poles and the full correlator can often be reconstructed from these singularities.  In contrast, correlators in de Sitter space  typically have a string of subleading singularities which require additional effort to fix.
As we will show in this section, in many cases the more complicated structure of singularities in de Sitter space can be obtained from the corresponding flat-space cases by applying certain ``transmutation" operations. 

\vskip4pt
One way to understand the connection between correlators in flat space and in de Sitter is to note that plane waves of massless fields in the two spacetimes are related by simple differential operations. As a consequence, the time integrals involved in the calculation of correlators are closely related. Keeping track of both the modified plane waves and time dependent vertex factors in de Sitter space then allows us to define operators that transform the flat-space answers to de Sitter space. From a purely boundary perspective, we can also derive and understand the relevant transmutation operators as objects that transform the cuts of flat-space correlators to their de Sitter counterparts. A conceptually useful aspect of this approach is that it is possible to account for the breaking of de Sitter symmetries by the bulk interactions at the level of the lifting operation.

\subsection{Three-Point Functions}

We will first illustrate the concept of transmutation for three-point functions. Since three-point functions are the building blocks for more complex correlators, we will be able to identify  transmutation operators that also appear in the lifting of higher-point functions.
For concreteness, we consider a variety of three-point functions involving conserved spin-$\ell$ currents and  scalar fields. In flat space, the scalar field will be massless, while in de Sitter it will be conformally coupled ($m^2 =2H^2$).  We are therefore interested in the following wavefunction coefficients\footnote{It is only for simplicity of presentation that we assume the currents $J_\ell$ to have the same spin $\ell$. Our approach can straightforwardly be generalized to more complicated situations.}
\beq
\psi_3 \in \big\{ \langle J_\ell\varphi\varphi\rangle\,,\, \langle J_\ell J_\ell \varphi\rangle\,,\,\langle J_\ell J_\ell J_\ell \rangle \big\}\, .
\label{equ:3pts}
\eeq
As we will show explicitly below, these three-point functions can be written as
\beq
\psi_3 = f_\ell \, \Pi_\ell\,,
\label{equ:psi3-ansatz}
\eeq
where $\Pi_\ell$ are polarization structures and $f_\ell$ are momentum-dependent {\it form factors}. The form factors are simple and universal in flat space, but are more complicated in de Sitter space. The goal is to find operators that transform the flat-space form factors into those in de Sitter space, $f_\ell^{({\rm dS})} = {\cal D} f_{\ell}^{({\rm flat})}$.
We will present two different ways to determine the transmutation operators~${\cal D}$. First, we will find these operators by comparing the bulk time integrals in flat space and de Sitter space, which suggests operators that connect the two computations {\it without} actually performing the integrals.
Second, we derive the transmutation operators without reference to the bulk by demanding that the ansatz~\eqref{equ:psi3-ansatz} satisfies the conformal Ward identities on the boundary.

\subsubsection{Form Factors From Bulk Transmutation}
\label{sec:3ptbulktransmutation}

Correlation functions involving conserved currents in four-dimensional de Sitter space are related in a simple way to their flat-space counterparts. At a technical level, this  is  because the bulk-to-boundary propagators for conserved currents are Bessel functions of half-integer order, which can be generated from the flat-space plane wave solutions by acting with differential operators. 
As a result, it is possible to generate de Sitter correlators involving conserved currents from the corresponding flat-space results.
Correlators of conserved currents in flat space are themselves simply related to scattering amplitudes. This provides a natural path from scattering amplitudes to de Sitter correlators (at least at the three-point level): first deform the amplitude to generate a flat-space correlation function, and then ``lift" it to de Sitter space by acting with transmutation operators.

\paragraph{Flat-space correlators}
At the three-point level, flat-space wavefunction coefficients are very closely related to scattering amplitudes. The main difference is that, for the wavefunction coefficients, the bulk vertex insertion points are integrated only up to the spatial boundary on which the correlations are defined. This is the origin of the total energy singularity at $K \equiv k_1+k_2+k_3 = 0$ discussed in Section~\ref{sec:singularities}. Flat-space three-point correlators can therefore be written as
\begin{align}
\langle J_\ell\varphi\varphi\rangle &= (\vec \xi_1 \cdot \vec k_2)^\ell \frac{1}{K} \,, \label{equ:JOO}\\
\langle J_\ell J_\ell \varphi\rangle &= \Big[ (k_3^2-k_2^2-k_1^2)(\vec \xi_1 \cdot \vec \xi_2)-2(\vec k_1 \cdot \vec \xi_2)(\vec k_2 \cdot \vec \xi_1) \Big]^\ell\frac{1}{K} \,,\\
\langle J_\ell J_\ell J_\ell \rangle^{({\rm n})} &=  \Big[ (\vec k_1\cdot \vec \xi_3)(\vec \xi_1\cdot \vec \xi_2)+(\vec k_3\cdot \vec \xi_2)(\vec \xi_1\cdot \vec \xi_3)+(\vec k_2\cdot \vec \xi_1)(\vec \xi_2\cdot \vec \xi_3)\Big]^\ell  \frac{1}{K} \,, \label{equ:JJJ1}\\
\langle J_\ell J_\ell J_\ell \rangle^{({\rm a})} &=  \Big[2 (\vec k_1\cdot \vec \xi_2)(\vec k_2\cdot \vec \xi_3)(\vec k_3\cdot \vec \xi_1) +\Big( k_1^2(\kxi{3}{1})(\xixi{2}{3})+ \text{2 perms} \Big)\Big]^\ell  \frac{1}{K} \,. \label{equ:JJJ2}
\end{align}
There are two conceptually different classes of correlators: one consists of correlators coming from ``minimal coupling" vertices, corresponding to $\langle J_\ell\varphi\varphi\rangle$ and $\langle J_\ell J_\ell J_\ell \rangle^{({\rm n})}$.
These vertices have the minimal number of derivatives in the bulk, and require a deformation of the gauge algebra (hence the label (n) for ``non-Abelian"). The flat-space three-point correlators following from these vertices are just the corresponding scattering amplitudes divided by $K$. The other class of 
 correlators are those that come from higher-derivative cubic vertices built from curvatures, which are invariant under linearized gauge transformations. These latter interactions give rise to $\langle J_\ell J_\ell \varphi\rangle $ and $\langle J_\ell J_\ell J_\ell \rangle^{({\rm a})} $ (where (a) stands for ``Abelian").\footnote{As an example, for spin-1 fields, the non-Abelian vertex is just the cubic Yang--Mills vertex ${\cal L}_{\rm YM} \sim A_\mu A^\nu\partial_\nu A^\mu$, while the Abelian vertex is ${\cal L} \sim F_\mu^\nu F_\nu^\rho F_\rho^\mu$, where in both cases we have suppressed the color factors required by Bose symmetry.} We will see later that these correlators are identically conserved, and are given by the corresponding flat-space amplitudes minus a contribution that makes the mixed-helicity correlator vanish, all divided by $K$.\footnote{Note that the vanishing of the mixed-helicity correlator is the natural extension of the flat-space scattering amplitude away from $K=0$, because the mixed helicity amplitudes coming from these vertices vanish. Interestingly, the analogous property of the non-Abelian vertices in flat space---that equal-helicity amplitudes vanish---is {\it not} preserved by the correlator. This is because the structures that we have written in~\eqref{equ:JOO} and~\eqref{equ:JJJ1} are unique, so there is no freedom to add a piece to cancel off the equal helicity contributions. It is possible to set either the $+++$ or the $---$ correlator to zero by adding a parity-violating coupling (but not both).}

\vskip4pt
We can also see that these are the correct expressions from a bulk computation. The vertex contractions in the amplitude and correlator computations are very closely related because we can perform both calculations in axial gauge for the bulk fields (which we denote by $A^{(\ell)}$), so that  $A^{(\ell)}_{0\mu_2\cdots\mu_\ell} = 0$. After this, a direct calculation verifies~\eqref{equ:JOO}--\eqref{equ:JJJ2}.

\vskip4pt
In de Sitter space, the polarization contractions are the same as in flat space, the only difference is that the $1/K$ factors become more complicated functions of the external particles' energies. In the following, we will show how these form factors can be generated in a simple way from the flat-space energy structures.

\paragraph{Spinor variables}
In order to make the helicity transformation properties of the wavefunction coefficients manifest, it is convenient to write 
them in terms of spinor variables. We will only introduce the minimal notation of the spinor formalism; further details can be found in~\cite{Maldacena:2011nz,Baumann:2020dch}. The main idea is to trade spatial momenta for SL$(2,{\mathbb C})$ spinors as\footnote{In the same way that it is useful to complexify spinor helicity variables in the context of scattering amplitudes, here we are complexifying the spatial rotation group in order to treat $\lambda$ and $\bar\lambda$ as independent.}
\beq
\lambda_\alpha \bar\lambda^\beta = k_i(\sigma^i)_\alpha^{~\beta}+k\,{\mathds 1}_\alpha^{~\beta}\,,
\eeq
where $(\sigma^i)_\alpha^{~\beta}$ are the usual Pauli matrices and $k$ is the magnitude of $\vec k$.  
The barred and un-barred spinors transform by opposite phases under the helicity subgroup of the rotation group. 
The invariant product between spinors is given by the epsilon symbol 
\beq
\begin{aligned}
\langle a b\rangle &= \epsilon^{\alpha\beta}\lambda_\alpha^a\lambda_\beta^b\,,\\
\langle \bar a \bar b\rangle &= \epsilon^{\alpha\beta}  \bar\lambda_\alpha^a \bar \lambda_\beta^b\,.
\end{aligned}
\label{eq:spinorbracket}
\eeq
 In contrast to the more familiar Lorentz spinors, we can also contract barred and un-barred spinors in the same way 
 \beq
\langle a \bar b\rangle = \epsilon^{\alpha\beta}\lambda_\alpha^a \bar \lambda_\beta^b\,.
\eeq
As a special case of this mixed bracket, we can contract a spinor with its barred counterpart, which extracts the energy associated to the momentum that defines the spinor: $\langle\bar\lambda\lambda\rangle = 2k$.  
In order to write~\eqref{equ:JOO}--\eqref{equ:JJJ2} in spinors, we also require a spinor representation of the polarization vectors:
\beq
\xi_{\alpha\beta}^+ = \frac{\bar\lambda_\alpha\bar\lambda_\beta}{2k}~~~{\rm and}~~~\xi_{\alpha\beta}^- = \frac{\lambda_\alpha\lambda_\beta}{2k}\,.
\label{eq:polarizationvecs}
\eeq
Unlike for scattering amplitudes, polarizations can be converted to spinors without the need for an auxiliary reference spinor.

\paragraph{Form factors in de Sitter space}
Since all of the polarization factors are the same in flat space and in de Sitter space, it is convenient to write the three-point functions as products of spinor bracket prefactors that contain the helicity information and form factors that encode the time dependence of the bulk spacetime. We can see this explicitly by substituting~\eqref{eq:polarizationvecs} into the polarization structures in~\eqref{equ:JOO}--\eqref{equ:JJJ2}:  
\begin{align}
\langle J^-_\ell\varphi\varphi\rangle &=\left[\frac{\AA{1}{2}\BA{2}{1}}{k_1}\right]^\ell f_{[\ell00]}\,, \label{equ:f-form}\\[4pt]
\langle J^-_\ell J^-_\ell \varphi\rangle &= \AA{1}{2}^{2\ell} \,f_{[\ell \ell 0]}\,, \label{equ:g-form}\\
\langle J^-_\ell J^-_\ell J_\ell^-\rangle^{({\rm n})} &=  \left[\frac{\AA{1}{2}\AA{2}{3}\AA{3}{1}\hat K}{k_1k_2k_3}\right]^\ell f_{[\ell\ell\ell]}^{({\rm n})}\,, \label{equ:h-form}\\[2pt]
\langle J^-_\ell J^-_\ell J_\ell^-\rangle^{({\rm a})} &=   \Big[\AA{1}{2} \AA{2}{3}\AA{3}{1}\Big]^\ell\,f_{[\ell\ell\ell]}^{({\rm a})} \, .  \label{equ:p-form}
\end{align}
Note that we have introduced the total energy written in terms of spinors, which is given by $\hat K = \tfrac{1}{2}\left(\langle\bar 1 1\rangle+\langle\bar 2 2\rangle+\langle\bar 3 3\rangle\right)$. The reason for doing this is that correlators with $+$ helicities can then be obtained from the above expressions by interchanging barred and un-barred spinors (which changes some signs in $\hat K$).\footnote{Note that $\langle J^+_\ell J^-_\ell J_\ell^-\rangle^{({\rm a})}=0$, i.e.~the mixed-helicity Abelian correlator vanishes.} 
Notice that the polarization structures in~\eqref{equ:JOO}--\eqref{equ:JJJ2} have generated additional energy factors when written in spinor variables. This selects a particular extension of the correlator away from the $K=0$ locus where the spinor structures are fixed by the flat-space helicity amplitudes.

\vskip4pt
Our goal is to find explicit expressions for the form factors, $f_{[\ell_1\ell_2\ell_3]}$, in~\eqref{equ:f-form}--\eqref{equ:p-form}. One way to do this would be to explicitly perform the bulk time integrals, but this turns out not to be necessary. Instead, we note that the bulk-to-boundary propagator for a massless spin-$\ell$ field in de Sitter space, ${\cal K}^{(\ell)}$, can be written in terms of the {\it scalar} bulk-to-boundary propagator~\eqref{eq:massivescalarbulkboundary}, ${\cal K}_{\ell-\frac{1}{2}}$, as\footnote{This bulk-to-boundary propagator arises for the fields $A^{(\ell)}_{\mu_1\cdots\mu_\ell}$ that one would write in the action. These fields have late-time fall-offs in de Sitter of the form $A^{(\ell)}\sim \eta^{\Delta-\ell}$. In many cases, it is convenient to re-define the bulk fields to absorb the $\eta^{-\ell}$ factor, so that the bulk-to-bulk propagator does not have the $\eta^{-\ell}$ prefactor. This does not change the time integrals that we will discuss in this section, because the additional time dependence effectively gets shuffled into the vertex factor.} 
\beq
({\cal K}^{(\ell)})^{j_1\cdots j_\ell}_{i_1\cdots i_\ell}(k,\eta) =  (\Pi_{\ell,\ell})^{j_1\cdots j_\ell}_{i_1\cdots i_\ell}\, \left(\frac{\eta}{\eta_*}\right)^{-\ell} {\cal K}_{\ell-\frac{1}{2}}(k,\eta) \,,
\label{eq:bulkboundarytransmute}
\eeq
where $(\Pi_{\ell,\ell})^{j_1\cdots j_\ell}_{i_1\cdots i_\ell}$ is the spin-$\ell$ transverse-traceless projector. We have  written only the transverse-traceless part of the bulk-to-boundary propagator because it is the only part that contributes to the fixed helicity correlators we are computing. In fact, we have already accounted for the polarization structures in the spinor brackets, so only the scalar bulk-to-boundary propagator factors are necessary to compute the form factors. 
For integer $\ell$, the scalar bulk-to-boundary propagators satisfy the following differential recursion relation in the limit $\eta_*\to 0$
\beq
\lim_{\eta_*\to0}{\cal K}_{\ell-\frac{1}{2}}(k,\eta) = \frac{1}{(2\ell-3)!!}\left(\frac{\eta}{\eta_*}\right)^{2-\ell}\,\prod_{j=1}^{\ell-1}\,{\cal S}_{k}^{[j]}\, e^{ik\eta}\,,
\label{eq:spinlscalarprop}
\eeq
where we have introduced the differential operator 
\begin{tcolorbox}[colframe=white,arc=0pt,colback=greyish2] 
\vspace{-3pt}
\beq
{\cal S}_{k}^{[j]} \equiv (2j-1)-k\partial_{k}\, .
\label{eq:spinraisingop}
\eeq
\end{tcolorbox}
As we will see below, the operator ${\cal S}_{k}^{[j]}$ acts like a spin-raising operator on the form factors.
The expression~\eqref{eq:spinlscalarprop} relates the de Sitter mode functions directly to flat-space plane waves, which will enable us to write the de Sitter integrals in terms of their flat-space counterparts.

\vskip4pt
\noindent
{\it Minimal coupling:} We first consider the correlators coming from the  minimal-coupling cubic vertices in the bulk: $\langle J_\ell\vp\vp\rangle$ and $\langle J_\ell J_\ell J_\ell\rangle^{({\rm n})}$.

\begin{itemize}
\item ${\boldsymbol \langle \boldsymbol J_{\boldsymbol \ell}\boldsymbol\vp\boldsymbol\vp\boldsymbol\rangle}${\bf :} 
The simplest case is the form factor associated to the correlator with a single current.
From the bulk Feynman rules, we infer the following integral representation of the form factor in~\eqref{equ:f-form}:\footnote{In these form factors we do not keep track of the overall factors of $i$ coming from the polarization structures, but these can be recovered by demanding that the total three point wavefunction coefficient has vanishing cut.}
\beq
f_{[\ell00]} =  i\int_{-\infty}^0\rd \eta\, \eta^{\ell-2}e^{ik_{23}\eta}{\cal K}_{\ell-\frac{1}{2}}(k_1, \eta)  \, .
\label{eq:joobulkint}
\eeq
The time dependence of the integrand is easy to understand: the external spin-$\ell$ particle contributes $\eta^{-\ell} {\cal K}_{\ell-\frac{1}{2}}(k_1,\eta)$, while the two external conformally coupled scalars give $\eta^2 e^{ik_{23}\eta}$. In addition, the measure of the action gives rise to a factor of $\eta^{-4}$. Finally,  the bulk vertex we are interested in has $2\ell$ indices ($\ell$ for the field $A^{(\ell)}$ and $\ell$ derivatives). We must raise $\ell$ of these indices, which leads to an additional factor of $\eta^{2\ell}$ from the inverse metrics. All together this gives the factor of $ \eta^{\ell-2}$ in~\eqref{eq:joobulkint}. Using~\eqref{eq:spinlscalarprop}, the factors of $\eta$ cancel, and (up to an overall factor) we get 
\beq
f_{[\ell00]} = \prod_{j=1}^{\ell-1}\,{\cal S}_{k_1}^{[j]}\ i\int^0_{-\infty}\rd \eta\,e^{iK\eta} \, .
\eeq
The time integral gives the flat-space form factor, $1/K$, so the de Sitter form factor can be written as
\begin{tcolorbox}[colframe=white,arc=0pt,colback=greyish2] 
\beq
f_{[\ell00]} = \prod_{j=1}^{\ell-1}{\cal S}_{k_1}^{[j]}\,\frac{1}{K} \, .
\label{eq:JOOtimeint2}
\eeq
\end{tcolorbox}

\item ${\boldsymbol \langle \boldsymbol J_{\boldsymbol \ell} \boldsymbol J_{\boldsymbol \ell} \boldsymbol J_{\boldsymbol \ell}\rangle}${\bf :} Next, we consider the three-point function involving three spin-$\ell$ currents. In the case of minimal coupling, the bulk Feynman rules lead to the time integral
\beq
f_{[\ell\ell\ell]}^{({\rm n})} = i\int^0_{-\infty}\rd \eta\,\eta^{\ell-4}{\cal K}_{\ell-\frac{1}{2}}(k_1, \eta)\, {\cal K}_{\ell-\frac{1}{2}}(k_2, \eta)\, {\cal K}_{\ell-\frac{1}{2}}(k_3, \eta)\,.
\label{eq:JJJntimeintegral}
\eeq
As before, each external spinning line contributes a factor of $ \eta^{-\ell}{\cal K}_{\ell-\frac{1}{2}}$, and we get a factor of $\eta^{-4}$ from the measure. The bulk vertex has three spin-$\ell$ fields and $\ell$ derivatives, so we require $2\ell$ inverse metrics to raise half of these indices, leading to a factor of $\eta^{4\ell}$. Putting all these factors together, we get~\eqref{eq:JJJntimeintegral}. Using~\eqref{eq:spinlscalarprop}, we can write the time integral as
\beq
f_{[\ell\ell\ell]}^{({\rm n})} =\prod_{j=1}^{\ell-1} {\cal S}_{k_1}^{[j]} {\cal S}_{k_2}^{[j]} {\cal S}_{k_3}^{[j]}\, i\int_{-\infty}^0\rd \eta\,\eta^{-2(\ell-1)}e^{iK\eta}\,.
\eeq
This time, the factors of $\eta$ in the integrand have not cancelled.
To deal with this, we {\it integrate} the flat-space form factor with respect to the total energy $K$ a number of times to produce the factor of $\eta^{-2(\ell-1)}$. This is philosophically similar to the integral representation~\eqref{eq:cc4ptint} of conformal scalar exchange in de Sitter.
The de Sitter form factor is then\footnote{Note that we have an indefinite integral in $K$. If necessary, the constants of integration can be fixed to match the exact result of the time integral given by 
\begin{align}
	i\int_{-\infty}^{\eta_*} \rd \eta\,\eta^{-2(\ell-1)}e^{iK\eta} =\frac{(-1)^{\ell+1}}{(2\ell-3)!} K^{2\ell-3}\Big[\log (-K\eta_*)+\gamma_E-\text{H}_{2\ell-3}-i\pi/2\Big] +\cdots\, ,
\end{align}
where $\text{H}_n$ is the $n$-th Harmonic number and we have dropped some 
$\eta_*$-dependent terms. However, only the $\log K$ piece inside the brackets survives when acted on by the spin-raising operators.
} 
\begin{tcolorbox}[colframe=white,arc=0pt,colback=greyish2]
\beq\label{fnlll}
f_{[\ell\ell\ell]}^{({\rm n})} = \prod_{j=1}^{\ell-1} {\cal S}_{k_1}^{[j]} {\cal S}_{k_2}^{[j]} {\cal S}_{k_3}^{[j]}   \left(\int\rd K\right)^{2(\ell-1)}\, \frac{1}{K}\,. 
\eeq
\end{tcolorbox}
To the best of our knowledge, these correlators for $\ell >2$ have not been written down in the literature before. For example, the $\ell = 3$ form factor is given by
\begin{align}
\nonumber
f_{[333]}^{({\rm n})} = \frac{1}{K^3}\Big(&3 K^6 - 9 K^4 (k_1k_2+k_1k_3+k_2k_3) + 3 K^3 k_1k_2k_3+ 3 K^2 (k_1k_2+k_1k_3+k_2k_3)^2 \\
& + 3 K (k_1k_2+k_1k_3+k_2k_3) k_1k_2k_3 + 
 2 (k_1k_2k_3)^2\Big)\,.
\end{align}
Given that these objects are unique, we expect that they may be of independent interest for other applications beyond cosmology.

\end{itemize}

\vskip4pt
\noindent
{\it Non-minimal coupling:} Next, we work out the case of non-minimal bulk couplings, 
which can be written in terms of (linear) gauge-invariant higher-spin Weyl tensors~\cite{Joung:2012qy,Henneaux:2015cda}.\footnote{Explicitly, these tensors in flat space are given by $\ell$ derivatives acting on the spin-$\ell$ bulk field as
\beq
W^{(\ell)}_{\mu_1\nu_1\cdots\mu_\ell\nu_\ell} = {\cal P}_{[\ell,\ell]}\,\partial_{\mu_1}\cdots\partial_{\mu_\ell}A^{(\ell)}_{\nu_1\cdots\nu_\ell}\,,
%\label{eq:weyltransform}
\eeq
where ${\cal P}_{[\ell,\ell]}$ is a Young projector onto a tensor with the symmetries of a (traceless) Young diagram with two rows of length $\ell$. This tensor is invariant under both linearized higher-spin gauge transformations and linearized higher-spin Weyl transformations, whose explicit forms can be found in~\cite{Joung:2012qy,Henneaux:2015cda}. The de Sitter analogues replace $\partial_\mu$ with covariant derivatives, and have subleading terms with fewer derivatives, which are uniquely fixed by the requirement of gauge invariance~\cite{Hinterbichler:2016fgl}. We will not require their explicit expressions.
}
For our purposes, the important property of these Weyl-like curvatures is that when evaluated on the bulk-to-boundary propagator~\eqref{eq:bulkboundarytransmute}, the de Sitter tensor is simply related to the flat-space Weyl tensor evaluated on an ordinary plane wave
\beq
W^{(\ell)}_{\rm dS}[{\cal K}^{(\ell)}]_{\mu_1\nu_1\cdots\mu_\ell\nu_\ell} \propto \left(\frac{k}{\eta}\right)^{\ell-1}W^{(\ell)}_{\rm flat}[e^{ik\eta}]_{\mu_1\nu_1\cdots\mu_\ell\nu_\ell}\,.
\label{eq:weyltransform}
\eeq
This is the higher-spin version of the relation used in~\cite{Maldacena:2011nz} to compute the graviton three-point function coming from a Weyl-cubed interaction, and with it we can easily determine the time integrals that produce de Sitter form factors.

\begin{itemize}

\item ${\boldsymbol \langle \boldsymbol J_{\boldsymbol \ell} \boldsymbol J_{\boldsymbol \ell} \boldsymbol \varphi\rangle}${\bf :} The simplest correlation function that arises from a non-minimal coupling in the bulk involves two massless spin-$\ell$ fields and a scalar.  For example, the spin-1 version of the bulk coupling is $\phi F_{\mu\nu}^2$, while the spin-2 version is built from the Weyl tensor, $\phi W_{\mu\nu\rho\sigma}^2$. Using the relation~\eqref{eq:weyltransform}, it is straightforward to determine the time integral that computes the de Sitter form factor
\beq
f_{[\ell \ell0]} = (k_1k_2)^{\ell-1}i\int_{-\infty}^0 \rd \eta\,\eta^{2\ell-1} e^{iK\eta}\, ,
\label{eq:JJOtimeint}
\eeq
where the factors of conformal time in the integral can be understood in the same way as in the previous examples.

\vskip4pt
There are several ways to write (\ref{eq:JJOtimeint}) as a differential operator acting on the flat-space form factor. The most obvious one is
\beq
f_{[\ell \ell0]} =  (k_1k_2)^{\ell-1}\left(\frac{\partial }{\partial K} \right)^{2\ell-1}\frac{1}{K}\,.
\eeq
However, for later applications this will not be the most useful way of writing the answer, because we often want to think of the transmutation operators as acting on the various external fields separately. A representation of the form factor with this property is 
\beq
f_{[\ell \ell0]} =  \partial_{k_3}\prod_{j=1}^{\ell-1}\Big({\cal S}_{k_1}^{[j]}-j\Big)\Big({\cal S}_{k_2}^{[j]}-j\Big)\frac{1}{K}\,.
\eeq
In this case, we can easily find a closed-form expression for the form factor, which we record for completeness
\begin{tcolorbox}[colframe=white,arc=0pt,colback=greyish2]
\beq
f_{[\ell \ell0]} =\left(\frac{k_1 k_2}{K^2}\right)^{\ell-1} \frac{1}{K^2}\,.\label{fll0sol}
\eeq
\end{tcolorbox}
Notice that in contrast to the minimal-coupling case, the $\ell=1$ correlator is different from its flat-space counterpart because the interaction is not Weyl invariant. 

\item ${\boldsymbol \langle \boldsymbol J_{\boldsymbol \ell} \boldsymbol J_{\boldsymbol \ell} \boldsymbol J_{\boldsymbol \ell}\rangle}${\bf :} As a final example, we consider the correlator of three massless spin-$\ell$ fields coming from a curvature-cubed coupling in the bulk. Following similar steps to before, we deduce that the form factor is given by 
\beq
f_{[\ell \ell \ell]}^{({\rm a})}  = (k_1k_2k_3)^{\ell-1}i\int_{-\infty}^0\rd \eta\,\eta^{3\ell-1} e^{iK\eta}\,,
\eeq
where $3(1-\ell)$ factors of $\eta$ come from converting the de Sitter bulk-to-boundary propagators to the flat-space propagators using~\eqref{eq:weyltransform}, $6\ell$ come from raising half of the $6\ell$ indices of the curvatures, and $\eta^{-4}$ comes from the measure. We can again write this form factor as the action of derivative operators on the flat-space form factor: 
\beq
f_{[\ell \ell \ell]}^{({\rm a})} =\prod_{j=1}^{\ell-1} \Big({\cal S}_{k_1}^{[j]}-j\Big)\Big({\cal S}_{k_2}^{[j]}-j\Big)\Big({\cal S}_{k_3}^{[j]}-j\Big)\frac{\partial^2}{\partial K^2}\frac{1}{K}\,.
\eeq
As in the previous case, this expression can be put in a closed form 
\begin{tcolorbox}[colframe=white,arc=0pt,colback=greyish2]
\beq
f_{[\ell \ell \ell]}^{({\rm a})} =  \left(\frac{k_1k_2k_3}{K^3}\right)^{\ell-1} \frac{1}{K^3}\,,\label{fllla}
\eeq
\end{tcolorbox}
which reduces to the known expressions for $\ell=1, 2$~\cite{Maldacena:2011nz,Bzowski:2013sza,Farrow:2018yni,Baumann:2020dch}. Note also that the correlators arising from curvature couplings have a double copy-like structure~\cite{Farrow:2018yni,Jain:2021qcl, Jain:2021vrv}.

\end{itemize}

\noindent
Interestingly, reintroducing factors of $i$, so that these three-point functions have vanishing cuts, we see that in some cases the wavefunction coefficients drop out of the square of the wavefunction. This causes the corresponding in-in correlation functions to vanish due to a cancellation between the two branches of the contour. An example where this happens is $\langle J_\ell J_\ell \vp\rangle$, for which the wavefunction coefficient is pure imaginary and the polarization structure is even under taking the complex conjugate.\footnote{More generally, the contribution to the in-in correlator will vanish whenever the form factor is invariant under flipping the sign of all its energies, as is elaborated on in~\cite{Cabass:2021fnw}.
}

\subsubsection{Form Factors From Conformal Symmetry}
The de Sitter form factors can also be found directly by utilizing the symmetries of the background de Sitter spacetime. 
On the boundary, these symmetries manifest themselves as conformal Ward identities that the correlation functions must obey. The form factors derived in the previous section then arise as solutions to these differential equations.
 The advantage of this approach is that it makes no reference to the bulk and therefore is more in the spirit of the bootstrap philosophy.

\vskip 4pt
There are two conceptually distinct kinds of Ward identities that boundary correlators arising from massless fields with spin must obey. The symmetries of the de Sitter spacetime translate directly into kinematic Ward identities, while the fact that the boundary operators dual to massless fields are conserved currents translates into current conservation Ward identities (which we will call Ward--Takahashi identities). Conserved operators must satisfy both of these simultaneously, which places highly nontrivial constraints on consistent interactions~\cite{Baumann:2020dch}.
The virtue of the spinor-helicity formalism is that the combined constraints of conformal symmetry and the Ward--Takahashi identity can be imposed with a single differential operator, namely the special conformal generator in spinor-helicity variables~\cite{Witten:2003nn}:
\begin{align}
	\tl K^i = 2{(\sigma^i)_\alpha}^\beta\frac{\partial^2}{\partial\lambda_\alpha\partial\overline\lambda^\beta}\, .\label{Ktl}
\end{align}
Applying this operator to the current dual to a massless spin-$\ell$ field gives
\beq
	\tl K^i \left(\frac{J_\ell}{k^{\ell-1}}\right) =  \frac{1}{k^{\ell-1}}\left( -\hs \xi^{i_1}\cdots \xi^{i_\ell} K^i + \frac{\ell(\ell+1)}{k^2}\,\xi^i k^{i_1}\xi^{i_2}\cdots\xi^{i_\ell}\right)J_\ell^{i_1\cdots i_\ell}\, ,\label{Ktildeaction}
\eeq
where $K^i$ is the ordinary special conformal generator, and $J_\ell \equiv \xi^{i_1} \cdots\xi^{i_\ell}J_\ell^{i_1\cdots i_\ell}$, which we have rescaled so that it has conformal weight 2, on which $\tl K^i$ acts nicely~\cite{Maldacena:2011nz, Mata:2012bx, Baumann:2020dch}. The first term on the right-hand side vanishes inside a correlator, while the second term gives the longitudinal part of the spinning field. The latter is needed to check the Ward--Takahashi identity, and we see that $\tl K$ can extract this from $J_\ell$ with definite helicity. It is hence sufficient to just consider form factors that multiply the transverse-traceless part of correlators.

\vskip 4pt
The divergence of the massless spinning correlators \eqref{equ:f-form}--\eqref{equ:p-form} is constrained by 
Ward--Takahashi identities.  For the correlators coming from minimal couplings in the bulk, the identities are
\begin{align}
	\xi_1^i k_1^j\LA J_\ell^{ij}\varphi\varphi\rangle & \propto \frac{(\AA{1}{2}\AB{1}{2})^{\ell-1}k_3+(-1)^{\ell}(\AA{1}{3}\AB{1}{3})^{\ell-1}k_2}{k_1^{\ell-1}} \, , \label{ward1}\\
	\xi_1^i k_1^j\langle J_\ell^{ij}J_\ell^- J_\ell^-\rangle^{\rm (n)} & \propto  \frac{(\AA{1}{2}\AA{2}{3}\AA{3}{1}K)^{\ell-1}}{(k_1k_2k_3)^{2\ell-1}}\frac{\AA{2}{3}^2}{k_1} (k_2^{2\ell-1}-k_3^{2\ell-1})\, . \label{div}
\end{align}
The correlators arising from non-minimal bulk couplings, on the other hand, are completely transverse and therefore obey
\begin{align}
		\xi_1^i k_1^j\LA J_\ell^{ij}J_\ell^-\varphi\RA  &=0 \, ,\\
		\xi_1^i k_1^j\langle J_\ell^{ij}J_\ell^- J_\ell^-\rangle^{\rm (a)} &=0\, .
\end{align}
The differential equations obeyed by the form factors can be obtained by acting with $\tl K$ on the three-point functions and then matching the longitudinal part with the expected structure from the above Ward identities; details of the procedure were given in~\cite{Baumann:2020dch}. (See also~\cite{Jain:2021vrv} for a related discussion.)

\vskip 4pt
As an example, let us see how this works for $\LA J_\ell^- \varphi\varphi\RA$. 
Acting with the special conformal generator on the rescaled correlator, we obtain
\begin{align}
	\sum_{a=1}^3  \k_1\cdot \tl K_a\left[\frac{\langle J_\ell^-\varphi\varphi\rangle}{k_1^{\ell-1}}\right] &= \frac{(\AA{1}{2}\AB{1}{2})^\ell}{k_1^{2\ell-1}} \Delta_{[\ell 00]} f_{[\ell 00]}=0\, ,\label{fl00eq}
\end{align}
where the longitudinal part of $J_\ell$ vanishes due to the contraction with $\k_1$, and we have defined the differential operator
\beq
	\Delta_{[\ell 00]}\equiv \sum_{a=1}^3(\k_1\cdot\k_a)\partial_{k_a}^2-2k_1(\ell-1)\partial_{k_1}\, .
	\label{equ:Dell00}
\eeq
Consider first the solution for $\ell=1$. In this case, the single derivative term in (\ref{equ:Dell00}) vanishes, and imposing permutation symmetry, together with dimensional analysis, fixes the solution to be $f_{[100]}=1/K$. 
The solution for $\ell>1$ can be found recursively by taking
\begin{align}
	f_{[\ell+1,00]}  = \left[(2\ell-1)-k_1\partial_{k_1}\right] f_{[\ell 00]} \equiv {\cal S}^{[\ell]}_{k_1}f_{[\ell 00]}\, . \label{eq:gell}
\end{align}
To see that this is indeed a solution, we note that
\begin{align}
	\Delta_{[\ell+1, 00]}{\cal S}^{[\ell]}_{k_1}f_{[\ell 00]} = {\cal S}^{[\ell+1]}_{k_1} \Delta_{[\ell 00]}f_{[\ell 00]}-(k_2^2-k_3^2)(\partial_{k_2}^2-\partial_{k_3}^2)f_{[\ell 00]} = 0\, ,
\end{align}
where we have used that $f_{[\ell 00]}$ is annihilated by $\partial_{k_2}^2-\partial_{k_3}^2$ because it is
symmetric in $k_2$ and $k_3$, given that both the initial solution $f_{[100]}$ and the recursive relation \eqref{eq:gell} have this permutation symmetry.
This shows how the transmutation operator~\eqref{eq:spinraisingop} arises from a purely boundary perspective.
Since $\LA J_\ell^-\varphi\varphi\RA$ is unique, it is sufficient to just consider the solution to the homogeneous differential equation \eqref{fl00eq}; it can be checked that the solution to this equation also satisfies the inhomogeneous Ward--Takahashi identity \eqref{ward1}.

\vskip 4pt
Following a similar procedure, we can also obtain the differential equations for the other form factors and find their solutions.  
To simplify the equations, it will be convenient to factor out some powers of momenta from the form factors and define
\begin{align}
	f_{[\ell \ell 0]}&\equiv (k_1k_2)^{\ell-1} \,g_{[\ell\ell0]}\, ,  \\
	f_{[\ell \ell \ell]}^{\rm (a)}&\equiv (k_1k_2k_3)^{\ell-1}\,g_{[\ell\ell\ell]}^{\rm (a)}\, , 	\\
	f_{[\ell \ell\ell ]}^{\rm (n)} &\equiv \hat K^{-\ell}(k_1k_2k_3)^{2\ell-1}\,g_{[\ell\ell\ell]}^{\rm (n)}\, ,  \label{gnlll}
\end{align}
where $\hat K$ was defined in  \eqref{equ:h-form}.
 The rescaled form factors then satisfy the following differential equations 
 \begin{align}
\vec \Delta_{[\ell\ell 0]} \,g_{[\ell \ell 0]} &\, = 0\, ,\label{Qll0}\\[4pt]
	\Delta_{[\ell\ell \ell]}\,g_{[\ell \ell \ell]}^{\rm (a)} &\,= 0\, , \label{Qllla}\\
	 \Delta_{[\ell\ell\ell]}\,g_{[\ell \ell\ell ]}^{\rm (n)}  &\,=\, \frac{K^{\ell-1}}{(k_1k_2k_3)^{2\ell+1}}\Big[k_2^2k_3^2(k_2^{2\ell-1}-k_3^{2\ell-1})+\text{cyc.}\Big]\, ,\label{Qlllm}
\end{align}
where the differential operators are given by
\begin{align}
	\vec\Delta_{[\ell\ell 0]} &\equiv \sum_{a=1}^3\k_a\partial_{k_a}^2 \, ,\\
	\Delta_{[\ell\ell\ell]} &\equiv(k_2-k_3)\partial_{k_1}^2+(k_3-k_1)\partial_{k_2}^2+(k_1-k_2)\partial_{k_3}^2\, .
\end{align}
The solutions to \eqref{Qll0} and \eqref{Qllla} are simply powers of $K$, and dilatation symmetry fixes each of these to be
\begin{align}
	g_{[\ell\ell 0]} &= \frac{1}{K^{2\ell}}\quad  \Rightarrow \quad f_{[\ell \ell0]} =\left(\frac{k_1 k_2}{K^2}\right)^{\ell-1} \frac{1}{K^2}\,,\\
	g_{[\ell\ell \ell]}^{\rm (a)} &=	\frac{1}{K^{3\ell}}\quad \Rightarrow \quad 	f_{[\ell\ell \ell]}^{\rm (a)} =  \left(\frac{k_1k_2k_3}{K^3}\right)^{\ell-1} \frac{1}{K^3}\,,
\end{align}
which  matches the results \eqref{fll0sol} and \eqref{fllla}.

\vskip 4pt
Lastly, we consider the inhomogeneous equation~\eqref{Qlllm}. For $\ell=1$, the source term simply becomes
\begin{align}
	\frac{k_2-k_3}{k_1^3k_2k_3} + \frac{k_3-k_1}{k_1k_2^3k_3}+\frac{k_1-k_2}{k_1k_2k_3^3}\, .
\end{align}
From the structure of the differential operator $\Delta_{[\ell\ell\ell]}$, 
it is easy to see that $g_{[111]}^{(\rm n)}=1/(k_1k_2k_3)$ is the solution. 
The solution for higher spins can again be written recursively: it can be checked by direct substitution that the recursive formula 
\beq
g_{[\ell\ell\ell]}^{({\rm n})} = \frac{K^\ell}{(k_1k_2k_3)^{2\ell-1}}\prod_{j=1}^{\ell-1} {\cal S}_{k_1}^{[j]} {\cal S}_{k_2}^{[j]} {\cal S}_{k_3}^{[j]}   \left(\int\rd K\right)^{2(\ell-1)}\, \frac{1}{K} \, ,
\eeq
solves the equation, which after using \eqref{gnlll} reproduces the form factor \eqref{fnlll} from the bulk.

\subsection{Four-Point Functions}
We next consider the transmutation of four-point functions. 
Four-point correlators arising from contact interactions have essentially the same properties as the three-point examples already considered and can be treated similarly. The conceptually new features at four points come from bulk exchanges. 
As before, we can take two complementary viewpoints on lifting these correlators from flat space to de Sitter. Inspection of the bulk time integrals giving rise to the exchange allows us to identify a set of operations that generate these integrals from their flat-space counterparts. However, we can also directly leverage the lessons learned from transmuting three-point correlators combined with information about the cuts of the four-point correlators. We will see that transmutation operators for exchange graphs can be understood as operations that transform the cut of a flat-space exchange diagram into the cut of the corresponding de Sitter correlator. Acting with this operator on the full flat-space correlator then lifts it to de Sitter space.

\subsubsection{Transmutation From Cuts}
\label{sec:transfromcuts}

We first illustrate the logic in a simple setting where only external lines need to be transmuted, before turning to  the more complicated example of spinning particle exchange, which requires transmuting internal lines as well.

\subsubsection*{Correlators with external spin}

Consider the four-point function of a single spin-$\ell$ current and three conformally coupled scalars, $\langle J_\ell \varphi\varphi\varphi\rangle$, where the exchanged particle is a scalar. For simplicity, we will focus on the $s$-channel contribution, but other channels are related to this by simple permutations. 

\vskip4pt
We first consider the case of an external spin-1 field, where the cutting rule \eqref{eq:2siteformula} reads
\be
{\rm Disc}[\psi_{J \vp^3}] = -2s\,\tl\psi_{J \vp^2}\,\tl{\psi}_{\vp^3}\,.
\ee
The idea is to infer the transmutation of the four-point function $\psi_{J \vp^3}$ from the transmutations of the two shifted three-point functions on the right-hand side.
 In flat space, it is easy to compute the full correlator directly, but 
in order to parallel the de Sitter treatment, we can also construct its cut from the available three-point data:
\beq
{\rm Disc}[\psi_{J\vp^3}^{({\rm flat})}]= i(\vec k_2\cdot\vec\xi_1)\frac{2s}{ (k_{12}^2-s^2)(k_{34}^2-s^2)}\,,
\label{eq:JOOOflat}
\eeq
where the right-hand side is the product of shifted versions of~\eqref{eq:flat3pt} and~\eqref{equ:JOO}. The cut in de Sitter space is very closely related. The $\psi_{J\vp^2}$ wavefunction is identical because the coupling is Weyl invariant, but the $\psi_{\vp^3}$ wavefunction is instead given by~\eqref{equ:psi3ds}. Putting these elements together, the cut of the de Sitter wavefunction is
\beq
{\rm Disc}[\psi_{J\vp^3}^{({\rm dS})}]=-(\vec k_2\cdot\vec\xi_1)\frac{1}{ (k_{12}^2-s^2)}\log\left(\frac{k_{34}-s}{k_{34}+s}\right) .
\label{eq:JOOOds}
\eeq
The only difference between~\eqref{eq:JOOOflat} and~\eqref{eq:JOOOds} is the factor coming from the shifted $\psi_{\vp^3}$ wavefunction, but it is straightforward to generate this structure from the shifted wavefunction in flat space:
\beq
\begin{aligned}
\widetilde \psi_{\vp^3}^{{\rm (dS)}} &=  -i\int_{k_{34}}^\infty\rd \tilde k_{34} \, \widetilde \psi_{\vp^3}^{{\rm (flat)}} \\[4pt]
&=  -i\int_{k_{34}}^\infty\rd \tilde k_{34}\, \frac{1}{\tilde k_{34}^2-s^2} = \frac{i}{2s} \log\left(\frac{k_{34}-s}{k_{34}+s}\right) ,
\end{aligned}
\label{equ:TransPhi3}
\eeq
which is the same relation we utilized in~\eqref{eq:integratecutflat4pt}. The idea is then to integrate the full flat-space four-point function in the same way: 
\begin{align}
\psi_{J\vp^3}^{({\rm dS})} &= -i \int_{k_{34}}^\infty\rd \tilde k_{34}\, \psi_{J\vp^3}^{({\rm flat})} \nonumber \\[4pt]
 &= - \int_{k_{34}}^\infty\rd \tilde k_{34}\, \frac{ (\vec k_2\cdot\vec\xi_1)}{(k_{12}+\tilde k_{34})(k_{12}+s)(\tilde k_{34}+s)} = \frac{ (\vec k_2\cdot\vec\xi_1)}{k_{12}^2-s^2}\log\left(\frac{k_{34}+s}{E}\right) , \label{eq:JOOOdslift}
 \end{align}
which is indeed the answer found in~\cite{Baumann:2020dch}. It is also straightforward to check that this expression reproduces the cut~\eqref{eq:JOOOds}. 

\vskip4pt
The spin-2 case is very similar. The cuts in flat space and in de Sitter are now given by
\begin{align}
{\rm Disc}[\psi_{T\vp^3}^{({\rm flat})}] &= -(\vec k_2\cdot\vec\xi_1)^2\frac{2s}{ (k_{12}^2-s^2)(k_{34}^2-s^2)}\,,
\label{eq:TOOOflat} \\[4pt]
{\rm Disc}[\psi_{T\vp^3}^{({\rm dS})}] &=-(\vec k_2\cdot\vec\xi_1)^2\frac{k_{12}^2+2k_{12}k_1-s^2}{ (k_{12}^2-s^2)^2}i\log\left(\frac{k_{34}-s}{k_{34}+s}\right) ,
\label{eq:TOOOds}
\end{align}
where the first factor in~\eqref{eq:TOOOds} is the shifted $\psi_{T\vp^2}$  wavefunction~\cite{Baumann:2020dch}. 
The $\psi_{T\vp^2}$  wavefunctions in flat space and de Sitter are related by the following transmutation
\beq
\begin{aligned}
\widetilde \psi_{T\vp^2}^{{\rm (dS)}} &=  {\cal S}_{k_1}^{[1]} \, \widetilde \psi_{T\vp^2}^{{\rm (flat)}} \\[4pt]
&=(\vec k_2\cdot\vec\xi_1)^2{\cal S}_{k_1}^{[1]}\frac{1}{k_{12}^2-s^2}=(\vec k_2\cdot\vec\xi_1)^2\frac{k_{12}^2+2k_{12}k_1-s^2}{ (k_{12}^2-s^2)^2}\,.
\end{aligned}
\label{eq:lifttpptodS}
\eeq
To transform the bulk $\phi^3$ vertex, we integrate with respect to $k_{34}$ in the same way as in (\ref{equ:TransPhi3}). Combining the transmutations of the left and right vertices, we can generate the full de Sitter correlator from the flat-space one
\begin{align}
\psi_{T\vp^3}^{({\rm dS})} &= -i {\cal S}_{k_1}^{[1]}\int_{k_{34}}^\infty\rd \tilde k_{34}\,\psi_{T\vp^3}^{({\rm flat})}\nonumber \\[4pt]
&= -i {\cal S}_{k_1}^{[1]}\int_{k_{34}}^\infty\rd \tilde k_{34}\,\frac{(\vec k_2\cdot\vec\xi_1)^2}{\tilde EE_L\tilde E_R} \nonumber \\[4pt]
&= i(\vec k_2\cdot\vec\xi_1)^2\left[\frac{k_{12}^2+2k_{12}k_1-s^2}{E_L^2(k_{12}-s)^2}\log\left(\frac{E_R}{E}\right)+\frac{k_1}{EE_L(k_{12}-s)}\right] , \label{eq:TOOOdslift}
\end{align}
where the energies with a tilde over them contain the variable being integrated, $\tilde k_{34}$. This
also reproduces the answer found in~\cite{Baumann:2020dch}.

\vskip4pt
Notice that both~\eqref{eq:JOOOdslift} and~\eqref{eq:TOOOdslift} apparently have singularities in the folded configuration $k_{12}=s$. However, these singularities are fictitious---if one tries to probe them, the other parts of the correlator vanish, so everything is actually regular. This is manifest in the flat-space expressions that these correlators are transmuted from, which only have physical singularities. Since the transmutation operations do not introduce any singularities, it then follows that the de Sitter correlators can only have singularities at the same locations (though possibly of higher order). 
An advantage of the transmutation approach therefore is that we see very explicitly how the complicated structure of subleading singularities in de Sitter space follows inevitably from the simpler flat-space structure.

\vskip4pt
The obvious generalization of the previous formulas to higher spin is\footnote{Although we have taken the approach of deriving the relevant transmutation operation by looking at the cuts of four-point functions, one can of course equally well parallel the treatment in \S\ref{sec:3ptbulktransmutation}. In this case, the relevant bulk time integral is of the form
\beq
\psi_{J_\ell\vp^3}^{({\rm dS})} =  (\vec\xi_1\cdot \vec k_2)^\ell \int \rd \eta\,\rd \eta'\,  \eta^{\ell-2}{\cal K}_{\ell- \frac{1}{2}}(k_1,\eta)e^{ik_{2}\eta}\frac{1}{\eta'}e^{ik_{34}\eta'}{\cal G}^{({\rm flat})}(s;\eta,\eta')\,,
\eeq
where we have performed similar manipulations to before, along with the relation between the conformally coupled and flat-space bulk-to-bulk propagators: ${\cal G}_{1/2}(s;\eta,\eta') =\eta\eta'{\cal G}^{({\rm flat})}(s;\eta,\eta')$. It is then straightforward to see that this time integral is related to the flat-space one by the action of the operator in~\eqref{eq:spinlschannelJOOO}.
} 
\beq
\psi_{J_\ell\vp^3}^{({\rm dS})} = -i \prod_{j=1}^{\ell-1} {\cal S}_{k_1}^{[j]}\int_{k_{34}}^\infty\rd\tilde k_{34}\, \psi_{J_\ell\vp^3}^{({\rm flat})}\,,
\label{eq:spinlschannelJOOO}
\eeq
where $\psi_{J_\ell\vp^3}^{({\rm flat})}$ is given by the simple expression
\beq
\psi_{J_\ell\vp^3}^{({\rm flat})} = \frac{(\vec k_2\cdot\vec\xi_1)^\ell}{EE_LE_R}\,.
\eeq
However, these higher-spin cases are not really physical. As noted before, we have only constructed the $s$-channel exchange part of the correlator, and in order for correlators involving external particles with spin to be consistent, they must also satisfy the current conservation Ward--Takahashi identities, which only hold for $\ell\leq 2$ in flat space~\cite{Baumann:2020dch}. The connection between the de Sitter and flat-space correlators implies that the same is true in de Sitter space---there is no way to combine exchanges in different channels to satisfy the Ward--Takahashi identities for correlators of the form $\langle J_\ell\vp\vp\vp\rangle$ if $\ell >2$~\cite{1866373}.

\subsubsection{Correlators With Internal Spin}
Next, we consider the transmutation of correlators involving the exchange of particles with spin.
One additional complication that arises when the internal field has spin 
is that there are contributions from all helicities of the exchanged particle (and constrained potential modes for massless fields). We can decompose such exchanges into their various helicity components. For example, the four-point function of conformally coupled scalars can be written as 
\beq
\psi_{\vp^4}^{(\ell)} = \sum_m f_{(\ell,m)}\, \tl\Pi_{\ell,m}\,,
\label{eq:helicitydecomp}
\eeq
where $\tl\Pi_{\ell,m}$ are polarization structures and $f_{(\ell,m)}$ are form factors.
Since only the highest-helicity piece, $\tl\Pi_{\ell,\ell}$, survives in the cutting equation, it will be most straightforward to lift these components. Transmuting the lower-helicity form factors will require a little more work.

\vskip4pt
As an illustration, we first consider the exchange of a massless spin-2 field between scalars. 
Using the cutting rule, it is easy to show that the cuts of the correlator in flat space and de Sitter space are
\begin{align}
{\rm Disc}[\psi^{(T)}_{(\rm flat)}] &= -2s \frac{  \tl\Pi_{2,2}}{6(k_{12}^2-s^2)(k_{34}^2-s^2)}\,,
\label{eq:flatTdisc} \\
{\rm Disc}[\psi^{(T)}_{(\rm dS)}] &= -2 s^3 \frac{2\tl\Pi_{2,2}}{3(k_{12}^2-s^2)^2(k_{34}^2-s^2)^2}\,.
\label{eq:dsTdisc}
\end{align}
The goal is to find an operator that transforms the cut of the flat-space exchange \eqref{eq:flatTdisc} into that of its de Sitter counterpart \eqref{eq:dsTdisc}.
The form of the two expressions suggests that we are looking for a differential operator that has single derivatives with respect to $k_{12}$ and $k_{34}$. One possibility that does the job would seem to be
\beq
\frac{s^2}{2k_{12}k_{34}}\partial_{k_{12}}\partial_{k_{34}}\,.
\eeq
However, this is not the correct operator---it does transmute the cut correctly, but it introduces spurious singularities at $k_{12}=0$ and $k_{34}=0$. We must impose the additional restriction that if our differential operator has singularities, it only has {\it physical} singularities. With this requirement, the following operator transforms the form factor of~\eqref{eq:flatTdisc} into that of~\eqref{eq:dsTdisc} 
\beq
{\cal I}^{[1]}_{s} \equiv \frac{2}{E}\left(\partial_{k_{12}}+\partial_{k_{34}}\right)+\partial_{k_{12}}\partial_{k_{34}}\,.
\label{eq:spin2exchI}
\eeq
An interesting feature of this operator is that it does not factorize into operators that act separately on the left and right vertex factors. This is to be expected because we are trying to transmute the internal line which connects the two vertices, so it is not surprising that the operator reflects this structure.

\vskip4pt
To demystify the origin of the operator~\eqref{eq:spin2exchI}, and to find generalizations that lift correlators with higher-spin exchange,
it is helpful to return to the bulk time integrals that give rise to these correlators. For simplicity, we consider first the time integral arising from the exchange of a massless spin-$\ell$ particle between conformally coupled scalars, leading to the following expression for the form factor
\beq
f_{(\ell,\ell)}^{({\rm dS})} = \int\rd\eta\, \rd\eta' e^{ik_{12}\eta}e^{ik_{34}\eta'}(\eta\eta')^{\ell-2}{\cal G}_{\ell- \frac{1}{2}}(s;\eta,\eta')\,.
\label{eq:spinldsexc}
\eeq
In order to relate this expression to its counterpart in flat space, it is useful to define 
\beq
G^{(\nu)}(k;\eta,\eta') \equiv (\eta\eta')^{\nu-3/2}{\cal G}_{\nu}(k;\eta,\eta')\Big\rvert_{\eta_* = 0}\,,
\eeq
which is precisely the combination appearing in~\eqref{eq:spinldsexc} with $\nu = \ell-\frac{1}{2}$.
 Note that we have taken the limit $\eta_* \to 0$ in the definition of $G^{(\nu)}$. The function $G^{(\nu)}$ satisfies the identity~\cite{Benincasa:2019vqr}
\beq
\left(\partial_{\eta}+\partial_{\eta'}\right)  G^{(\nu)}(\eta,\eta') = (\eta+\eta') 2(\nu-1)G^{(\nu-1)}(\eta,\eta') - \eta \eta' \left(\partial_{\eta}+\partial_{\eta'}\right)G^{(\nu-1)}(\eta,\eta')\,,
\label{eq:bbrecursion}
\eeq
which relates bulk-to-bulk propagators with $\nu$ shifted by 1. This relation is particularly useful for massless fields with spin precisely because $\nu = \ell- \frac{1}{2}$, so the highest-helicity parts of their bulk-to-bulk propagators have indices that differ by integers.

\vskip4pt
Using~\eqref{eq:bbrecursion}, it is relatively straightforward to derive relations between $\psi_{4}^{(\ell)}$ with different spins. To do so, we consider the slightly modified time integral:
\beq
E\,f_{(\ell,\ell)}^{({\rm dS})}  = i \int\rd\eta \,\rd\eta' e^{ik_{12}\eta}e^{ik_{34}\eta'}\left(\partial_{\eta}+\partial_{\eta'}\right)G^{(\ell- {1/2})}(s;\eta,\eta')\,,
\eeq
where the factor of $E$ on the left-hand side can be understood by integrating the right-hand side by parts. Using the identity~\eqref{eq:bbrecursion}, and integrating by parts to trade factors of $\eta$ for derivatives with respect to energies, we get 
\beq
f_{(\ell,\ell)}^{({\rm dS})} = \left[2(\ell-1)\frac{1}{E}\left(\partial_{k_{12}}+\partial_{k_{34}}\right)+\partial_{k_{12}}\partial_{k_{34}}\right]f_{(\ell-1,\ell-1)}^{({\rm dS})} \, .
\eeq
 This equation provides a recursion relation between the four-point function of conformally coupled scalars arising from the  highest-helicity part of massless spin-$\ell$ exchange. From this, we can extract the differential operator 
\begin{tcolorbox}[colframe=white,arc=0pt,colback=greyish2]
\beq
{\cal I}^{[j]}_{s} \equiv \frac{2 j}{E}\left(\partial_{k_{12}}+\partial_{k_{34}}\right)+\partial_{k_{12}}\partial_{k_{34}}\,,
\label{eq:intspinraise}
\eeq
\end{tcolorbox}
\noindent
which  raises the spin to $j+1$ when acting on an exchange of spin $j$, and where the derivatives are appropriate for $s$-channel exchange. Note also that the same operator can be used to generate correlation functions involving the exchange of scalar fields with half-integer $\nu$ (corresponding to integer $\Delta$)---e.g.~massless scalars---from flat-space exchanges. Ideas along this line were explored in~\cite{Benincasa:2019vqr}.

\vskip4pt
For $\ell=1$, the time integral~\eqref{eq:spinldsexc} is just the usual flat-space four-point correlator $1/(EE_LE_R)$, because 
spin-$1$ fields are conformally coupled.\footnote{Note that the $\ell=0$ scalar field that fits into this series of exchanges is the conformally coupled scalar on de Sitter space.} We can therefore write the highest-helicity form factor in the decomposition~\eqref{eq:helicitydecomp} as
\beq
f_{(\ell,\ell)}^{({\rm dS})} = \prod_{j=1}^{\ell-1}{\cal I}^{[j]}_s\, \frac{1}{EE_LE_R}\, .
\eeq
We could, of course, have derived~\eqref{eq:intspinraise} by looking at the cuts of higher-spin exchanges in the same way that we did for spin-2, but the approach we have taken is algebraically simpler because it does not require explicit expressions for the shifted wavefunctions appearing in the cut.

\vskip4pt
Since the lower-helicity components of the exchange drop out of the cut, their uplifting to de Sitter space requires more care. However, for conformally coupled external fields it is relatively straightforward, because the parts of the propagator responsible for potential modes are related in a simple way to the highest-helicity piece, and the vertex factors have a simple relation to their flat-space counterparts. The propagator for potential modes in de Sitter is just the $s\to 0$ limit of the highest-helicity propagating mode, which implies that the form factors satisfy
\beq
f_{(\ell, m)}^{({\rm dS})} = \lim_{s\to 0} (-1)^{\ell-m}k_{12}k_{34} f_{(\ell, \ell)}^{({\rm dS})}\,,
\label{eq:limitform}
\eeq
where the factor of $k_{12}k_{34}$ is a consequence of the way we have defined the polarization sums. 
If we want to generate the lower-helicity form factors by acting with a differential operator on the flat-space form factors, we can use the following operator
\beq
\hat {\cal I}^{[j]}_s \equiv k_{12}k_{34}\,{\cal I}^{[j]}_s \frac{1}{k_{12}k_{34}}\,,
\eeq
where derivatives act on everything to their right. 
Note that this operator becomes equivalent to $\partial_E^2$ when acting on the $1/E$ form factors that accompany the lower-helicity polarization sums in flat space.

\vskip4pt
For external fields that are not conformally coupled, the lifting of the longitudinal polarization sums is more complicated. The essential complication is that there is no longer a simple relation between the vertex factors for the lowest-helicity components in flat space and those in de Sitter space. We will see an avatar of this complication in Section~\ref{sec:bootstrapping} when we lift the flat-space graviton Compton correlator to de Sitter space.

\subsubsection*{Examples}
 We now illustrate the transmutation procedure explicitly for some simple cases, first treating a few low-spin exchanges before generalizing to arbitrary spin exchange between conformally coupled scalars. 

\paragraph*{Spin-2 exchange} We first consider massless spin-2 exchange in de Sitter. As an input, we take the four-point function of massless scalars in flat-space exchanging a massless spin-2 particle  
\beq
\psi_{4,\,{\rm flat}}^{(T)} =  \frac{1}{6E E_L E_R}\tl\Pi_{2,2}- \frac{1}{6E}\tl\Pi_{2,1}+ \frac{1}{6E }\tl\Pi_{2,0} +\cdots\,,
\label{eq:flatliftspin2exc}
\eeq
where the ellipses  denote the second line appearing in the explicit answer~\eqref{eq:flatspacespin2exc}. While these additional terms are necessary to make the flat-space correlator vanish in the soft limit, it turns out that they do not contribute to the uplifting---instead they produce a de Sitter-invariant contact solution, 
which is degenerate with higher-dimensional contact interactions. This can be understood from the difference in the structure of the longitudinal part of the graviton propagators in de Sitter space and in flat space.
Comparing~\eqref{eq:flatliftspin2exc} to~\eqref{eq:helicitydecomp}, we identify the following form factors 
\beq
f_{(2,2)}^{\rm (flat)} = \frac{1}{6E E_L E_R}\,, \qquad f_{(2,1)}^{\rm (flat)}  = -\frac{1}{6E }\,,\qquad f_{(2,0)}^{\rm (flat)}  = \frac{1}{6E }\,.
\eeq
Acting on these form factors with the relevant transmutation operators, we obtain
\begin{align}
f_{(2,2)}^{({\rm dS})} &={\cal I}^{[1]}_s f_{(2,2)}^{\rm (flat)}   =-\frac{sE+E_LE_R}{3E^3E_L^2E_R^2}\,,\\
f_{(2,1)}^{({\rm dS})} &=\hat{\cal I}^{[1]}_s f_{(2,1)}^{\rm (flat)}   =\frac{1}{3E^3}\,,\\
f_{(2,0)}^{({\rm dS})} &=\hat {\cal I}^{[1]}_s f_{(2,0)}^{\rm (flat)}   =-\frac{1}{3E^3}\, ,
\end{align}
which, combined with the associated polarization structures, leads to the expression
\beq
\psi^{(T)}_{4,\,{\rm dS}} = -\frac{sE+E_LE_R}{3E^3E_L^2E_R^2} \tl\Pi_{2,2} + \frac{1}{3E^3}\tl\Pi_{2,1}- \frac{1 }{3E^3 }\widetilde \Pi_{2,0}\,.
\label{eq:transmutedspin2cc}
\eeq
This is exactly the answer obtained in~\eqref{eq:spin2dsrecurse} from recursion, and computed in~\cite{Arkani-Hamed:2018kmz,Baumann:2019oyu,Baumann:2020dch}. The benefit of the transmutation approach is that we see the commonalities between the de Sitter and flat-space correlators.

\paragraph*{Spin-3 exchange} Next, we consider massless spin-3 exchange.
Much like for spin-1 exchange, this requires multiple flavors of scalars in order to have a non-vanishing correlator, but we suppress this flavor structure in the following. In flat space, the spin-3 exchange correlator is
\beq
	\psi^{(\ell=3)}_{4,\,{\rm flat}} =\frac{1}{24EE_LE_R}\tl\Pi_{3,3}-\frac{1}{24E}\tl\Pi_{3,2}+\frac{1}{24E}\tl\Pi_{3,1}  - \frac{1}{24E}\tl\Pi_{3,0} +\cdots\,,
	\label{equ:spin3ex}
\eeq
where we have defined
\beq
\begin{aligned}
\tl\Pi_{3,3} &\equiv s^6\Pi_{3,3}\,,  \qquad\qquad~\tl\Pi_{3,2}  \equiv s^4\Pi_{3,2}\,, &\quad\tl\Pi_{3,1} \equiv (E_LE_R-sE)s^2\Pi_{3,1}\,,\\\tl\Pi_{3,0} &\equiv \left[ (E_LE_R-sE)^2-s^2E^2/3 \right] \Pi_{3,0} \, .
 \end{aligned}
\eeq
As in the spin-2 case, this result can be obtained by demanding that the correlator vanishes in the soft limits and reduces to a Lorentz-invariant amplitude on the total energy singularity (up to contact contributions).
The polarization sums $\Pi_{3,m}$ can be found in Appendix~\ref{app:pols}.
The form factors in \eqref{equ:spin3ex} are
 $1/EE_LE_R$ for the helicity-3 component and $\pm 1/E$ for the other helicities. Acting on these form factors with the same transmutation operators as before, we obtain the following result in de Sitter space:
\beq
\psi^{(\ell=3)}_{4,\,{\rm dS}} = \frac{3E_LE_R(E_LE_R+sE)+2s^2E^2}{3E^5 E_L^3E_R^3} \tl\Pi_{3,3} - \frac{1}{E^5}\tl\Pi_{3,2} + \frac{1}{E^5} \tl\Pi_{3,1} - \frac{1}{E^5}\tl\Pi_{3,0} \, .
\label{equ:Spin3Exchange}
\eeq
It can be checked that this answer is de Sitter invariant and captures the exchange of a massless spin-3 particle in the bulk.

\paragraph*{Spin-${\boldsymbol \ell}$ exchange} The above results can be generalized to arbitrary spin exchange.
The spin-$\ell$ exchange correlator in flat space takes the form
\beq
	\psi^{(\ell)}_{4,\text{flat}} \propto  \frac{1}{E}\left[\frac{1}{E_L E_R}\tl\Pi_{\ell,\ell}+ \sum_{m=0}^{\ell-1} \tl \Pi_{\ell,m}\right]+\cdots\,,\label{psi4flat}
\eeq
where the ellipses denote terms that can be fixed by demanding that  the correlator has vanishing soft limits. These terms are not necessary to produce a de Sitter-invariant correlator for the same reason as in the spin-2 case.
The polarization sums are given by (see Appendix~\ref{app:pols})
\begin{align}
	\tl\Pi_{\ell,\ell} &= s^{2\ell}\Pi_{\ell,\ell}\, , \\
	\tl\Pi_{\ell,m} &= s^{2m}(sE)^{\ell-m-1}\tl P_{\ell-1}^m\bigg(\frac{E_LE_R}{s E}-1\bigg)\,\Pi_{\ell,m}\quad (m\ne \ell)\, ,
\end{align}
where we have defined a modified version of the associated Legendre polynomial
\beq
		\tl P_{\ell-1}^m(y) \equiv\frac{(-2)^{\ell-2}(\ell-2)!(\ell-m-1)!}{(2\ell-3)!} \frac{P_{\ell-1}^m(y)}{(1-y^2)^{m/2}} = (-1)^m y^{\ell-1} +\cdots\, .\label{eq:assocLeg}
\eeq
The flat-space form factors take the simple form
\beq
	f_{(\ell,\ell)}^{\rm (flat)}=\frac{1}{EE_LE_R}\, ,\qquad f_{(\ell,m\ne\ell)}^{\rm (flat)}=\frac{(-1)^{\ell-m}}{E}\, .
\eeq
Lifting these form factors using the operator \eqref{eq:intspinraise}, we find the corresponding result for massless spin-$\ell$ exchange in de Sitter space\footnote{We may also express the correlator as
\beq
	\psi^{(\ell,\Delta=\ell+1)}_{4,\text{dS}} =  \sum_{m=0}^{\ell} \frac{s^{\ell-m-1} }{E^{\ell+m}}\tl P_{\ell-1}^{(m,-m)}\bigg(\frac{E_LE_R}{s E}-1\bigg)\,\Pi_{\ell,m}\,.
\eeq
This representation is nice in that it treats all helicities on an equal footing, but it makes the energy singularities less manifest.
} 
\begin{tcolorbox}[colframe=white,arc=0pt,colback=greyish2]
\vspace{-0.5cm}
\begin{align}
	\psi^{(\ell,\Delta=\ell+1)}_{4,\text{dS}} \propto  \frac{1}{E^{2\ell-1}}\left[\frac{p_\ell}{E_L^\ell E_R^\ell}\tl \Pi_{\ell,\ell}+ \sum_{m=0}^{\ell-1} \tl \Pi_{\ell,m}\right] , \label{psi4dS}
\end{align}
\end{tcolorbox}
\noindent
where the coefficient of the highest-helicity contribution,
\beq
	p_\ell \equiv (sE)^{\ell-1}\tl P^{(\ell,-\ell)}_{\ell-1}\bigg(\frac{E_LE_R}{s E}-1\bigg)\, ,
\eeq
is defined in terms of a modified version of the Jacobi polynomial
\beq
	\tl P_{\ell-1}^{(\ell,-\ell)}(y) \equiv \frac{2^{\ell-2}(\ell-2)!(\ell-1)!}{(2\ell-3)!}P_{\ell-1}^{(\ell,-\ell)}(y) = y^{\ell-1} + \cdots\, .
\eeq
It can be checked that \eqref{psi4dS} is consistent with the general spin-$\ell$ exchange result \cite{Arkani-Hamed:2018kmz, Baumann:2019oyu}, and is thus conformally invariant.

\subsubsection{Boost-Breaking From Transmutation}
\label{sec:dotphi3bbreaking}

An appealing feature of the transmutation procedure is that it does not rely strongly on de Sitter symmetry and can therefore be extended to cases where the bulk interactions break the symmetry. Since the interactions in inflationary models necessarily break de Sitter symmetry~\cite{Creminelli:2006xe,Cheung:2007st,Green:2020ebl}, this situation is particularly relevant phenomenologically, especially in models with large non-Gaussianities~\cite{Creminelli:2003iq,Alishahiha:2004eh,Chen:2006nt}.

\vskip 4pt
As an illustrative example, we again consider the four-point function arising from a $\dot\phi^3$ interaction in de Sitter space, given by the action~\eqref{eq:dot3lagrangian}; see also~\cite{Chu:2018ovy, Chu:2018kec}.
The three-point wavefunction coefficient in this theory is fairly simple:
\beq
\psi_{3, \dot\phi^3}^{({\rm dS})}(k_1,k_2,s) = -\frac{2(k_1k_2s)^2}{(k_1+k_2+s)^3}\,,
\eeq
and we can use it to generate the cut of the  four-point function (in the $s$-channel) 
\begin{align}
{\rm Disc}[\psi_{4,\dot\phi^3}^{({\rm dS})}] &= -2 s^3\,\tl\psi^{({\rm dS})}_{3,\dot\phi^3}(k_1,k_2,s)\times \tl\psi^{({\rm dS})}_{3, \dot\phi^3}(s,k_3,k_4) \nonumber \\[4pt]
&= -2s^3\,\frac{2k_1^2k_2^2(3k_{12}^2+s^2)}{(k_{12}^2-s^2)^3}\frac{2k_3^2k_4^2(3k_{34}^2+s^2)}{(k_{34}^2-s^2)^3}\,.
\label{equ:dphi3cut}
\end{align}
We want to find a differential operator that transforms the cut of the flat-space exchange $1/(EE_LE_R)$, given by
\beq
{\rm Disc}[\psi_{4,\phi^3}^{({\rm flat})}]  = -2s \frac{1}{(k_{12}^2-s^2)(k_{34}^2-s^2)}\,,
\label{eq:flatspace4ptcutbbreak}
\eeq
into (\ref{equ:dphi3cut}). A very simple such operator is\hskip 2pt\footnote{It is relatively easy to see that this operator will work as desired because of the relation
\beq
 \partial_{k_{12}}^2 \frac{1}{k_{12}^2-s^2} = \frac{2(3k_{12}^2+s^2)}{(k_{12}^2-s^2)^3}\,,
\eeq
which transforms the shifted three-point function of a massless scalar in flat space into that of $\dot\phi^3$ theory in de Sitter space.}
\beq
{\cal B}_{s} \equiv (k_1 k_2 k_3 k_4 )^2\,s^2 \partial_{k_{12}}^2\partial_{k_{34}}^2\,.
\label{eq:boostbreaking1}
\eeq
This operator does not introduce any spurious singularities, so we can act with it on the full flat-space exchange solution to obtain
\beq
\begin{aligned}
\widehat\psi_{4, \dot\phi^3}^{({\rm dS})} \equiv{\cal B}_s\psi_{4,\phi^3}^{({\rm flat})}\,=\, 2s(k_1k_2k_3k_4)^2\bigg[\frac{6}{E^5}\left(\frac{1}{E_L}+\frac{1}{E_R}\right)&+\frac{3}{E^4}\left(\frac{1}{E_L^2}+\frac{1}{E_R^2}\right) \\
%&\hspace{4cm}
&+\frac{1}{E^3}\left(\frac{1}{E_L^3}+\frac{1}{E_R^3}\right)-\frac{1}{E_L^3E_R^3}\bigg]\,.
\end{aligned}
\eeq
Interestingly, this is not quite the full answer~\eqref{eq:dotphi34ptfunction}, which is
\beq
\psi_{4, \dot\phi^3}^{({\rm dS})}=\widehat\psi_{4, \dot\phi^3}^{({\rm dS})}-\frac{24(k_1k_2k_3k_4)^2}{E^5}\,.
\eeq
We see that the solution generated by the operator~\eqref{eq:boostbreaking1} differs from the result of a bulk calculation by a contact solution. This is completely consistent, because we inferred the structure of the operator~\eqref{eq:boostbreaking1} by transmuting the cut of the four-point function, which precisely projects out the contributions from contact solutions. This solution also coincides with the structure that arises from a $\dot\phi^4$ bulk interaction.
By manipulating the bulk time integral, we can find another operator that transmutes~\eqref{eq:flatspace4ptcutbbreak} into~\eqref{equ:dphi3cut}
\beq
{\cal B}_s^{\hskip 1pt \prime} \equiv (k_1k_2k_3k_4)^2\partial_{k_{12}}\partial_{k_{34}}\left(k_{12}k_{34}{\cal I}^{[1]}_s\right) ,
\eeq
where the derivatives act on everything to their right. This operator generates exactly the four-point function~\eqref{eq:dotphi34ptfunction} by acting on the flat-space exchange:
\beq
\left({\cal B}_s^{\hskip 1pt \prime} -{\cal B}_s\right)\psi_{4,\phi^3}^{(\rm flat)} = -\frac{24(k_1k_2k_3k_4)^2}{E^5}\,.
\eeq
As expected, the combination ${\cal B}_s^{\hskip 1pt \prime} -{\cal B}_s$ annihilates the cut~\eqref{eq:flatspace4ptcutbbreak}. This suggests that there is an organization of the bulk effective field theory from the boundary point of view, where operators that transmute the cuts of flat-space correlators generate the corresponding contributions from exchanges in de Sitter space, while combinations of operators that annihilate flat-space cuts generate the contributions from contact diagrams in de Sitter. It would be interesting to investigate this more systematically.

\vskip4pt
We have, so far, focused on the simplest correlation functions that arise from de Sitter boost-breaking interactions, but the general philosophy can be readily extended to more complicated situations. Since this procedure takes as an input the three-point wavefunction, it synergizes naturally with recent progress in bootstrapping these building blocks~\cite{Pajer:2020wxk,Jazayeri:2021fvk}.
As an example, in Appendix~\ref{app:EFTapp}, we apply this approach to the EFT of inflation and generate the contribution to the trispectrum arising from the $\dot\pi(\nabla\pi)^2$ operator.
In this case, the result of a direct bulk computation is very complicated, but we show how it arises from the transmutation of a simple seed in flat space.

\subsection{Summary: Transmutation Rules}

We have seen in a number of examples how flat-space correlation functions can be transformed to de Sitter space. We now wish to distill some general lessons from these examples that are more widely applicable.

\begin{figure}[t!]
   \centering
           \begin{tabular}{cc}
\scalebox{1}{
\begin{tikzpicture}[line width=1. pt, scale=2,
sines/.style={
        line width=1pt,
        line join=round, 
        draw=blue3, 
        decorate, 
        decoration={complete sines, number of sines=4, amplitude=4pt}
    }
]
\draw[color=gray, fill=gray!2,, line width=1.pt] (.2,0) -- (0,0.7);
\draw[color=gray, fill=gray!2,, line width=1.pt] (-1.2,0) -- (-1,0.7);
\filldraw[color=gray, fill=gray!2, very thick](-1,0.7) circle[radius=0.1] ;
\filldraw[color=gray, fill=gray!2, very thick](0,0.7) circle[radius=0.1] ;
\draw[black, line width=1.pt] (-1,0.7) -- (0,0.7);
\draw[black, line width=1.pt] (-1,0.7) -- (-0.75,1.3);
\path[postaction={sines}] (-1,0.7) -- (-1.25,1.3);
\draw[black, line width=1.pt] (0,0.7) -- (0.25,1.3);
\draw[black, line width=1.pt] (0,0.7) -- (-0.25,1.3);
\draw[lightgray, line width=2.pt] (-1.8,1.3) -- (.8,1.3);
\filldraw[color=gray, fill=gray!5, very thick](-1.2,-0.1) circle[radius=0.45] node[black]{ $~\displaystyle\prod_{j=1}^{\ell-1}{\cal S}_{k_1}^{[j]}$};
\filldraw[color=gray, fill=gray!5, very thick](.2,-0.1) circle[radius=0.45] node[black]{$~\displaystyle{\int\rd k_{34}}$};
\draw  (-1,0.7) node[blue3, fill, circle, scale=0.25]  {};
\draw  (0,0.7) node[black, fill, circle, scale=0.25]  {};
\node[scale=1] at (-1.25,1.45) {$J_\ell$};
\node[scale=1] at (-0.25,1.45) {$\varphi$};
\node[scale=1] at (-0.75,1.45) {$\varphi$};
\node[scale=1] at (0.25,1.45) {$\varphi$};
\draw[white, line width=0.7pt] (0.8,-0.31) -- (1.2,-0.31);
\end{tikzpicture}}&~~
\scalebox{1}{
\begin{tikzpicture}[line width=1. pt, scale=2,
sines/.style={
        line width=1pt,
        line join=round, 
        draw=green3,
        decorate, 
        decoration={complete sines, number of sines=4, amplitude=4pt}
    },
    sines2/.style={
        line width=1pt,
        line join=round, 
        draw=blue3, 
        decorate, 
        decoration={complete sines, number of sines=6, amplitude=4pt}
    }
]
\draw[color=gray, fill=gray!2,, line width=1.pt] (-0.5,0.1) -- (-0.5,0.7);
\draw[white,postaction={sines2}]  (-1,0.7) -- (0,0.7);
\draw[black, line width=1.pt] (-1,0.7) -- (-0.75,1.3);
\draw[black, line width=1.pt] (-1,0.7) -- (-1.25,1.3);
\draw[black, line width=1.pt] (0,0.7) -- (0.25,1.3);
\draw[black, line width=1.pt] (0,0.7) -- (-0.25,1.3);
\draw[lightgray, line width=2.pt] (-1.8,1.3) -- (.8,1.3);
\filldraw[color=gray, fill=gray!5, very thick](-0.5,-0.1) circle[radius=0.45] node[black]{$~\displaystyle\prod_{j=1}^{\ell-1}{\cal I}_s^{[j]}$};
\node[scale=1] at (-1.25,1.45) {$\varphi$};
\node[scale=1] at (-0.25,1.45) {$\varphi$};
\node[scale=1] at (-0.75,1.45) {$\varphi$};
\node[scale=1] at (0.25,1.45) {$\varphi$};
\draw  (-1,0.7) node[blue3, fill, circle, scale=0.25]  {};
\draw  (0,0.7) node[blue3, fill, circle, scale=0.25]  {};
\draw[white, line width=0.7pt] (0.8,-0.31) -- (1.2,-0.31);
\end{tikzpicture}}
\end{tabular}
   \caption{Illustration of the transmutation operators for $\langle J_\ell\varphi\varphi\varphi\rangle$ ({\it left}) and $\langle \varphi \varphi \varphi \varphi \rangle^{(\ell)}$ ({\it right}).}
  \label{fig:Transmutation}
\end{figure}
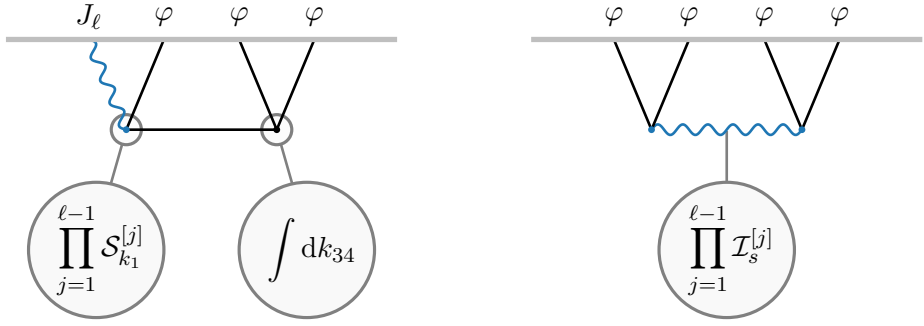

\vskip4pt
First, we introduced a differential operator that maps an external spin-$j$ field to a spin-$(j+1)$ field in de Sitter:
\beq
{\cal S}_{k}^{[j]} \equiv (2j-1)-k\partial_{k}\, ,
\label{eq:spinraisingop2}
\eeq
where $k$ is the energy of the field.
Applying this operator $\ell - 1$ times allows us to transmute a flat-space correlator with an external spin-$\ell$ field to de Sitter space:
\beq
{\cal D}_{[\ell 00]}  \equiv \prod_{j=1}^{\ell-1}{\cal S}_{k_1}^{[j]}\,.
\label{eq:JOOOoperator}
\eeq
In the left panel of Fig.~\ref{fig:Transmutation}, we illustrate how this operator arises in the transmutation of $\langle J_\ell\varphi\varphi\varphi\rangle$.  In this example, we furthermore have to transmute the scalar three-point vertex of massless fields in flat space to conformally coupled fields in de Sitter. This involves integration with respect to $k_{34}$, the sum of the energies of the external legs attached to the scalar vertex.
This integration accounts for the differences in the time dependence of the vertex factors for massless fields in flat space and conformally coupled fields in de Sitter.
 It is also worth reiterating that the simplicity of the operators in this example is due in part to the appearance of conformally coupled scalars in the de Sitter correlator. In cases with fields of other masses---for example, massless fields or fields with spin---additional operations involving integration  will  be needed to transmute these external lines.

\vskip4pt
Another important operation corresponds to the lifting of a spinning internal line. In this case, the operator that effects this transformation is 
\beq
{\cal I}^{[j]}_{s} \equiv \frac{ 2j}{E}\left(\partial_{k_{12}}+\partial_{k_{34}}\right)+\partial_{k_{12}}\partial_{k_{34}}\,,
\label{eq:internalspinup}
\eeq
which maps a spin-$j$ exchange to a spin-$(j+1)$ exchange between conformally coupled scalars in de Sitter, as depicted in the right panel of~Fig.~\ref{fig:Transmutation}. Since spin-$1$ fields are conformally coupled, their transmutation to de Sitter space is trivial. The exchange of fields with higher spins $\ell > 1$ is therefore transmuted by acting $\ell-1$ times with the operator in \eqref{eq:internalspinup}.  In much the same way as for~\eqref{eq:JOOOoperator}, in cases where the external fields are not conformally coupled---which we will encounter in the next section---additional operations are needed to transform the vertices to de Sitter space.

\section{Bootstrapping Using Cuts and Transmutation}
\label{sec:bootstrapping}
To illustrate the power of the bootstrap approach---and the utility of the tools collected in this paper---we will now apply it to a more complex example: the correlator associated with the Compton scattering of gravitons. 
(An even more complicated case, the four-point function of gravitons, will be presented elsewhere~\cite{InProgress}.) 
The expression for this wavefunction coefficient is known in the literature: in \cite{Baumann:2020dch}, we derived it 
in de Sitter space as a solution to the conformal Ward identities utilizing weight-shifting operators to generate the solution from simpler seed solutions involving only scalar fields. 
Even with this technology, the derivation was somewhat involved---in fact, this example arguably pushed the weight-shifting formalism near its limits, which motivates looking for a more streamlined approach to move forward.
Moreover,  the
final answer obtained in~\cite{Baumann:2020dch} was not very illuminating, involving fairly complicated subleading singularities that were fixed by conformal symmetry, but whose connection to other physical principles was obscure.
We will see that---much like in the other examples we have considered so far---the precise structure of singularities follows in a straightforward way from the singularities of the corresponding flat-space process. This gives us some additional insight into the structure of the final answer.

\subsection{Inspiration: Graviton Compton Scattering} 
Our goal is to construct the graviton Compton correlator using only simple physical principles like locality and unitarity.
As a useful motivational success story of this philosophy, it is illuminating to determine the gravitational Compton scattering amplitude from a similar perspective, which will have parallels in the correlator construction. Aside from serving as inspiration, the flat-space scattering amplitude will also be an input to our construction of the correlator.

\vskip 4pt
We have seen that the cuts of correlators provide useful constraints on their analytic structure, and the same is true for scattering amplitudes.
An important constraint on scattering amplitudes is that they must obey the optical theorem
\beq
2\,{\rm Im}[{A}(i \to f)] = i \sum_X\int\rd{\rm LIPS}\,(2\pi)^4\delta(p_i-p_X) \,{A}(i\to X){A}^*(X\to f) \, ,
\label{eq:Smatrixoptical}
\eeq
which relates the imaginary part of a scattering process to the sum over all possible intermediate states between factorized lower-point amplitudes. In this expression $\rd{\rm LIPS}$ indicates that the integral is over the Lorentz-invariant phase space accessible to the on-shell intermediate states~$X$. The famous Cutkosky rules systematize the perturbative expansion of the relation~\eqref{eq:Smatrixoptical}. We now show how the tree-level graviton Compton S-matrix can be derived from these cuts.

\begin{figure}[t!]
   \centering
            \includegraphics[width=.9\textwidth]{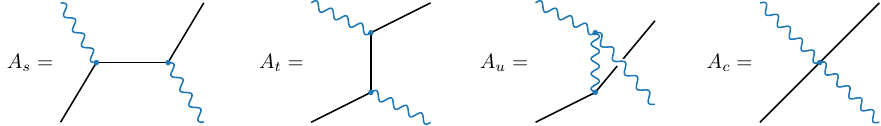}
   \caption{Illustration of the different contributions to gravitational Compton scattering. }
  \label{fig:ComptonAmp}
\end{figure}

\paragraph*{$\boldsymbol{s/t}$-channel} 
Three distinct channels  contribute to the scattering between two gravitons and two scalar particles (see Fig.~\ref{fig:ComptonAmp}). 
In two of them, a scalar is exchanged (in our conventions, the $s$ and $t$-channels), much like in ordinary Abelian vector Compton scattering in scalar QED. In the third channel (the $u$-channel), a graviton is exchanged. The channels with scalar exchange are simpler, so we treat them first. 
For the $s$-channel contribution, the optical theorem reads
\beq
i{A}_{s} - i{A}_{s}^* = -2\pi\delta(S) A_{g\phi\phi}\,A_{\phi g\phi}\,,
\label{eq:SchannelcutGCompton}
\eeq
which expresses the imaginary part of the four-point amplitude in the $s$-channel in terms of (on-shell) three-point amplitudes involving two scalars and a graviton. These three-particle amplitudes have a very simple form: 
\beq
 iA_{g\phi\phi}(p_1,p_2,p_3) =  (\epsilon_1\cdot p_2)^2\,.
 \label{eq:gpp3pt}
\eeq
Substituting this into~\eqref{eq:SchannelcutGCompton} and using a standard distribution identity for the delta function yields
\beq
i{A}_{s} - i{A}_{s}^* = -\left[\frac{1}{S+i \varepsilon} - \frac{1}{S-i\varepsilon} \right] (\epsilon_1\cdot p_2)^2\,(\epsilon_3\cdot p_4)^2 \, .
\eeq
It is easy to extract from this expression
an amplitude with the correct cut:
\beq
i{A}_{s} = -\frac{ (\epsilon_1\cdot p_2)^2\,(\epsilon_3\cdot p_4)^2}{S+i\varepsilon}\, .
\label{eq:gcomptonampSchan}
\eeq
 The $t$-channel result, ${A}_{t}$, is obtained by interchanging $2\leftrightarrow 4$.

\paragraph*{$\boldsymbol{u}$-channel} Deriving the $u$-channel contribution is only slightly more elaborate. This involves graviton exchange, so the optical theorem now gives
\beq
i{A}_{u} - i{A}_{u}^*  =-2\pi\delta(U) {A}^{\mu\nu}_{ggg} \,G_{\mu\nu,\alpha\beta}\,{A}_{g\phi\phi}^{\alpha\beta}\,,
\eeq
where the right-hand side implicitly contains a sum over internal polarizations through the propagator\footnote{Here, we are using the simplified graviton propagator of~\cite{Bern:1999ji}. Using the graviton propagator in a different gauge does not change the result.} $G_{\mu\nu,\alpha\beta} = - i\eta_{\mu\alpha}\eta_{\nu\beta}$.
The graviton-scalar-scalar on-shell three-point amplitude is just a permutation of~\eqref{eq:gpp3pt}, while the graviton three-point amplitude is given by 
\beq
i{A}_{ggg}(p_1,p_2,p_3) =  \Big[ (p_1\cdot \epsilon_3)(\epsilon_1\cdot \epsilon_2)+(p_3\cdot \epsilon_2)(\epsilon_1\cdot \epsilon_3)+(p_2\cdot \epsilon_1)(\epsilon_2\cdot \epsilon_3)\Big]^2\,.
\eeq
Putting these elements together, and performing similar steps as for the $s$-channel, we find that the $u$-channel amplitude with the correct cut is
\beq
i{A}_{u} =- \frac{1}{U+i\varepsilon}\Big[ -ST (\epsilon_1\cdot\epsilon_3)^2+2(T-S)(\epsilon_1\cdot\epsilon_3)(\epsilon_1\circ\epsilon_3)+4(\epsilon_1\circ\epsilon_3)^2\Big]\,,
\eeq
where $\epsilon_1\circ\epsilon_3$ denotes the combination
\beq
\epsilon_1\circ\epsilon_3 \equiv  (\epsilon_1\cdot p_2)(\epsilon_3\cdot p_4)-(\epsilon_1\cdot p_4)(\epsilon_3\cdot p_2)\,.
\label{eq:epsiloncirc}
\eeq
In addition to the $s, t$ and $u$-channels, the amplitude has one further piece.

\paragraph*{Contact term}   The sum of the above contributions is not  yet the full answer because it isn't gauge invariant. Restoring gauge invariance requires a contact term contribution. 
The most general possibility arising from a two-derivative vertex, and with the correct permutation symmetry, is
\beq
{A}_{c} = c_1(\epsilon_1\cdot\epsilon_3)^2+c_2(\epsilon_1\cdot \epsilon_3)\Big[  (\epsilon_1\cdot p_2) (\epsilon_3\cdot p_4)+  (\epsilon_1\cdot p_4)(\epsilon_3\cdot p_2)\Big]\,.
\eeq
Demanding gauge invariance of the total amplitude
\beq
{A} \equiv c_s{A}_{s}+c_t{A}_{t}+c_u {A}_{u}+{A}_{c}\,,
\eeq
then fixes all coefficients up to an overall normalization 
\beq
c_s = c_t= 4 c_u\,,~~~~~c_1 = 0\,,~~~~~c_2 = -\frac{c_s}{2}\,.
\eeq
The full amplitude can then be written in the extremely simple KLT-like form 
\beq
iA = -\frac{ST}{4U}\left[\frac{4(\epsilon_1\cdot p_2)(\epsilon_3\cdot p_4)}{S}+\frac{4(\epsilon_1\cdot p_4)(\epsilon_3\cdot p_2) }{T} - 2 (\epsilon_1\cdot \epsilon_3)\right]^2\,,
\eeq
where the structure appearing in the brackets is the Abelian vector Compton scattering amplitude.
For our later purposes, it will be convenient to write the amplitude in a somewhat less elegant, but equivalent, form
\beq
\begin{aligned}
iA &=- \frac{4(\epsilon_1\hskip -1pt \cdot \hskip -1pt  p_2)^2 (\epsilon_3\hskip -1pt \cdot \hskip -1pt  p_4)^2}{S} -\frac{4(\epsilon_1 \hskip -1pt \cdot \hskip -1pt  p_4)^2 (\epsilon_3 \hskip -1pt \cdot \hskip -1pt  p_2)^2}{T} \\
&~~~-\frac{ 1 }{U} \left[ \frac{1}{6}U^2P_2\left(1+\frac{2S}{U}\right)(\epsilon_1 \hskip -1pt \cdot \hskip -1pt \epsilon_3)^2  -2UP_1\left(1+\frac{2S}{U}\right)(\epsilon_1\cdot\epsilon_3)(\epsilon_1\circ\epsilon_3) + 4 (\epsilon_1 \circ \epsilon_3)^2   \right]\\
&~~~+\frac{1}{6}U(\epsilon_1 \hskip -1pt \cdot \hskip -1pt \epsilon_3)^2+2(\epsilon_1\cdot \epsilon_3)\Big[(\epsilon_1\cdot p_2)(\epsilon_3\cdot p_4)+(\epsilon_1\cdot p_4)(\epsilon_3\cdot p_2)\Big]\, ,
\end{aligned}
 \label{eq:ComptonSA}
\eeq
where $P_1$ and $P_2$ are the first and second Legendre polynomials.
This result agrees with that obtained by an explicit computation in~\cite{Holstein:2006bh}. It is worth appreciating, however, how much simpler this procedure was than computing the answer directly using Feynman rules.

\subsection{Bootstrapping the Flat-Space Correlator}
Inspired by the simplicity of the bootstrap approach for the graviton Compton scattering amplitude, we now revisit the derivation of the graviton Compton correlator from a similar perspective.
Our strategy will be to first bootstrap this correlator in flat space and then lift the answer to de Sitter using the transmutation procedure discussed in Section~\ref{sec:lifting}.

\subsubsection{Correlator From Cutting} \label{subsec:bootstrappingcutting}
The benefit of first constructing the correlator in flat space is that nearly all of its structure is fixed by singularities in a transparent way. Our strategy will be same as for the amplitude: first construct the cut of the wavefunction in its various exchange channels and then infer the contributions away from the cut by demanding that the answer has the correct singularities.

\paragraph*{$\boldsymbol{s/t}$-channel} 
It is again simplest to start with the $s$-channel.  
We can build the cut of the correlator in this channel using the wavefunction coefficient $\psi_{T\vp\vp}$, found in~\eqref{equ:JOO}. 
We obtain
\beq \label{cuteqComs}
{\rm Disc}[\psi_{T\vp T\vp}^{(s)}]
= -2 s\frac{4(\vec k_2\cdot\vec\xi_1)^2(\vec k_4\cdot\vec\xi_3)^2}{(k_{12}^2-s^2)(k_{34}^2-s^2)}\, .
\eeq
Recall that this is an expression of the form $\psi_{T\vp T\vp}^{(s)}(k_{12},k_{34})+\psi_{T\vp T\vp}^{(s)}(-k_{12},-k_{34})$, from which we want to infer $\psi_{T\vp T\vp}^{(s)}(k_{12},k_{34})$. It is clear that the polarization structure must be $(\vec k_2\cdot\vec\xi_1)^2(\vec k_4\cdot\vec\xi_3)^2$, so we only need to determine the structure of the energy variables. 
Locality dictates that the $s$-channel has singularities when $E$, $E_L^{(s)}$ and $E_R^{(s)}$ go to zero. In flat space, these singularities are simple poles. 
The unique object that reproduces~\eqref{cuteqComs} and only has these singularities is 
\beq
\psi_{T\vp T\vp}^{(s)} = \frac{4(\vec k_2\cdot\vec\xi_1)^2(\vec k_4\cdot\vec\xi_3)^2}{EE_L^{(s)}E_R^{(s)}}\,.
\eeq
Notice that we are forced to have a total energy singularity, whose residue reproduces~\eqref{eq:gcomptonampSchan}. The $t$-channel result is obtained from this by the permutation $2\leftrightarrow 4$.

\paragraph*{$\boldsymbol{u}$-channel}  
The $u$-channel is a bit more involved. In order to construct the cut, we need the tensor three-point function
\beq
\psi_{TTT}  = \frac{1}{2(k_{13}+u)}\Big[\big[(\vec k_1 - \vec k_3)  \cdot \vec \xi_u\big] \, \vec\xi_1\cdot \vec\xi_3+2(\vec k_3\cdot \vec\xi_1)(\vec\xi_3\cdot \vec\xi_u)-2(\vec k_1 \cdot \vec\xi_3)(\vec\xi_u\cdot \vec\xi_1)\Big]^2\,.
\eeq
Combining this with the three-point function~\eqref{equ:JOO}, we get 
\beq \label{cuteqComu}
{\rm Disc}[\psi^{(u)}_{T\varphi T\varphi}] 
= -2u\, \frac{\frac{1}{6}\tl\Pi_{2,2}^{(u)}(\vec\xi_1\cdot\vec\xi_3)^2+2\tl\Pi_{1,1}^{(u)} (\vec\xi_1\cdot\vec\xi_3)(\vec \xi_1\circ\vec\xi_3)+4(\vec \xi_1\circ\vec\xi_3)^2}{(k_{13}^2-u^2)(k_{24}^2-u^2)},
\eeq
where $\vec\xi_1\circ\vec\xi_3$ is defined in the same way as in~\eqref{eq:epsiloncirc}:
\beq
\vec\xi_1\circ\vec\xi_3 \equiv (\vec\xi_1\cdot \vec k_2)(\vec\xi_3\cdot \vec k_4) - (\vec\xi_1\cdot \vec k_4)(\vec\xi_3\cdot \vec k_2) \, .
\eeq
We want to reconstruct the full $u$-channel contribution to the correlator from this cut. 
As before, the correct answer must have simple poles when either $E$, $E_L^{(u)}$ or $E_R^{(u)}$ vanish.
Our strategy will be to make an ansatz that has these singularities and the correct polarization structure, and then fix its coefficient functions by requiring consistency with the cut~\eqref{cuteqComu} and the correct residues on the singularities. Our ansatz is
\beq
\psi^{(u)}_{T\varphi T\varphi} = \frac{g_2(\vec\xi_1\cdot\vec\xi_3)^2+g_1(\vec\xi_1\cdot\vec\xi_3)(\vec \xi_1\circ\vec\xi_3)+g_0(\vec \xi_1\circ\vec\xi_3)^2}{EE_L^{(u)}E_R^{(u)}} \, ,
\label{eq:gcompuchanansatz}
\eeq
where $g_2$, $g_1$, and $g_0$ are functions of the energies with mass dimension 4, 2 and 0, respectively. Plugging this ansatz into~\eqref{cuteqComu}  fixes $g_0=4$, which is also consistent with the $E\to0$ limit where it must match the scattering amplitude. For $g_1$ and $g_2$, the cut \eqref{cuteqComu} implies
\beq 
\label{f1f2ans}
g_1 = 2\tl\Pi_{1,1}^{(u)} + E_L^{(u)}E_R^{(u)}g_1^{(h)} \quad \text{and} \quad g_2 = \frac{1}{6}\tl \Pi_{2,2}^{(u)} + E_L^{(u)}E_R^{(u)}g_2^{(h)} \, ,
\eeq
where $g_{1,2}^{(h)}(k_a)=g_{1,2}^{(h)}(-k_a)$ are even functions of the external energies, which parametrize solutions with vanishing cut, i.e.~homogeneous solutions to the cutting equation.
To fix the functions $g_{1,2}$, we therefore require additional information. We know that in the $E\to 0$ limit,~\eqref{eq:gcompuchanansatz} should match the middle line of~\eqref{eq:ComptonSA}, which implies that we should have
\beq
g_1 \, \xrightarrow{\ E\to 0\ } \, -2U P_1\left(1+\frac{2S}{U}\right) \quad \text{and} \ \quad g_2 \, \xrightarrow{\ E\to 0\ } \, \frac{1}{6} U^2P_2\left(1+\frac{2S}{U}\right)  .
\eeq
The natural extensions away from the $E=0$ limit are provided by the following functions~\cite{Baumann:2020dch} (see also Appendix~\ref{app:pols}),   
\beq \label{eq:GenLegendre}
\begin{aligned}
	{\cal P}^{(u)}_1 &\equiv \tl\Pi_{1,1}^{(u)} - E_L^{(u)}E_R^{(u)} \tl\Pi_{1,0}^{(u)} \, , \\
	{\cal P}^{(u)}_2 &\equiv \tl\Pi_{2,2}^{(u)} - E_L^{(u)}E_R^{(u)} \tl\Pi^{(u)}_{2,1} + E_L^{(u)}E_R^{(u)} \tl\Pi^{(u)}_{2,0} \, .
\end{aligned}
\eeq
Up to regular terms of the form $EE^{(u)}_L E^{(u)}_R (\ldots)$, the coefficient functions therefore  are
\beq
g_1 = 2{\cal P}^{(u)}_1 \quad \text{and} \quad g_2 = \frac{1}{6}{\cal P}^{(u)}_2 \, .
\eeq
Putting everything together, the final $u$-channel correlator is
\beq
\psi^{(u)}_{T\varphi T\varphi} =\frac{1}{EE_L^{(u)}E_R^{(u)}}\left[ \frac{1}{6}{\cal P}_2^{(u)} (\vec\xi_1\cdot\vec\xi_3)^2+2{\cal P}_1 (\vec\xi_1\cdot\vec\xi_3)(\vec \xi_1\circ\vec\xi_3)+4(\vec \xi_1\circ\vec\xi_3)^2 \right]  ,
\eeq
which, by construction, reproduces~\eqref{cuteqComu} and has the correct residues on its singularities.

\paragraph*{Contact term}  Much like in the amplitude case, the sum of $s$, $t$ and $u$-channels is not the full answer. The clearest way to see this is to note that the sum does not reproduce the flat-space scattering amplitude on the total energy pole because we are missing a contact term. 
Like before, we can  parametrize the most general contact term consistent with dimensional analysis, the presence of only a total energy pole and the symmetries of the problem:
\beq
\psi^{(c)}_{T\varphi T\varphi} = \frac{g^{(c)}_1}{E} (\vec\xi_1\cdot\vec\xi_3)^2+\frac{g^{(c)}_{2}}{E}(\vec\xi_1\cdot\vec\xi_3)\left[(\vec\xi_1\cdot\vec k_2)(\vec\xi_3\cdot\vec k_4)+(\vec\xi_1\cdot\vec k_4)(\vec\xi_3\cdot\vec k_2)
\right]  .
\eeq
where $g^{(c)}_1$ and $g^{(c)}_2$ are functions of dimension 2 and 0, respectively. Matching the $E\to 0$ limit of this expression to the last line of the scattering amplitude~\eqref{eq:ComptonSA}, we find that $g^{(c)}_2=2$ and
\beq \label{fc1fsl}
g^{(c)}_1 \, \xrightarrow{\ E\to 0\ } \, \frac{U}{6} \, .
\eeq
We further know that $\psi^{(c)}_{T\varphi T\varphi}$ must have a vanishing cut. These requirements fix the correlator up to terms that are regular in the flat-space limit. In other words, we can choose $g^{(c)}_1$ to be any function that is symmetric under 
$k_a \to -k_a$ and has the limit \eqref{fc1fsl}. As a further physical input, we can demand that the correlator vanishes when the momenta of either of the scalar lines ($2$ 	or $4$) are taken to be soft. A function that has these properties is
\beq
g^{(c)}_1= - \frac{k_{13}k_{24}+u^2}{6}-\frac{E}{4}\left[\left(\frac{k_{13}\beta_u^2}{u^2}+\frac{k_{24}\alpha_u^2}{u^2}\right)-E\right]  ,
\eeq
where $\alpha_u \equiv k_1-k_3$ and $\beta_u \equiv k_2-k_4$. There is some leftover freedom to add terms without singularities that identically vanish in the soft limit.
Putting all of this together,  the flat-space graviton Compton correlator takes the form 
\begin{tcolorbox}[colframe=white,arc=0pt,colback=greyish2]
\vspace{-0.2cm}
\beq
\label{eq:ComptonC}
\begin{aligned}
\psi_{T\varphi T\varphi}^{(\rm flat)}  &=  \frac{4(\vec\xi_1\hskip -1pt \cdot \hskip -1pt  \vec k_2)^2 (\vec\xi_3\hskip -1pt \cdot \hskip -1pt  \vec k_4)^2}{EE_L^{(s)}E_R^{(s)}} + \frac{4(\vec\xi_1 \hskip -1pt \cdot \hskip -1pt  \vec k_4)^2 (\vec\xi_3 \hskip -1pt \cdot \hskip -1pt  \vec k_2)^2}{EEE_L^{(t)}E_R^{(t)}}\\
&~~~~+\psi_{4,\,{\rm flat}}^{(T,u)}(\vec\xi_1\cdot\vec\xi_3)^2+\frac{1}{E E_L^{(u)}E_R^{(u)}}\left[ 2{\cal P}_1^{(u)} (\vec\xi_1\cdot\vec\xi_3)(\vec \xi_1\circ\vec\xi_3)+4(\vec \xi_1\circ\vec\xi_3)^2
\right]\\
&~~~~+\frac{2}{E}(\vec\xi_1\cdot\vec\xi_3)\bigg[(\vec\xi_1\cdot\vec k_2)(\vec\xi_3\cdot\vec k_4)+(\vec\xi_1\cdot\vec k_4)(\vec\xi_3\cdot\vec k_2)
\bigg]\,,
\end{aligned}
\eeq
\end{tcolorbox}
\noindent
where $\psi_{4,\,{\rm flat}}^{(T,u)}$ is the four-point scalar correlator arising from exchange of a massless spin-2 field; cf.~\eqref{eq:flatspacespin2exc} permuted to the $u$-channel. The appearance of this scalar correlator is a useful consistency check on the final answer, since from the bulk perspective it is easy to see that the part of the three-graviton Feynman rule that is proportional to $(\vec\xi_1\cdot\vec\xi_3)^2$ is identical to that of two scalars and a graviton.

\vskip 4pt
By construction, the result~\eqref{eq:ComptonC} reproduces the correct cuts in all of its channels and has the correct residues on its partial and total energy singularities. However, the result is not completely unambiguous, since we are, in principle, still free to add to it any expression that does not have any singularities and has vanishing cuts.  In analogy with the ``cut-constructible" parts of amplitudes, the singular part of the answer could be called the ``cut + singularity-constructible" part of the correlator. We have fixed the additional  regular terms by demanding that the correlator vanishes in the soft limit for either of the scalar particles, leading to the last line in~\eqref{eq:ComptonC}. This property follows from the shift symmetry of the bulk interactions.\footnote{Equivalently, these pieces can be fixed by demanding that the wavefunction coefficient is proportional to the scalar four-point function when we choose either $\vec \xi_1$ or $\vec \xi_3$ to be transverse to both $\vec k_2$ and $\vec k_4$.}

\subsubsection{Correlator From Recursion}

The expression~\eqref{eq:ComptonC} is completely fixed by its singularities, up to terms that are regular and have vanishing cuts. To make this more explicit, we will now construct the correlation function following the recursive procedure outlined in Section~\ref{sec:gluing}.

\vskip4pt
Consider the following complex shift of the energy of the first leg
\beq \label{eq:Comptonshift}
k_1\mapsto k_1+z\,.
\eeq
The poles of the deformed correlator $\psi_{T\vp T\vp}(z)$ are at 
\begin{align}
z_E &= -E\,,\\
z_s &= -E_L^{(s)}\,,\\
z_t &= -E_L^{(t)}\,,\\
z_u &= -E_L^{(u)}\,.
\end{align}
The goal is to write the original correlator, $\psi_{T\vp T\vp}(z=0)$, as a sum over the residues of these singularities.

\vskip4pt
The residue of the total energy pole is given by the scattering amplitude~\eqref{eq:ComptonSA}.
We need to write this residue in terms of the energy variables of the correlator. 
As we discussed in the previous section, the natural way to 
extend the Legendre polynomials away from $E=0$ is to write them as~\eqref{eq:GenLegendre}, which reduce back to the Legendre polynomials when $E=0$. Since we have deformed $k_1$,  the residue does not depend on $k_1$, so we introduce
\beq \label{eq:modGenLeg}
\begin{aligned}
	\tl{\cal P}^{(u)}_1 &\equiv \tl\Pi^{(u)}_{1,1}-(-k_{24}+u)E_R^{(u)}\tl\Pi^{(u)}_{1,0}\,,  \\[4pt]
	\tl{\cal P}^{(u)}_2 &\equiv \tl\Pi^{(u)}_{2,2}  - (-k_{24}+u) E_R^{(u)}\, \tl\Pi^{(u)}_{2,1} + (-k_{24}+u) E_R^{(u)}\tl\Pi^{(u)}_{2,0} \,.
\end{aligned}
\eeq
Recall that the polarization sums $\tl\Pi_{\ell,m}^{(u)}$ come from contracting momentum vectors and polarization tensors, so they are not affected by the complex shift~\eqref{eq:Comptonshift}. The residue at $z=z_E$ then is  
\begin{align}
	-\underset{z=z_E}{\rm Res} \left(\frac{\psi(z)}{z}\right)  =&  -4\frac{(\vec\xi_1\hskip -1pt \cdot \hskip -1pt  \vec k_2)^2 (\vec\xi_3\hskip -1pt \cdot \hskip -1pt  \vec k_4)^2}{E(k_{34}^2-s^2)} -4\frac{(\vec\xi_1\hskip -1pt \cdot \hskip -1pt  \vec k_4)^2 (\vec\xi_3\hskip -1pt \cdot \hskip -1pt  \vec k_2)^2}{E(k_{23}^2-t^2)} \nn \\
	&\hspace{-1pt} -\frac{1}{E(k_{24}^2-u^2)} \bigg[ \frac{1}{6} \tl{\cal P}^{(u)}_2 (\vec\xi_1\cdot\vec\xi_3)^2 + 2\tl{\cal P}^{(u)}_1 (\vec\xi_1\cdot\vec\xi_3)(\vec \xi_1\circ\vec\xi_3)+4(\vec \xi_1\circ\vec\xi_3)^2 \bigg] \\
	&\hspace{-1pt} +\frac{k_{24}^2-u^2}{6E} (\vec\xi_1\cdot\vec\xi_3)^2 +\frac{2}{E}(\vec\xi_1\cdot\vec\xi_3)\bigg[(\vec\xi_1\cdot\vec k_2)(\vec\xi_3\cdot\vec k_4)+(\vec\xi_1\cdot\vec k_4)(\vec\xi_3\cdot\vec k_2)
	\bigg]\,.  \nn
\end{align}
We can derive the residues of the  partial energy singularities from the available three-point data~\cite{Baumann:2020dch}
\begin{align}
-\underset{z=z_s}{\rm Res}\left(\frac{\psi(z)}{z}\right) &=\frac{A_{T\varphi \varphi} \, \tl \psi_{\varphi T  \varphi}}{E_L^{(s)}} =   \frac{4(\vec \xi_1 \cdot \vec k_2)^2(\vec \xi_3 \cdot \vec k_4)^2}{ E_L^{(s)}(k_{34}-s)(k_{34}+s) }\, ,\\
-\underset{z=z_t}{\rm Res}\left(\frac{\psi(z)}{z}\right)   &=\frac{A_{T\varphi \varphi} \, \tl \psi_{\varphi T  \varphi}}{E_L^{(t)}} =  \frac{4(\vec \xi_1 \cdot \vec k_4)^2(\vec \xi_3 \cdot \vec k_2)^2 }{ E_L^{(t)}(k_{23}-t)(k_{23}+t) }\, ,\\
-\underset{z=z_u}{\rm Res}\left(\frac{\psi(z)}{z}\right)  &=\frac{A_{TTT} \otimes \tl \psi_{T\varphi   \varphi}}{E_L^{(u)}} = \frac{ \frac{1}{6} \tl\Pi_{2,2}^{(u)} (\vec \xi_1\cdot\vec\xi_3)^2+ 2\tl\Pi_{1,1}^{(u)} (\vec\xi_1\cdot \vec\xi_3) (\vec\xi_1\circ \vec\xi_3) + 4(\vec\xi_1\circ \vec\xi_3)^2 }{E_L^{(u)}(k_{24}-u)(k_{24}+u)} \, .
\end{align}
Summing up all residues according to~\eqref{eq:residue}, we get exactly the result~\eqref{eq:ComptonC}, except for the regular terms on the last line. It isn't surprising that we don't capture these regular terms since they don't contribute to the residues of the deformed correlator.
Of course, we could reproduce them by imposing the same soft behavior as in the previous section.

\subsection{Transmutation to de Sitter Space}
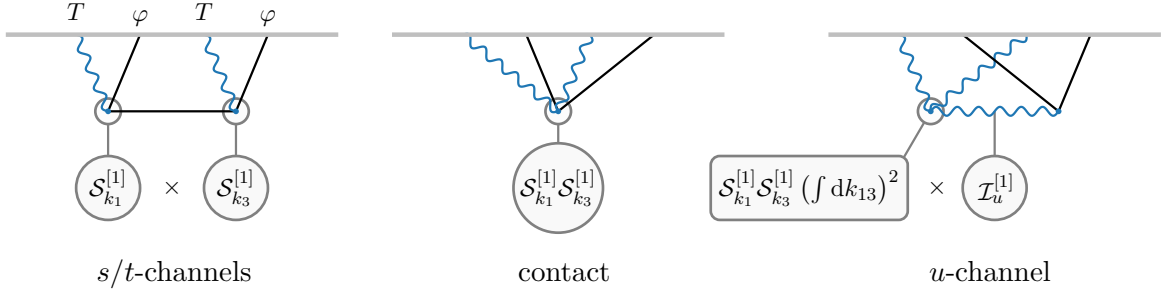
\begin{figure}[t!]
   \centering
           \begin{tabular}{c@{\hskip -0.5cm}c@{\hskip -1cm}c}
\scalebox{0.85}{
\begin{tikzpicture}[line width=1. pt, scale=2,
sines/.style={
        line width=1pt,
        line join=round, 
        draw=blue3, 
        decorate, 
        decoration={complete sines, number of sines=4, amplitude=4pt}
    }
]
\draw[color=gray, fill=gray!2,, line width=1.pt] (0,0) -- (0,0.7);
\draw[color=gray, fill=gray!2,, line width=1.pt] (-1,0) -- (-1,0.7);
\filldraw[color=gray, fill=gray!2, very thick](-1,0.7) circle[radius=0.1] ;
\filldraw[color=gray, fill=gray!2, very thick](0,0.7) circle[radius=0.1] ;
\draw[black, line width=1.pt] (-1,0.7) -- (0,0.7);
\draw[black, line width=1.pt] (-1,0.7) -- (-0.75,1.3);
\path[postaction={sines}] (-1,0.7) -- (-1.25,1.3);
\draw[black, line width=1.pt] (0,0.7) -- (0.25,1.3);
\path[postaction={sines}] (0,0.7) -- (-0.25,1.3);
\draw[lightgray, line width=2.pt] (-1.8,1.3) -- (.8,1.3);
\filldraw[color=gray, fill=gray!5, very thick](-1,0.1) circle[radius=0.25] node[black]{ ${\cal S}_{k_1}^{[1]}$};
\filldraw[color=gray, fill=gray!5, very thick](0,0.1) circle[radius=0.25] node[black]{${\cal S}_{k_3}^{[1]}$};
\node[scale=1] at (-0.5,0.1) {$\times$};
\node[scale=1] at (-1.25,1.48) {$T$};
\node[scale=1] at (-0.25,1.48) {$T$};
\node[scale=1] at (-0.75,1.43) {$\varphi$};
\node[scale=1] at (0.25,1.43) {$\varphi$};
\draw  (-1,0.7) node[blue3, fill, circle, scale=0.25]  {};
\draw  (0,0.7) node[blue3, fill, circle, scale=0.25]  {};
\draw[white, line width=0.7pt] (0.8,-0.31) -- (1.2,-0.31);
\end{tikzpicture}}&~~
\scalebox{0.85}{
\begin{tikzpicture}[line width=1. pt, scale=2,
sines/.style={
        line width=1pt,
        line join=round, 
        draw=blue3, 
        decorate, 
        decoration={complete sines, number of sines=4, amplitude=4pt}
    },
    sines2/.style={
        line width=1pt,
        line join=round, 
        draw=blue3, 
        decorate, 
        decoration={complete sines, number of sines=6, amplitude=4pt}
    }
]
\draw[color=gray, fill=gray!2,, line width=1.pt] (-0.5,0) -- (-0.5,0.7);
\filldraw[color=gray, fill=gray!2, very thick](-0.5,0.7) circle[radius=0.1] ;
\draw[black, line width=1.pt] (-0.5,0.7) -- (-0.75,1.3);
\path[postaction={sines2}] (-0.5,0.7) -- (-1.25,1.3);
\path[postaction={sines}] (-0.5,0.7) -- (-0.25,1.3);
\draw[black, line width=1.pt] (-0.5,0.7) -- (0.25,1.3);
\draw[lightgray, line width=2.pt] (-1.8,1.3) -- (.8,1.3);
\filldraw[color=gray, fill=gray!5, very thick](-0.5,0.1) circle[radius=0.35] node[black]{${\cal S}_{k_1}^{[1]}{\cal S}_{k_3}^{[1]}$};
\draw  (-.5,0.7) node[blue3, fill, circle, scale=0.25]  {};
\draw[white, line width=0.7pt] (0.8,-0.31) -- (1.2,-0.31);
\end{tikzpicture}}&
\scalebox{0.85}{
\begin{tikzpicture}[line width=1. pt, scale=2,
sines/.style={
        line width=1pt,
        line join=round, 
        draw=blue3,
        decorate, 
        decoration={complete sines, number of sines=4, amplitude=4pt}
    },
    sines2/.style={
        line width=1pt,
        line join=round, 
        draw=blue3, 
        decorate, 
        decoration={complete sines, number of sines=6, amplitude=4pt}
    }
 ]
\draw[color=gray, fill=gray!2,, line width=1.pt] (-1.35,0.1) -- (-1,0.7);
\filldraw[color=gray, fill=gray!2, very thick](-1,0.7) circle[radius=0.1] ;
\draw[color=gray, fill=gray!2,, line width=1.pt] (-0.5,0.1) -- (-0.5,0.7);
\path[postaction={sines2}] (-1,0.7) -- (-0.25,1.3);
\path[postaction={sines}] (-1,0.7) -- (-1.25,1.3);
\path[postaction={sines2}]  (-1,0.7) -- (0,0.7);
\draw[black, line width=1.pt] (0,0.7) -- (0.25,1.3);
\draw[black, line width=1.pt] (0,0.7) -- (-0.75,1.3);
\draw[lightgray, line width=2.pt] (-1.8,1.3) -- (.8,1.3);
\node at (-1.95,0.1) [rectangle, rounded corners, minimum width=2cm, minimum height=1cm,text centered, draw=gray, fill=gray!5,very thick] {${\cal S}_{k_1}^{[1]}{\cal S}_{k_3}^{[1]} \left(\int \rd k_{13}\right)^2$};
\filldraw[color=gray, fill=gray!5, very thick](-0.5,0.1) circle[radius=0.25] node[black]{$\,{\cal I}_u^{[1]}$};
\node[scale=1] at (-0.95,0.1) {$\times$};
\draw[white, line width=0.7pt] (0.8,-0.31) -- (1.2,-0.31);
\draw  (-1,0.7) node[blue3, fill, circle, scale=0.25]  {};
\draw  (0,0.7) node[blue3, fill, circle, scale=0.25]  {};
\end{tikzpicture}}\\
\hspace{-0.5cm}$s/t$-channels & contact &\hspace{0.75cm} $u$-channel
\end{tabular}
   \caption{Illustration of the transmutation operators for the graviton Compton correlator. }
  \label{fig:Transmutation2}
\end{figure}

Having generated the graviton Compton correlator in flat space, the next step is to lift it to de Sitter space following the procedure described in Section~\ref{sec:lifting}. 
The required transmutation operators are summarized in Fig.~\ref{fig:Transmutation2}.

\vskip4pt
First, it is helpful to write the correlator~\eqref{eq:ComptonC} as 
\beq
\psi_{T\vp T\vp}^{(\rm flat)}  = f_s^{(\rm flat)}{\cal P}_s + f_t^{(\rm flat)} {\cal P}_t +f_c^{(\rm flat)} {\cal P}_c + f_{u,T}^{(\rm flat)} {\cal P}_u^T+ f_{u,\perp}^{(\rm flat)} {\cal P}_u^\perp + f_{u,\parallel}^{(\rm flat)} {\cal P}_u^\parallel \,,
\label{equ:PsiSing}
\eeq
which splits the various quantities appearing into polarization structures,
\beq
\begin{aligned}
{\cal P}_s &\equiv  (\vec\xi_1\hskip -1pt \cdot \hskip -1pt  \vec k_2)^2 (\vec\xi_3\hskip -1pt \cdot \hskip -1pt  \vec k_4)^2 \,,\\
{\cal P}_t &\equiv (\vec\xi_1 \hskip -1pt \cdot \hskip -1pt  \vec k_4)^2 (\vec\xi_3 \hskip -1pt \cdot \hskip -1pt  \vec k_2)^2\,, \\
{\cal P}_c &\equiv (\vec\xi_1\cdot\vec\xi_3)\left[(\vec\xi_1\cdot\vec k_2)(\vec\xi_3\cdot\vec k_4)+(\vec\xi_1\cdot\vec k_4)(\vec\xi_3\cdot\vec k_2)\right]  , \\[4pt]
 {\cal P}_u^T &\equiv (\vec \xi_1\cdot\vec\xi_3)^2\,, \\
 {\cal P}_u^\perp &\equiv 2\tl\Pi_{1,1}^{(u)}(\vec \xi_1\cdot\vec\xi_3)(\vec \xi_1\circ\vec\xi_3)+4(\vec \xi_1\circ\vec\xi_3)^2\,,\\
{\cal P}_u^\parallel &\equiv 2\tl\Pi_{1,0}^{(u)} (\vec \xi_1\cdot\vec\xi_3)(\vec \xi_1\circ\vec\xi_3) \,,
\end{aligned}
\eeq
%\beq
%\begin{aligned}
%{\cal P}_s &\equiv  (\vec\xi_1\hskip -1pt \cdot \hskip -1pt  \vec k_2)^2 (\vec\xi_3\hskip -1pt \cdot \hskip -1pt  \vec k_4)^2 \,,& {\cal P}_u^T &\equiv (\vec \xi_1\cdot\vec\xi_3)^2\,,\\
%{\cal P}_t &\equiv (\vec\xi_1 \hskip -1pt \cdot \hskip -1pt  \vec k_4)^2 (\vec\xi_3 \hskip -1pt \cdot \hskip -1pt  \vec k_2)^2\,, & {\cal P}_u^\perp &\equiv 2\tl\Pi_{1,1}^{(u)}(\vec \xi_1\cdot\vec\xi_3)(\vec \xi_1\circ\vec\xi_3)+4(\vec \xi_1\circ\vec\xi_3)^2\,,\\
%{\cal P}_c &\equiv (\vec\xi_1\cdot\vec\xi_3)\left[(\vec\xi_1\cdot\vec k_2)(\vec\xi_3\cdot\vec k_4)+(\vec\xi_1\cdot\vec k_4)(\vec\xi_3\cdot\vec k_2)\right]  , &
%{\cal P}_u^\parallel &\equiv 2\tl\Pi_{1,0}^{(u)} (\vec \xi_1\cdot\vec\xi_3)(\vec \xi_1\circ\vec\xi_3) \,,
%\end{aligned}
%\eeq
and form factors, 
\beq
\begin{aligned}
f_s^{(\rm flat)}  &\equiv - \frac{4}{E{\cal S}} \,, \hspace{20pt} &
f_t^{(\rm flat)}  &\equiv - \frac{4}{E{\cal T}} \,,   \hspace{20pt}&  f_c^{(\rm flat)}  &\equiv \frac{2}{E}\,,\\ 
 f_{u,T}^{(\rm flat)} &\equiv  \psi_{4,\,{\rm flat}}^{(T,u)}\,,
 &
f_{u,\perp}^{(\rm flat)}  &\equiv - \frac{1}{E\,{\cal U}}\,, &
f_{u,\parallel}^{(\rm flat)} &\equiv - \frac{1}{E}\, ,
\end{aligned}
\eeq
where we have introduced the variables 
\beq
{\cal S} \equiv -E_L^{(s)} E_R^{(s)}\,, \qquad {\cal T} \equiv -E_L^{(t)} E_R^{(t)}\,, \qquad {\cal U} \equiv -E_L^{(u)} E_R^{(u)}\, .
\eeq
In the limit $E \to 0$, these reduce to the ordinary Mandelstam variables. 
Similar to the example considered in~\S\ref{sec:transfromcuts}, we have separated the contribution coming from graviton exchange into several components. Aside from separating out the polarization structure that is proportional to a scalar correlator, we have isolated the helicity-1 piece, denoted by $f_{u,\perp}$, and the lower-helicity contributions, $f_{u,\parallel}$. Each of these pieces will have to be transformed individually.
It is useful to treat the $(\vec \xi_1\cdot\vec\xi_3)^2$ polarization structure separately because its split into exchange and contact contributions is gauge-dependent, and thus ambiguous. Instead, we will see that we can lift this contribution all at once.\footnote{Alternatively, we could make some choice of assignment into the $u$-channel and contact contributions. The subsequent lifting procedure would then have an ambiguity at ${\cal O}(E^{-1})$, which can then be corrected systematically, as we will explain below.}

\paragraph*{$\boldsymbol{s/t}$-channel}  It is easiest to first consider the transmutation of the $s$ and $t$-channels. Looking at Fig.~\ref{fig:Transmutation2}, we see that these processes involve two copies of the vertex~\eqref{eq:lifttpptodS} connected by the exchange of a scalar. We can therefore lift the $s$ and $t$-channel contributions, by acting with the operator ${\cal S}^{[1]}$ defined in \eqref{eq:spinraisingop2} on each of the spinning external lines:
\begin{align}
f_s^{({\rm dS})} &= {\cal S}_{k_1}^{[1]}{\cal S}_{k_3}^{[1]}f_s^{({\rm flat})} \,, \\
 f_t^{(\rm dS)}  &= {\cal S}_{k_1}^{[1]}{\cal S}_{k_3}^{[1]} f_t^{\rm (flat)}\,.
\end{align}
This produces the following de Sitter form factors
\begin{align}
f_s^{({\rm dS})} &= -4 \left[\frac{1}{{\cal S}} \left( \frac{2k_1k_3}{E^3} +\frac{k_{13}}{E^2} + \frac{1}{E}\right) -\, \frac{1}{{\cal S}^2} \left( \frac{2s k_1 k_3 }{E^2} + \frac{2k_1k_3+ k_3E_L^{(s)} + k_1E_R^{(s)}}{E} \right) \right] , \\
f_t^{(\rm dS)}  &= -4 \left[\frac{1}{{\cal T}} \left( \frac{2k_1k_3}{E^3} +\frac{k_{13}}{E^2} + \frac{1}{E}\right) -\, \frac{1}{{\cal T}^2} \left( \frac{2t k_1 k_3 }{E^2} + \frac{2k_1k_3+ k_3E_L^{(t)} + k_1E_R^{(t)}}{E} \right) \right] .
\end{align}
These expressions precisely reproduce the complicated analytic structure of the form factors found in~\cite{Baumann:2020dch}, from the transmutation of a very simple flat-space seed.

\vskip4pt
\paragraph*{Contact term}  The contact contribution can be lifted to de Sitter space in precisely the same way as the scalar exchange channels, because it involves the same spinning external lines. Explicitly, the de Sitter result is
\beq
f_c^{({\rm dS})} = {\cal S}_{k_1}^{[1]}{\cal S}_{k_3}^{[1]} f_c^{({\rm flat})} = 2\left(\frac{2 k_1 k_3}{E^3}+\frac{k_{13}}{E^2}+\frac{1}{E}\right) ,
\eeq
which has a set of subleading singularities that come along with the leading total energy singularity.

\vskip4pt
\paragraph*{$\boldsymbol{u}$-channel} The $u$-channel contribution is a little more complex. We start with the $f_{u,\perp}$ form factor. Since the interaction vertex involves three spinning fields, we must transform both the external lines with  ${\cal S}_{k_1}^{[1]}{\cal S}_{k_3}^{[1]}$ and also the internal line with ${\cal I}_u^{[1]}$ defined in (\ref{eq:internalspinup}). Additionally, the fact that all the particles are massless requires us to integrate with respect to energies, as in~\eqref{fnlll}. A small subtlety is that we integrate with respect to 
$k_{13}$ rather than $E_L$ because we only want to act on the external lines.
We can therefore write the helicity-2 form factor in de Sitter as (see Fig.~\ref{fig:Transmutation2})
\beq
\begin{aligned}
f_{u,\perp}^{({\rm dS})} 
&= -{\cal S}_{k_1}^{[1]}{\cal S}_{k_3}^{[1]} \left(\int \rd k_{13}\right)^2 {\cal I}_u^{[1]} f_{u,\perp}^{\rm (flat)}=- \frac{1}{{\cal U}}  \left(\frac{2k_1k_3}{E^3} + \frac{k_{13}}{E^2} \right) + \frac{u }{{\cal U}^2} \left(\frac{2k_1k_3}{E^2} + \frac{E_L^{(u)}}{E} \right) .
\end{aligned}
\eeq
The longitudinal form factor is most simply written as $-k_{13}k_{24}$ times the $u\to0$ limit of this form factor: 
\beq
f_{u,\parallel}^{({\rm dS})} =  -\frac{2k_1k_3}{E^3} - \frac{k_{13}}{E^2}\, .
\eeq
Finally, we have to lift $f_{u,T}^{(\rm flat)}$. This is done most simply  by replacing $\psi_{4,\,{\rm flat}}^{(T,u)}$ with its de Sitter counterpart $\psi_{\phi\vp\phi\vp}^{(T,u)}$, which is a scalar exchange correlator involving two massless scalars and two conformally coupled scalars where a graviton is exchanged in the $u$-channel.\footnote{This procedure can be justified from several different perspectives. First, this is the unique de Sitter-invariant object whose $E\to 0$ residue is the correct scattering amplitude. It is also fixed by its cuts and amplitude, along with the manifestly local test of~\cite{Jazayeri:2021fvk} for the massless lines. Finally, the coefficient of the $(\vec\xi_1\cdot\vec\xi_3)^2$ term in $\psi_{T\vp T\vp}$ must take this form in order for it to satisfy the leading and subleading graviton soft theorems of~\cite{Maldacena:2002vr,Hinterbichler:2013dpa}.}
Explicitly, this correlator is
\beq
\psi^{(T,u)}_{\phi\vp\phi\vp} = -\frac{1}{6}\bigg[g_{(2,2)} \tl\Pi_{2,2}^{(u)} + g_{(2,1)}\tl\Pi_{2,1}^{(u)}+g_{(2,0)}\,\tl \Pi_{2,0}^{(u)}+g_c\bigg]\,,
\eeq
where the form factors are given by
\begin{align}
g_{(2,2)} &\equiv \frac{1}{{\cal U}}  \left(\frac{2k_1k_3}{E^3} + \frac{k_{13}}{E^2} \right)-\frac{u }{{\cal U}^2} \left(\frac{2k_1k_3}{E^2} + \frac{E_L^{(u)}}{E} \right)\,,\\
g_{(2,1)} &= -g_{(2,0)} \equiv \frac{2k_1k_3}{E^3} + \frac{k_{13}}{E^2} \,,\\[6pt]
%f^\vp_{(2,0)} &= -\frac{2k_1k_3}{E^3} - \frac{k_{13}}{E^2}\,,\\
\nonumber
g_c &\equiv \frac{\left[(k_1-k_3)^2+u^2\right]\left[3(k_2-k_4)^2-u^2\right]}{4 E u^2}\\
& ~~~~-\frac{2k_1k_3(k_{13}^2-u^2)}{E^3}-(k_1^2+k_3^2-u^2)\left(\frac{k_{13}}{E^2}+\frac{1}{E}\right)\,.
\end{align}
This correlator can itself either be taken as given, or can be bootstrapped using the techniques described in this paper.

\vskip4pt
Having lifted each of these form factors to de Sitter space, we can assemble the full correlation function as
\beq
\psi_{T\vp T\vp}^{(\rm dS)}  = f_s^{(\rm dS)}{\cal P}_s + f_t^{(\rm dS)} {\cal P}_t +f_c^{(\rm dS)} {\cal P}_c + f_{u,T}^{(\rm dS)} {\cal P}_u^T+ f_{u,\perp}^{(\rm dS)} {\cal P}_u^\perp + f_{u,\parallel}^{(\rm dS)} {\cal P}_u^\parallel \, ,
\label{equ:tptpdsschematic}
\eeq
which, more explicitly, can be written as
\begin{align}
\psi_{T\varphi T\varphi}^{(\rm dS)}  &= -4 (\vec\xi_1\hskip -1pt \cdot \hskip -1pt  \vec k_2)^2 (\vec\xi_3\hskip -1pt \cdot \hskip -1pt  \vec k_4)^2{\cal S}_{k_1}^{[1]}{\cal S}_{k_3}^{[1]}\frac{1}{E\,{\cal S}} - 4(\vec\xi_1 \hskip -1pt \cdot \hskip -1pt  \vec k_4)^2 (\vec\xi_3 \hskip -1pt \cdot \hskip -1pt  \vec k_2)^2{\cal S}_{k_1}^{[1]}{\cal S}_{k_3}^{[1]}\frac{1}{E\,{\cal T}}+ \psi^{(T)}_{\phi\vp\phi\vp}(\vec\xi_1\cdot\vec\xi_3)^2 \nonumber \\
&~~~~-\left[  \frac{1}{{\cal U}}  \left(\frac{2k_1k_3}{E^3} + \frac{k_{13}}{E^2} \right)-\frac{u }{{\cal U}^2} \left(\frac{2k_1k_3}{E^2} + \frac{E_L^{(u)}}{E} \right)\right]\bigg[ 2u^2\Pi_{1,1}^{(u)} (\vec\xi_1\cdot\vec\xi_3)(\vec \xi_1\circ\vec\xi_3)+4(\vec \xi_1\circ\vec\xi_3)^2
\bigg] \nonumber \\
&~~~~- \left(\frac{2k_1k_3}{E^3} + \frac{k_{13}}{E^2} \right)2\Pi_{1,0}^{(u)} (\vec\xi_1\cdot\vec\xi_3)(\vec \xi_1\circ\vec\xi_3) \nonumber \\
&~~~~+2\left(\frac{2k_1k_3}{E^2}+\frac{k_{13}}{E}+\frac{1}{E}\right)(\vec\xi_1\cdot\vec\xi_3)\bigg[(\vec\xi_1\cdot\vec k_2)(\vec\xi_3\cdot\vec k_4)+(\vec\xi_1\cdot\vec k_4)(\vec\xi_3\cdot\vec k_2)\bigg]\,. \label{eq:totolifted}
\end{align}
It is easy to verify that this expression agrees exactly with the answer found in~\cite{Baumann:2020dch}.

\vskip4pt
Given the relative simplicity of the calculations involved in generating~\eqref{eq:totolifted} compared to the weight-shifting approach of~\cite{Baumann:2020dch}, we expect that these techniques will have wider applicability. Note that the organization of part of the answer into a scalar correlator is both computationally and conceptually useful---for example, it is straightforward following this approach to reproduce the graviton Compton correlator involving massless scalars also derived in~\cite{Baumann:2020dch}. We also anticipate that this approach will be useful to generate the four-graviton wavefunction.

\section{Conclusion}
\label{sec:conclusions}

An important insight into the structure of cosmological correlation functions is that they contain scattering amplitudes within them, residing at the singularities of the correlators. The symmetries of the bulk interactions that give rise to these correlations are encoded in the symmetries of the amplitudes appearing at their singularities. Interestingly, much of the structure of correlators is determined by these singularities. The challenge of computing cosmological correlation functions can then be translated into the problem of extending the correlators away from their singular limits.

\vskip4pt
In this paper, we have explored ways of  connecting the singularities of cosmological correlators, which are particularly powerful for constructing correlators that are rational functions of the energies. This includes many cases of interest, for example Yang--Mills theory, gravity, and inflationary correlators.
These correlators have a simplified singularity structure consisting only of poles---in contrast to the branch cuts that can arise from the exchange of massive particles~\cite{Arkani-Hamed:2018kmz}---allowing them to be bootstrapped efficiently. 
For example, the constraints of bulk unitarity---manifested as a set of cutting rules for correlators---provide additional information about the structure away from kinematic singularities that in many cases is enough to completely reconstruct rational correlators. In particular, in de Sitter space these cutting rules completely fix the structure of subleading partial energy singularities. This, combined with the physical input that the final correlator should not have any singularities in folded configurations, can be used to reconstruct rational correlators via energy recursion relations. We have also explored a somewhat surprising and intriguing connection between rational correlators in de Sitter space and those in flat space. In many cases, the de Sitter correlators of interest in cosmology can be ``transmuted" from their flat-space counterparts. Since flat-space correlators are much simpler to bootstrap, this provides an efficient way to generate de Sitter correlators.

\vskip4pt
Although there has been
much recent progress in uncovering the structures underlying cosmological correlators, many mysteries still remain, and  our work suggest a number of interesting directions for further investigation.

\begin{itemize}

\item We have illustrated the bootstrapping of rational correlators with a number of nontrivial examples.
The relative simplicity of these calculations compared to, for example, the weight-shifting approach taken in~\cite{Baumann:2020dch} suggests that these methods will also be useful to compute more complicated correlators.
An interesting test case would be the wavefunction for pure graviton scattering in de Sitter. It would also be interesting to push the computation of Yang--Mills correlators to higher multiplicity. The Weyl invariance of the theory in four dimensions means that the singularities at tree level will only be simple poles, as in flat space, making the recursive approach particularly powerful. It is easy to compute the four-point Yang--Mills correlator (see Appendix~\ref{app:YMapp}), and we expect that higher-multiplicity calculations will help to reveal underlying structure, much as it did for the S-matrix.

\item  Our derivation of the transmutation operators was not completely systematic. We identified specific transmutation operators by comparing the wavefunction calculations in flat space and de Sitter space in selected examples and then showed how these operators can be combined in more complicated cases. It would be interesting to derive these transmutation operators (and their generalizations) more systematically. The current situation is similar to the construction of the weight-shifting operators in~\cite{Arkani-Hamed:2018kmz}. In that case, there was a deeper interpretation of these operators developed independently in the context of conformal field theory~\cite{Baumann:2019oyu,Costa:2011dw,Karateev:2017jgd}. It is therefore natural to speculate that there is a similarly unifying explanation for the existence of the transmutation operators. One interesting possibility is that the two concepts are actually related: the flat-space seeds that we lifted are often conformal in other dimensions (for example $\phi^3$ theory is conformal in $D=6$), so it could be that  some generalization of the weight-shifting procedure that changes dimensions is connected to the transmutation operators.\footnote{It is interesting to speculate that this is related to recent observations of connections between correlators in different dimensions in~\cite{Albayrak:2020fyp,Alday:2021odx}.}

\item The deepest insights coming from the modern amplitudes program have been the discovery of unexpected hidden structures that are not apparent in the traditional approaches~\cite{Cheung:2017pzi}. Since correlators are in a precise sense deformations of scattering amplitudes, it is natural to expect that the remarkable structures found in amplitudes will also have manifestations in correlators. A particularly surprising and useful example is the double copy structure that relates gravitational amplitudes to their Yang--Mills counterparts~\cite{Kawai:1985xq,Bern:2008qj,Bern:2010yg}. Aside from the deep connection between apparently unrelated theories, the double copy has enabled remarkable perturbative gravity computations. Finding the extension of these relations to the cosmological context holds the promise of similar insights, and some recent progress has been made~\cite{Armstrong:2020woi,Albayrak:2020fyp}. A natural strategy to pursue is to find the correlator analogue of the double copy for flat-space correlation functions, and then lift these correlators to de Sitter space via transmutation operations. Since the double copy reorganizes the various contributions, the transmutation procedure may become simpler in this representation. 

\item One deficiency of the cutting rules that we have presented, compared to unitarity methods in the context of scattering amplitudes, is that they are still tied to the Feynman diagram expansion. In particular, the cuts in various channels that we performed are essentially cuts of the corresponding bulk Feynman diagrams. Similarly, the transmutation of the correlators from flat space to de Sitter was performed channel by channel. It would be nice to find a presentation of cosmological correlators that combines their various pieces in a more irreducible way. The recursive construction presented in Section~\ref{sec:gluing} is already a step in this direction, because the residue of the total energy singularity links the various exchange channels together. This way of writing the correlators is suggestive that there is an even more invariant way of presenting the correlators, and one may hope that the transmutation is particularly natural in this presentation.

\item We have seen that the cuts of correlators impose important constraints already at tree level, but we also know that, for scattering amplitudes, unitarity methods really display their utility at loop level. It is therefore natural to use these tools to study loop correlators in cosmological spacetimes. Very little is known about the analytic structure of the wavefunction beyond tree level. 
Even at tree level, correlators can have
branch cut singularities if the theory involves massive particles. We therefore expect the analytic structure of loop correlators to be rather rich.
A natural way to make progress is to compute explicit examples and see what abstract lessons can be learned from them. There is already a body of work approaching this question from a modern perspective~\cite{Benincasa:2018ssx,Hillman:2019wgh,Cohen:2020php,Meltzer:2020qbr,Albayrak:2020saa,Melville:2021lst}, and we expect that there are many more insights to be uncovered using the tools described here.

\end{itemize}

\noindent
At a fundamental level, the motivation for studying and measuring cosmological correlation functions is to understand the evolution of the early universe. Understanding fully how this time evolution is encoded in observables is therefore paramount. In this regard, it is interesting that there are two somewhat different ways in which the bulk time evolution manifests itself in the boundary correlators. For heavy fields, the time evolution is encoded in a differential equation whose solution relates a branch cut singularity in the unphysical region to the full correlator in the physical region. However, in theories involving only light fields---like the graviton or the inflaton---correlators are completely rational functions. This makes linking their singularities comparatively simpler,
but at face value these objects seem to have ``less" information about the bulk time dependence, which we typically associate with the particle production encoded in branch cuts. It is important to understand these features in more detail, with the ultimate goal of extracting information about the universe's earliest moments.

\vspace{0.5cm}
%=======================================
\paragraph{Acknowledgements} 
%=======================================
Thanks to Tarek Anous, Nima Arkani-Hamed, Paolo Benincasa, Noah Bittermann, James Bonifacio, Garrett Goon, Tanguy Grall, Aaron Hillman, Kurt Hinterbichler, Yu-tin Huang, Lam Hui, Sadra Jazayeri, Manuel Loparco, Paul McFadden, Scott Melville, Alberto Nicolis, Enrico Pajer, Diederik Roest, Rachel Rosen, Luca Santoni, David Stefanyszyn, John Stout, Mark Trodden, Dong-Gang Wang, and Sam Wong for helpful discussions. 
DB and CDP~are supported by a VIDI grant of the Netherlands Organisation for Scientific Research~(NWO) that is funded by the Dutch Ministry of Education, Culture and Science~(OCW). DB is supported by a Jade Mountain Professorship at National Taiwan University funded by the Ministry of Science and Technology (Taiwan). HL is supported by DOE grant DE-SC0019018. The work of DB, AJ and GP is part of the Delta-ITP consortium.

\newpage
\appendix

%==================================
\section{Polarization Tensors and Sums}
\label{app:pols}
%==================================
%
%
In many places in the text, we consider internal lines involving spinning particles. In order to efficiently express the contributions coming from the different polarizations of the exchanged fields we require some formalism. In this appendix, we review the relevant polarization tensors and sums that we use. Further details can be found in~\cite{Baumann:2019oyu,Baumann:2020dch}.

\subsection{Polarization Tensors}
On a given spatial slice---either in the bulk spacetime or on the boundary---it is convenient to decompose a spatial tensor into its irreducible components with respect to the (spatial) rotation group. This can be achieved by introducing a basis of projectors for symmetric, traceless, $\ell$-index tensors. We will only require these projectors for spins 1 and 2, but see~\cite{Baumann:2019oyu} for the generalization to higher spin.

\begin{itemize}
\item {\bf Spin 1:} 
For a spin-1 field, we define the basis of projectors
\begin{align}
\label{eq:spin1pol1}
(\Pi_{1,1})^{i}_{j}  &= \pi^i_j\,,\\
(\Pi_{1,0})^{i}_{j}  &=\hat k^i\hat k_j\, ,
\label{eq:spin1pol2}
\end{align}
where  $\pi^i_j$ is the transverse projector
\beq
\pi^i_j \equiv \delta^i_j -\hat k^i\hat k_j\,, \label{equ:pij}
\eeq
with $\hat k_i \equiv k_i/k$. The tensors $\pi_{ij}$ and $\hat k_i \hat k_j$ are orthonormal and form a complete basis of projectors for a vector.

\item {\bf Spin 2:} For a spin-2 field, we use the following basis of projectors
\begin{align}
(\Pi_{2,2})^{i_1i_2}_{j_1j_2} &= \pi^{(i_1}_{(j_1} \pi^{i_2)}_{j_2)}-\frac{1}{2}\pi^{i_1i_2}\pi_{j_1j_2}\,,\\
(\Pi_{2,1})^{i_1i_2}_{j_1j_2} &= 2\hat k^{(i_1}\hat k_{(j_1}\pi^{i_2)}_{j_2)}\,,\\
(\Pi_{2,0})^{i_1i_2}_{j_1j_2} &= \frac{3}{2}\left (\hat k^{i_1}\hat k^{i_2}-\frac{1}{3}\delta^{i_1i_2}\right )\left (\hat k_{j_1}\hat k_{j_2}-\frac{1}{3}\delta_{j_1j_2}\right) .\label{piS}
\end{align}
These projectors are orthonormal, complete and transverse in the sense:
\be
{\rm orthonormality}:& \hspace{0.4cm} (\Pi_{2,m})^{i_1i_2}_{j_1j_2}(\Pi_{2,m'})^{j_1j_2}_{l_1l_2} = \delta_{mm'}(\Pi_{2,m})^{i_1i_2}_{l_1l_2}\, , \\[4pt]
{\rm completeness}:& \hspace{0.4cm}  (\Pi_{2,2})^{i_1i_2}_{j_1j_2}+(\Pi_{2,1})^{i_1i_2}_{j_1j_2}+(\Pi_{2,0})^{i_1i_2}_{j_1j_2} = \delta^{(i_1}_{(j_1}\delta^{i_2)_T}_{j_2)_T}\, , \\[4pt]
{\rm transversality}:&\hspace{0.4cm}   k^{j_1}(\Pi_{2,2})^{i_1i_2}_{j_1j_2} = k^{j_1} k^{j_2}(\Pi_{2,1})^{i_1i_2}_{j_1j_2} =0\, ,
\ee
where $(\cdots)_T$ denotes the traceless symmetric part of the enclosed indices. 
These tensors therefore provide a complete basis for traceless, symmetric tensors.

\end{itemize}
These projectors appear both in the expansion of the two-point function of spinning fields and in their bulk-to-bulk propagators. For our purposes, the latter appearance will be most relevant.

\subsection{Polarization Sums}
\label{equ:PolarSums}
In most cases, the polarization tensors introduced in the previous section will be contracted with vertex factors which will combine into scalar polarization sums that capture the exchanges of the various helicity components of spinning fields. These polarization sums are therefore useful building blocks, and in the de Sitter context have convenient conformal transformation properties~\cite{Arkani-Hamed:2018kmz,Baumann:2019oyu}. In this section, we collect formulas involving these polarization sums that are relevant for the calculations in the main text.

\vskip4pt
For simplicity, we present the polarization sums as they would appear for the $s$-channel exchange of a spinning particle, but the momentum arguments can be permuted to obtain the analogous expressions in different exchange channels.
In the $s$-channel, the relevant combinations of external momenta in vertex factors are typically $\vec\alpha \equiv \vec k_1-\vec k_2$ and $\vec\beta \equiv \vec k_3-\vec k_4$, and the internal momentum that the polarization tensors depend on is $\vec s \equiv \vec k_1+\vec k_2$.

\begin{itemize}
\item {\bf Spin 1:} In the spin-1 case, there are two polarization sums, defined as
\be
\Pi_{1,1} &\equiv   \frac{1}{s^2}\alpha^i(\Pi_{1,1})_{ij}\beta^j  =   \frac{k_{12}k_{34}\alpha  \beta-s^2(t^2-u^2)}{s^4}\,, \\
\Pi_{1,0} &\equiv -\frac{1}{k_{12}k_{34}}\alpha^i(\Pi_{1,0})_{ij}\beta^j=\frac{\alpha \beta}{s^2}\,,
\ee
which consist of contractions of the projectors~\eqref{eq:spin1pol1} and~\eqref{eq:spin1pol2} with external momenta. Recall that $\alpha \equiv k_1-k_2$ and $\beta\equiv k_3-k_4$, which are not the magnitudes of $\vec\alpha$ and $\vec\beta$.

\item {\bf Spin 2:} The relevant spin-2 polarization sums are 
\begin{align}
\Pi_{2,2} &\equiv \frac{3}{2s^4}\, \alpha_i \alpha_j (\Pi_{2,2})^{ij}_{lm}  \beta^l \beta^m\,, \\[4pt]
\Pi_{2,1} &\equiv -\frac{s^2}{k_{12}k_{34}} \frac{3}{2 s^4}\alpha_i \alpha_j (\Pi_{2,1})^{ij}_{lm}  \beta^l \beta^m=3 \frac{ \alpha \beta}{s^2} \frac{\alpha^i \pi_{ij}  \beta^j}{s^2}\, , \\[4pt]
\Pi_{2,0} &\equiv \frac{1}{4}\left(1-3\frac{\alpha^2}{s^2} \right)\left(1-3 \frac{ \beta^2}{s^2}\right) .
\end{align}
Note that the helicity-1 and 2 polarization sums are defined in a natural way via contraction with $\vec\alpha$ and $\vec\beta$, while the helicity-0 polarization sum $\Pi_{2,0}$ does not have such an obvious definition. Instead it is defined in such a way that it depends only on $\alpha$ and $\beta$, and has good conformal transformation properties~\cite{Arkani-Hamed:2018kmz,Baumann:2019oyu,Baumann:2020dch}.
\item {\bf Spin $\boldsymbol\ell$:} For general spin $\ell$, the polarization sums can be expressed in closed form as~\cite{Arkani-Hamed:2018kmz, Baumann:2019oyu}
\begin{align}
	\Pi_{\ell,m}=(2-\delta_{m0})(-\hat L)^m\cos(m\psi)\tl P^m_\ell(\alpha/s)\tl P^m_\ell(\beta/s)\, ,
\end{align}
where $\tl P^\ell_m$ was defined in \eqref{eq:assocLeg} and $\cos\psi \equiv \Pi_{1,1}/\hat L$ is the angle between $\k_1$ and $\k_3$ projected onto the plane perpendicular to $\vec s$, with 
\begin{align}
	\hat L^2 &\equiv  \frac{1}{s^4}\alpha^i(\Pi_{1,1})_{ij}\alpha^j\beta^k(\Pi_{1,1})_{kl}\beta^l = \frac{(k_{12}^2-s^2)(k_{34}^2-s^2)(\alpha^2-s^2)(\beta^2-s^2)}{s^8}\,.
\end{align}
The polarization sums defined in this way capture the angular dependence of a correlator that arises from the spin exchange. An additional nice property of the polarization sums 
is that $s^{-1} \Pi_{\ell,0}$ solves the conformal  Ward identities with $\Delta = 2$. 
\end{itemize}
For a given spin $\ell$, the helicity components $0\le m\le \ell$ must combine in a way  consistent with the amplitude limit and the cutting equation (as well as conformal symmetry for de Sitter correlators). As we explain below, there is a natural modification of the polarization sums that achieves this.

\subsubsection*{Amplitude limit}
The $E\to 0$ singularities of correlators play an important role in our discussion. In this limit, the coefficients of singular correlation functions are the corresponding flat-space scattering amplitudes. We therefore expect the internal polarization structures to simplify in this limit. Each individual polarization sum does not produce a Lorentz-invariant structure in the $E\to 0$ limit, but certain specific combinations do. These  combinations therefore appear in the natural extensions of flat-space scattering amplitudes away from the $E = 0$ locus.

\vskip4pt
The following limits are particularly useful
\be
{\cal P}_1 \equiv s^2\Pi_{1,1}-E_LE_R\Pi_{1,0}  &\xrightarrow{\ E \to 0\ } -S\,P_1\left(1+\frac{2U}{S}\right) , \label{eq:P1polsum} \\
{\cal P}_2 \equiv s^4\Pi_{2,2}  - E_L E_R\, s^2\Pi_{2,1} + E_L E_R(E_LE_R-sE)\,\Pi_{2,0}  &\xrightarrow{\ E \to 0\ } S^2\,P_2\left(1+\frac{2U}{S}\right) ,
\label{eq:P2polsum}
\ee
where $P_\ell$ is a Legendre polynomial and recall that $E_L \equiv k_{12}+s$ and $E_R \equiv k_{34}+s$, while $S$ and $U$ are the ordinary flat-space Mandelstam variables. These particular combinations of polarization sums therefore turn into Legendre polynomials of the Mandelstam invariants in the $E\to 0$ limit, which are Lorentz invariant and reproduce the expected angular structure of a spin-$\ell$ exchange.

\subsubsection*{Rescaled polarization sums}
Notice that in~\eqref{eq:P2polsum} we have included a term with $sE$ multiplying $\Pi_{2,0}$, which obviously does not contribute to the $E\to 0$ limit. The motivation for including this additional piece is that when the polarization sum multiplies a form factor that changes sign under the flip of external energies (for example $1/E$ in flat space), it causes the $\Pi_{2,0}$ structure to drop out of the cut, as it must. With this in mind, it is convenient to introduce some modified polarization sums 
\begin{align}
	\tl\Pi_{\ell,\ell} &\equiv s^{2\ell}\Pi_{\ell,\ell}\, ,\\
	\tl\Pi_{\ell,m} &\equiv s^{2m}(sE)^{\ell-m-1}\tl P_{\ell-1}^m\bigg(\frac{E_LE_R}{s E}-1\bigg)\,\Pi_{\ell,m}\quad (m\ne \ell)\, ,
\end{align}
which have mass dimensions $2\ell$ and $2(\ell-1)$ for $m=\ell$ and $m\ne\ell$, respectively.
A nice feature of this basis is that it gives a well-defined flat-space limit and, being functions of $E_LE_R-sE=k_{12}k_{34}+s^2$, the longitudinal components manifestly vanish when taking the cut of a correlator if they are multiplied by form factors with odd powers of $1/E$. For reference, we list some lower-spin structures that are frequently used in the main text:
\be
\tl\Pi_{1,1} &=s^2\Pi_{1,1}\,,& \tl\Pi_{1,0} &= \Pi_{1,0}\,,\\
\tl\Pi_{2,2} &=s^4\Pi_{2,2}\,,& \tl\Pi_{2,1} &=s^2\Pi_{2,1}\,, &\tl\Pi_{2,0} &=(E_LE_R-sE)\Pi_{2,0}\,.
\ee
Not only will this simplify the appearance of many expressions, but all of these sums now have the property that they drop out of the cut of an exchange correlator if they multiply a form factor that changes sign when we flip the external energies, which happens for lower-helicity components.

\newpage
%==================================
\section{Wavefunction Cutting Rules}
\label{app:cutproof}
%==================================

In the main text, our primary focus was on the construction of four-point correlation functions at tree level. Consequently, we only required the simplest cutting formulas presented in Section~\ref{sec:cuttingsec}, involving graphs with a single internal line. However, cutting rules are far more general, as we explain in this appendix. See~\cite{Meltzer:2020qbr,Melville:2021lst,Goodhew:2021oqg} for related discussions.

\subsection{Preliminaries}

We first introduce some notation that will simplify the description of the cutting rules. The main objects of interest will be wavefunctional coefficients $\psi_n$.  As was explained in Section~\ref{sec:basic}, these can be computed via a Feynman diagram expansion involving two types of propagator, a bulk-to-boundary propagator, ${\cal K}$, for external lines, and a bulk-to-bulk propagator, ${\cal G}$, for internal lines. In flat space, these propagators are given by~\eqref{eq:bboundaryflat} and~\eqref{eq:bbulkflat}, respectively, while in de Sitter space, they are \eqref{eq:massivescalarbulkboundary} and \eqref{eq:massivebulkbulkprop}. In addition to these propagators, we associate to each vertex a factor $iV$, obtained in the usual way.

\vskip4pt
It will also prove useful to introduce conjugate wavefunction coefficients, $\bar\psi_n$. These are computed in nearly the same way, except that we associate an anti-bulk-to-bulk propagator, $\bar{\cal G} = {\cal G}^*$, to internal lines, and use the sign-flipped Feynman rule for the vertices, $-iV$. 
The conjugated wavefunction coefficients are then simply related to the complex conjugate of the ordinary wavefunction coefficients. For example, in flat space, the relation in a unitary theory is given by~\eqref{eq:unitarityflat}.

\vskip4pt
At their core, the cutting formulas that we will present follow from a simple identity involving bulk-to-bulk propagators:
\beq
{\cal G}(k;\eta, \eta') +\bar{\cal G}(k;\eta, \eta')  = \tl{\cal G}(k;\eta, \eta') \,,
\label{eq:ABCid}
\eeq
where the {\it cut propagator}, $\tl{\cal G}$, can be written in terms of bulk-to-boundary propagators as 
\beq
\widetilde {\cal G}(k;\eta, \eta') = -P(k)\Big({\cal K}(-k,\eta)-{\cal K}(k,\eta)\Big)\Big({\cal K}(-k,\eta')-{\cal K}(k,\eta')\Big)\,,
\label{eq:cutpropappendix}
\eeq
with $P(k)$ being the power spectrum of the field associated to the internal line that has been cut. The utility of this expression is that, while the individual bulk-to-bulk and anti-bulk-to-bulk propagators involve time-ordered pieces, the cut propagator is completely un-time-ordered. This factorizes the computation of bulk time integrals, and effectively splits the relevant diagram in two, simplifying the evaluation of this combination of objects.

\vskip4pt
For the flat-space wavefunctional, a simplification occurs because the bulk-to-boundary propagator is a pure exponential. Wavefunction coefficients then depend only on the sum of the energies that flow from a given vertex to the boundary. It is therefore convenient to denote these sums by $x_a$, where $a$ labels the corresponding vertex. Once we have associated an energy variable, $x_a$ to a vertex, we can remove all the lines that connect this vertex to the boundary and consider a truncated graph that only displays the relationships between internal lines~\cite{Arkani-Hamed:2017fdk}. Consequently, it will also be convenient in the following to introduce some notation for the wavefunction coefficient arising from a tree graph with $n$ vertices: $\psi_{(n)}$.\footnote{For sufficiently many vertices, there are graphs that are topologically inequivalent, making it difficult to give them a uniform notation, but we will deal with those cases as they arise. Similarly, it is difficult to give a uniform notation for graphs involving loops, but we will only encounter isolated examples.} In de Sitter space, one must instead keep track of the individual energies flowing to the boundary. However, since all of our manipulations will involve internal lines in graphs, the relevant combinatorics are the same in de Sitter space. Nevertheless, in what follows we will mostly give explicit formulas for flat-space wavefunction coefficients in a theory with $\phi^n$ interactions, for simplicity. The generalization to more complicated situations does not require any conceptually new ingredients.

\vskip4pt
Both the relation between the wavefunction and its conjugate~\eqref{eq:unitarityflat} and the differences of bulk-to-boundary propagators in~\eqref{eq:cutpropappendix} involve sums and differences between wavefunctions evaluated at positive and negative energies. It is therefore useful to define some notation to simplify these operations. First, we define the ``discontinuity"
\beq
\underset{x_a}{{\rm Disc}} \left[\psi(x_a, x_b, s_{ab})\right] \equiv \psi(x_a, x_b, s_{ab})+\psi^*(-x_a, x_b, s_{ab})  \,,
\label{equ:Disc}
\eeq
where $x_a$, $x_b$ are external energies and $s_{ab}$ are internal energies, with $x_a$ being the subset of the energies that are flipped.  
This function is the same as that defined in~\eqref{eq:discdef}. This operation is useful for taking the cut of wavefunction coefficients. In cases where cut propagators appear, it is useful to define the related operation\footnote{This difference of wavefunction coefficients can be written in terms of~\eqref{equ:Disc} by introducing factors of $i$ as
\beq
\underset{s_{ab}}{\tl{\rm Disc}} \left[\psi(x_a, s_{ab})\right] = -i \underset{s_{ab}}{{\rm Disc}}  \left[i\psi(x_a, s_{ab})\right] ,
\eeq
but it is convenient to introduce this notation to avoid a proliferation of these factors.}
\beq
\underset{s_{ab}}{\tl{\rm Disc}} \left[\psi(x_a, s_{ab})\right] \equiv \psi(x_a, s_{ab})-\psi^*(x_a,-s_{ab})\,,
\label{eq:tldisc}
\eeq
which will typically involve internal energies. As an example, we can write the shifted wavefunction~\eqref{eq:shiftedWF} as
\beq
\tl\psi_3(k_1,k_2,s) = -\frac{1}{2s}\underset{s}{\tl{\rm Disc}} [\psi_3(k_1,k_2,s)]\,.
\eeq
As we will see, these two operations will simplify the identities derived from the cutting rules. With these preliminaries out of the way, we can now state the cutting rules for an arbitrary graph.

\subsection{Cutting Rules}
\label{app:cutrules}

It is simplest to first state the cutting rules for a general graph, and then illustrate them with a few simple examples. We then turn to their proof at tree level in Section~\ref{sec:treeproof}, and comment on the extension to loops in Section~\ref{sec:loops}. Given a Feynman--Witten diagram that computes a tree-level wavefunction coefficient, the procedure is the following:
\begin{itemize}
\item Pick a direction (arbitrarily) to move through the graph associated to $\psi_n$. One can think of this as a partial ordering of the vertices in the graph.
\vspace{-3pt}
\item Consider cuts that separate the graph into two halves.
\vspace{-3pt}
\item  On one side of the cut---for concreteness the left---compute the wavefunction associated to the relevant graph, but use the cut propagator for the cut line.
\vspace{-3pt}
\item  On the other side of the cut, compute the wavefunction at negative energy, again using the  cut propagator
for internal lines and then take its complex conjugate. Alternatively, compute the conjugate wavefunction. 
\vspace{-3pt}
\item  Sum over all cuts of this type. 
%\vspace{-.3cm}
\end{itemize}
At the diagrammatic level, it is useful to think of each cut as partitioning the diagram into two types of vertices that we can color differently:  $\bullet$ for vertices to the left of a cut and $\circ$ for vertices to the right of the cut. For the $\bullet$ vertices, we compute the wavefunction coefficients as normal, while we flip the energies of the $\circ$ vertices and take the complex conjugate at the end. Any internal lines that are cut should be replaced with the cut propagator.

\vskip4pt
The claim is that the sum over all cuts (including the trivial cuts completely to the left or right of the graph) will vanish. This implies the following schematic cutting rule identity 
\begin{tcolorbox}[colframe=white,arc=0pt,colback=greyish2]
%\vspace{-6pt}
\beq
\psi_n(X)+ \psi_n^*(-X) = -\sum_{\rm cuts} \psi_n\,,
\label{eq:corrcutting}
\eeq
\end{tcolorbox}
\noindent
where $X$ is a multi-index that accounts for all the external energy arguments of the wavefunction coefficient, $\psi_n$. We have separated the trivial cuts out and put them on the left-hand side, so that
the sum on the right runs over all nontrivial cuts of the diagram---that is, those that pass through at least one internal line. In practice, the cuts of the diagrams appearing on the right-hand side are computed using the $\tl{\rm Disc}$ operation~\eqref{eq:tldisc}, while the left-hand side is precisely the discontinuity~\eqref{equ:Disc}.

\subsection{Examples}
This is still somewhat abstract, so it is useful to illustrate the procedure on a number of simple examples, which will make the proof more intuitive.

\subsubsection*{One-site graph}
The simplest possible example is provided by a one-site graph without any internal lines. There are still two 
``trivial" cuts, which we can think of as completely to the right or left of the lone vertex. Pictorially, we have
\beq
\raisebox{-13pt}{
\begin{tikzpicture}[line width=1. pt, scale=2]
\draw[fill=black] (0,0) circle (.03cm);
\node[scale=1] at (0,-.15) {$x_1$};
\draw[red3, line width=1.5pt,opacity=.85] (.2,-.25) -- (0.2,0.25);
\end{tikzpicture}
}\ \ +\ \ 
\raisebox{-14pt}{
\begin{tikzpicture}[line width=1. pt, scale=2]
\draw[fill=black] (0,0) circle (.03cm);
\draw[fill=white] (0,0) circle (.03cm);
\node[scale=1] at (0,-.15) {$-x_1$};
\draw[red3, line width=1.5pt,opacity=.85] (-.2,-.25) -- (-0.2,0.25);
\end{tikzpicture}
} = \,0\,,
\nonumber
\eeq
which as an equation reads $\psi_{(1)}(x_1)+\psi_{(1)}^*(-x_1) = 0$. In the flat-space context, we have $\psi_{(1)}(x_1) = 1/x_1$, so it is straightforward to verify that this relation is satisfied. In de Sitter space, one simple example is provided by the contact interaction~\eqref{eq:cphi33pt}. Other examples can be found in~\cite{Goodhew:2020hob}.

\subsubsection*{Two-site graph}
Next, we consider a two-site graph with a single internal line.
The corresponding cutting rule was used extensively in a variety of contexts in the main text, but
 for completeness we briefly review it here. The cutting rule  pictorially reads
\beq
\raisebox{-16pt}{
\begin{tikzpicture}[line width=1. pt, scale=2]
\draw[fill=black] (0,0) -- (1,0);
\draw[fill=black] (0,0) circle (.03cm);
%\draw[fill=lightgray] (1,0.21) circle (.2cm);
\draw[fill=black] (1,0) circle (.03cm);
\node[scale=1] at (0,-.15) {$x_1$};
\node[scale=1] at (1,-.15) {$x_2$};
\node[scale=1] at (0.5,.12) {$s_{12}$};
\draw[red3, line width=1.5pt,opacity=.85] (1.25,-.3) -- (1.25,0.3);
\end{tikzpicture}
}+
\raisebox{-16pt}{
\begin{tikzpicture}[line width=1. pt, scale=2]
\draw[fill=black] (0,0) -- (1,0);
\draw[fill=black] (0,0) circle (.03cm);
\draw[fill=white] (0,0) circle (.03cm);
%\draw[fill=lightgray] (1,0.21) circle (.2cm);
\draw[fill=black] (1,0) circle (.03cm);
\draw[fill=white] (1,0) circle (.03cm);
\node[scale=1] at (0,-.15) {$-x_1$};
\node[scale=1] at (1,-.15) {$-x_2$};
\node[scale=1] at (0.5,.12) {$s_{12}$};

\draw[red3, line width=1.5pt,opacity=.85] (-.25,-.3) -- (-0.25,0.3);
\end{tikzpicture}
}+
\raisebox{-16pt}{
\begin{tikzpicture}[line width=1. pt, scale=2]

\draw[fill=black,dashed] (0,0) -- (1,0);
\draw[fill=black] (0,0) circle (.03cm);
%\draw[fill=lightgray] (1,0.21) circle (.2cm);
\draw[fill=black] (1,0) circle (.03cm);
\draw[fill=white] (1,0) circle (.03cm);
\node[scale=1] at (0,-.15) {$x_1$};
\node[scale=1] at (1,-.15) {$-x_2$};
\draw[red3, line width=1.5pt,opacity=.85] (.5,-.3) -- (.5,0.3);
\end{tikzpicture}
} = 0\,,
\label{eq:app2sitecutpic}
\nonumber
\eeq
which translates into a formula for $\psi_{(2)}$ in terms of $\psi_{(1)}$:
\beq
\underset{x_1,x_2}{{\rm Disc}}[\psi_{(2)}(x_1,x_2,s_{12})]= - P(s_{12})\, \underset{s_{12}}{\tl{\rm Disc}}[\psi_{(1)}(x_1+s_{12})]\, \underset{s_{12}}{\tl{\rm Disc}}[\psi_{(1)}(-x_2+s_{12})]\,.
\label{eq:2sitecutformula}
\eeq
In flat space, explicit expressions for the relevant wavefunction coefficients are easy to obtain for a theory with polynomial interactions:
\beq
\begin{aligned}
\psi_{(1)}(x_1)  &= \frac{1}{x_1}\,,\\
\psi_{(2)}(x_1,x_2,s_{12})  &= \frac{1}{(x_1+x_2)(x_1+s_{12})(x_2+s_{12})}\,,
\end{aligned}
\label{eq:1and2sitegraphs}
\eeq
along with $P(k) = 1/(2k)$.
Using these expressions, it is straightforward to verify~\eqref{eq:2sitecutformula}.

\subsubsection*{Three-site graph}

As a more complex example, we can consider the three-site graph, which introduces a couple novelties compared to the two-site graph. In particular, there are now multiple internal cuts that will appear in the sum over cuts. Another interesting aspect is that there are now two inequivalent ways of cutting the diagram, leading to different identities, which are both satisfied.
The truncated three-site Feynman graph takes the form
\beq
\psi_{(3)} = 
\raisebox{-13pt}{
\begin{tikzpicture}[line width=1. pt, scale=2]
\draw[fill=black] (0,0) -- (1.2,0);
\draw[fill=black] (0,0) circle (.03cm);
%\draw[fill=lightgray] (1,0.21) circle (.2cm);
\draw[fill=black] (.6,0) circle (.03cm);
\draw[fill=black] (1.2,0) circle (.03cm);
\node[scale=1] at (0,-.15) {$x_1$};
\node[scale=1] at (.3,.12) {$s_{12}$};
\node[scale=1] at (.6,-.15) {$x_2$};
\node[scale=1] at (.9,.12) {$s_{23}$};
\node[scale=1] at (1.2,-.15) {$x_3$};
\end{tikzpicture}
},
\label{eq:3sitechain}
\eeq
where we have labeled both the energies associated to the vertices and the internal lines.

\vskip4pt 
We can assign a bulk time integral expression to~\eqref{eq:3sitechain}, which will involve two bulk-to-bulk propagators that must be integrated over.  
These integrals were evaluated in~\cite{Arkani-Hamed:2017fdk}. One of the benefits of cutting rules is that we don't actually need to perform these integrations in order to obtain a formula for part of the answer in terms of lower-point objects.

\vskip4pt 
Following the recipe described above, we sum up the following cuts of the three-site graph: 
%%
%%%%%%%%%%%%%%%%%%%%%%%%%%%%%%%%%%%%%%%%%
\beq
%\begin{aligned}
%\hspace{0.55cm} 
\raisebox{-15pt}{
\begin{tikzpicture}[line width=1. pt, scale=2]
\draw[fill=black] (0,0) -- (1.2,0);
\draw[fill=black] (0,0) circle (.03cm);
%\draw[fill=lightgray] (1,0.21) circle (.2cm);
\draw[fill=black] (.6,0) circle (.03cm);
\draw[fill=black] (1.2,0) circle (.03cm);
\node[scale=1] at (0,-.15) {$x_1$};
\node[scale=1] at (.6,-.15) {$x_2$};
\node[scale=1] at (1.2,-.15) {$x_3$};
\draw[red3, line width=1.5pt,opacity=.85] (1.4,-.3) -- (1.4,0.3);
\end{tikzpicture}
}
%\hspace{2pt}+\hspace{2pt}
+
\raisebox{-15pt}{
\begin{tikzpicture}[line width=1. pt, scale=2]
\draw[fill=black] (0,0) -- (1.2,0);
\draw[fill=black] (0,0) circle (.03cm);
\draw[fill=white] (0,0) circle (.03cm);
\draw[fill=black] (.6,0) circle (.03cm);
\draw[fill=white] (.6,0) circle (.03cm);
\draw[fill=black] (1.2,0) circle (.03cm);
\draw[fill=white] (1.2,0) circle (.03cm);
\node[scale=1] at (0.025,-.15) {$-x_1$};
\node[scale=1] at (.6,-.15) {$-x_2$};
\node[scale=1] at (1.2,-.15) {$-x_3$};
\draw[red3, line width=1.5pt,opacity=.85] (-.2,-.3) -- (-0.2,0.3);
\end{tikzpicture}
} 
\hspace{1pt}+\hspace{1pt}
\raisebox{-15pt}{
\begin{tikzpicture}[line width=1. pt, scale=2]
\draw[fill=black,dashed] (0,0) -- (.6,0);
\draw[fill=black] (.6,0) -- (1.2,0);
\draw[fill=black] (0,0) circle (.03cm);
%\draw[fill=lightgray] (1,0.21) circle (.2cm);
\draw[fill=black] (.6,0) circle (.03cm);
\draw[fill=white] (.6,0) circle (.03cm);
\draw[fill=black] (1.2,0) circle (.03cm);
\draw[fill=white] (1.2,0) circle (.03cm);
\node[scale=1] at (0,-.15) {$x_1$};
\node[scale=1] at (.6,-.15) {$-x_2$};
\node[scale=1] at (1.2,-.15) {$-x_3$};
\draw[red3, line width=1.5pt,opacity=.85] (.3,-.3) -- (0.3,0.3);
\end{tikzpicture}
}
%\hspace{1pt}+\hspace{1pt}
+
\raisebox{-15pt}{
\begin{tikzpicture}[line width=1. pt, scale=2]
\draw[fill=black] (0,0) -- (.6,0);
\draw[fill=black,dashed] (.6,0) -- (1.2,0);
\draw[fill=black] (0,0) circle (.03cm);
%\draw[fill=lightgray] (1,0.21) circle (.2cm);
\draw[fill=black] (.6,0) circle (.03cm);
\draw[fill=black] (1.2,0) circle (.03cm);
\draw[fill=white] (1.2,0) circle (.03cm);
\node[scale=1] at (0,-.15) {$x_1$};
\node[scale=1] at (.6,-.15) {$x_2$};
\node[scale=1] at (1.2,-.15) {$-x_3$};
\draw[red3, line width=1.5pt,opacity=.85] (.9,-.3) -- (.9,0.3);
\end{tikzpicture}
}
%\end{aligned}
=0\,. \nonumber
\eeq
%%%%%%%%%%%%%%%%%%%%%%%%%%%%%%%%%%%%%%%%%
%
%
We see that this case is fairly similar to the two-site graph. In particular, there are again two terms where the cuts do not pass through any of the internal lines of the graph, which produce the left-hand side of~\eqref{eq:corrcutting}.
In addition, there are  two terms with cuts through internal lines. We should replace these internal lines with cut propagators, which will shift the energies of the one and two-site graphs that they connect to. Finally, we flip the energies of nodes that appear to the right of a cut. Translating this picture into a formula, we obtain the following identity satisfied by the three-site graph
\beq
\begin{aligned}
\underset{x_1,x_2,x_3}{{\rm Disc}}[\psi_{(3)}(x_1,x_2,x_3)]= &- P(s_{12})\,\underset{s_{12}}{\tl{\rm Disc}}[\psi_{(1)}(x_1+s_{12})]\,\underset{s_{12}}{\tl{\rm Disc}}[{\psi}_{(2)}(-x_2+s_{12},-x_3,s_{23})]\\
&-P(s_{23})\,\underset{s_{23}}{\tl{\rm Disc}}[\psi_{(2)}(x_1,x_2+s_{23},s_{12})]\,\underset{s_{23}}{\tl{\rm Disc}}[\widetilde{\psi}_{(1)}(-x_3+s_{23})]\,.
\end{aligned}
\label{eq:3siteID1}
\eeq
This expression relates part of the three-site graph wavefunction to shifted versions of one and two-site wavefunctions. 
Given the explicit expressions for the one, two, and three-site graphs in flat space, it is straightforward to verify that this identity is indeed satisfied.\footnote{In addition to~\eqref{eq:1and2sitegraphs}, this requires the following formula for  the three-site graph
%\beq
%\begin{aligned}
%\psi^{(3)}(x_1,x_2,x_3)= &\frac{1}{x_1+x_2+x_3}\frac{1}{(x_1+x_2+s_{23})(s_{23}+x_3)(x_1+s_{12})(s_{12}+x_2+s_{23})}\\
%&+\frac{1}{x_1+x_2+x_3}\frac{1}{(s_{12}+x_2+x_3)(x_1+s_{12})(s_{12}+x_2+s_{23})(s_{23}+x_3)}\,.
%\end{aligned}
%\eeq
\beq
\psi^{(3)}(x_1,x_2,x_3)= \frac{1}{x_1+x_2+x_3}\frac{1}{(x_1+s_{12})(x_3+s_{23})(s_{12}+x_2+s_{23})}\bigg[\frac{1}{x_1+x_2+s_{23}}+\frac{1}{x_2+x_3+s_{12}}\bigg]\,.
\eeq
}

\vskip4pt
In the way that we have cut the three-site graph, we have taken one of the vertices that has a single internal line emanating from it as the ``rightmost" vertex. This leads to the set of cuts presented above. However, we could just as well have taken the middle vertex, which connects to two internal lines, as the rightmost vertex, in which case we would draw the graph as:
\beq
\psi_{(3)}=
\raisebox{-29pt}{
\begin{tikzpicture}[line width=1. pt, scale=2]
\draw[fill=black] (0,0) -- (150:.6);
\draw[fill=black] (0,0) -- (210:.6);
\draw[fill=black] (150:.6) circle (.03cm);
\draw[fill=black] (210:.6) circle (.03cm);
\draw[fill=black] (0,0) circle (.03cm);
\node[scale=1] at (0,-.15) {$x_2$};
\node[scale=1] at (-.52,.45) {$x_1$};
\node[scale=1] at (-.52,-.45) {$x_3$};
\end{tikzpicture}
}\,.
\eeq
Of course, this graph and~\eqref{eq:3sitechain} are topologically equivalent, so it would seem that this is merely a cosmetic choice. However, from the perspective of the cutting rules, this presentation of the graph is essentially a different choice of partial ordering on the vertices of the graph, and so leads us to cut the graph in a different way. This feature first appears for the three-site graph and, interestingly, implies that there are two distinct identities satisfied by the wavefunction. 

\vskip4pt
In this case, the set of cuts that we should sum over is
%
%%%%%%%%%%%%%%%%%%%%%%%%%%%%%%%%%%%%%%%%%
\beq
\raisebox{-30pt}{
\begin{tikzpicture}[line width=1. pt, scale=2]
\draw[fill=black] (0,0) -- (150:.6);
\draw[fill=black] (0,0) -- (210:.6);
\draw[fill=black] (150:.6) circle (.03cm);
\draw[fill=black] (210:.6) circle (.03cm);
\draw[fill=black] (0,0) circle (.03cm);
\node[scale=1] at (0,-.15) {$x_2$};
\node[scale=1] at (-.52,.45) {$x_1$};
\node[scale=1] at (-.52,-.45) {$x_3$};
\draw[red3, line width=1.5pt,opacity=.85] (.2,-.5) -- (.2,0.5);
\end{tikzpicture}
}
~+~
\raisebox{-30pt}{
\begin{tikzpicture}[line width=1. pt, scale=2]
\draw[fill=black] (0,0) -- (150:.6);
\draw[fill=black] (0,0) -- (210:.6);
%\draw[fill=black] (0,0) -- (150:.6) circle (.03cm);
\draw[fill=white] (150:.6) circle (.03cm);
%\draw[fill=black] (0,0) -- (210:.6) circle (.03cm);
\draw[fill=white] (210:.6) circle (.03cm);
%\draw[fill=black] (0,0) circle (.03cm);
\draw[fill=white] (0,0) circle (.03cm);
\node[scale=1] at (0,-.15) {$-x_2$};
\node[scale=1] at (-.52,.45) {$-x_1$};
\node[scale=1] at (-.52,-.45) {$-x_3$};
\draw[red3, line width=1.5pt,opacity=.85] (-.74,-.5) -- (-.74,0.5);
\end{tikzpicture}
}
~+~
\raisebox{-30pt}{
\begin{tikzpicture}[line width=1. pt, scale=2]
\draw[fill=black,dashed] (0,0) -- (150:.6);
\draw[fill=black] (0,0) -- (210:.6);
\draw[fill=black] (150:.6) circle (.03cm);
%\draw[fill=black] (0,0) -- (210:.6) circle (.03cm);
\draw[fill=white] (210:.6) circle (.03cm);
%\draw[fill=black] (0,0) circle (.03cm);
\draw[fill=white] (0,0) circle (.03cm);
\node[scale=1] at (0,-.15) {$-x_2$};
\node[scale=1] at (-.52,.45) {$x_1$};
\node[scale=1] at (-.52,-.45) {$-x_3$};
\draw[red3, line width=1.5pt,opacity=.85] (-.27,0) -- (-.27,0.5);
\end{tikzpicture}
}
~+~
\raisebox{-30pt}{
\begin{tikzpicture}[line width=1. pt, scale=2]
\draw[fill=black] (0,0) -- (150:.6);
\draw[fill=black,dashed] (0,0) -- (210:.6);
\draw[fill=white] (150:.6) circle (.03cm);
\draw[fill=black] (210:.6) circle (.03cm);
\draw[fill=white] (0,0) circle (.03cm);
\node[scale=1] at (0,-.15) {$-x_2$};
\node[scale=1] at (-.52,.45) {$-x_1$};
\node[scale=1] at (-.52,-.45) {$x_3$};
\draw[red3, line width=1.5pt,opacity=.85] (-.27,-.5) -- (-.27,0);
\end{tikzpicture}
}
~+~
\raisebox{-30pt}{
\begin{tikzpicture}[line width=1. pt, scale=2]
\draw[fill=black,dashed] (0,0) -- (150:.6);
\draw[fill=black,dashed] (0,0) -- (210:.6);
\draw[fill=black] (150:.6) circle (.03cm);
\draw[fill=black]  (210:.6) circle (.03cm);
\draw[fill=white] (0,0) circle (.03cm);
\node[scale=1] at (0,-.15) {$-x_2$};
\node[scale=1] at (-.52,.45) {$x_1$};
\node[scale=1] at (-.52,-.45) {$x_3$};
\draw[red3, line width=1.5pt,opacity=.85] (-.27,-.5) -- (-.27,.5);
\end{tikzpicture}
}
=0\,.
\label{eq:2nd3sitecuts}
\nonumber
\eeq
%%%%%%%%%%%%%%%%%%%%%%%%%%%%%%%%%%%%%%%%%
%
The main difference with the presentation~\eqref{eq:3sitechain} is that now one of the cuts passes through two lines, separating the graph into three one-site graphs, all with shifted energies. In particular, the $x_2$ node is shifted with respect to {\it both} the $s_{12}$ and $s_{23}$ internal energies. We again translate this picture into a formula as
\beq
\begin{aligned}
\underset{x_1,x_2,x_3}{{\rm Disc}}[\psi_{(3)}(x_1,x_2,x_3)] = &- P(s_{12})\,\underset{s_{12}}{\tl{\rm Disc}}[\psi_{(1)}(x_1+s_{12})]\,\underset{s_{12}}{\tl{\rm Disc}}[\psi_{(2)}(-x_2+s_{12},-x_3,s_{23})]\\
&- P(s_{23})\,\underset{s_{23}}{\tl{\rm Disc}}[\psi_{(2)}(-x_1,-x_2+s_{23},s_{12})]\,\underset{s_{23}}{\tl{\rm Disc}}[\psi_{(1)}(x_3+s_{23})]\\
&-P(s_{12})\,P(s_{23})\, 
\underset{s_{12}}{\tl{\rm Disc}}[\psi_{(1)}(x_1+s_{12})]\underset{s_{23}}{\tl{\rm Disc}}[\psi_{(1)}(x_3+s_{23})]  \\
&\hspace{2.35cm}\times \underset{s_{12}}{\tl{\rm Disc}}\big[\underset{s_{23}}{\tl{\rm Disc}}[\psi_{(1)}(-x_2+s_{12}+s_{23})]\big]
\,.
\end{aligned}
\label{eq:3siteID2}
\eeq
Notice that the vertex connected to two cut lines involves a double shifting of the associated wavefunction coefficient, which for concreteness, is given by
\beq
\begin{aligned}
\underset{s_{a}}{\tl{\rm Disc}}\big[\underset{s_{b}}{\tl{\rm Disc}}[\psi_{(1)}(x+s_a+s_b)]\big] = ~ &
\psi_{(1)}(x-s_a-s_b)- \psi^*_{(1)}(x+s_{a}-s_{b})\\
&-\psi^*_{(1)}(x-s_{b}+s_{a})+\psi_{(1)}(x+s_{a}+s_{b})\,.
\end{aligned}
\label{eq:ddiscexp}
\eeq
It is again straightforward to verify that the identity~\eqref{eq:3siteID2} is satisfied in flat space, by using the explicit expressions for the various wavefunction coefficients.

\vskip4pt
The fact that the cutting rules generate true identities irrespective of the presentation of the graph suggests that there is some invariant notion of cutting the wavefunctional coefficients, which then gets presented in various ways depending on how one chooses to order the graph, in this case leading either to~\eqref{eq:3siteID1} or~\eqref{eq:3siteID2}. It is tempting to speculate that this might have a natural interpretation in terms of the cosmological polytopes that compute these wavefunction coefficients~\cite{Arkani-Hamed:2017fdk}.

\vskip4pt
It is not difficult to extend the above discussion to general tree graphs. Once one has chosen an ordering of the various vertices in the graph, it is just a matter of summing over the cuts, which can then be translated into an identity satisfied by the wavefunction as above.

\subsection{Tree-Level Proof}
\label{sec:treeproof}
We now want to describe how one proves these identities, at least for tree-level wavefunction coefficients. As we describe in the next section, the fact that the cutting rules hold at the level of the time integrand allows us to derive cutting identities satisfied by loops from the tree cutting rules.

\vskip4pt
At their core, the cutting rules follow from the identity~\eqref{eq:ABCid}, which relates the three relevant propagators: the bulk-to-bulk, anti-bulk-to-bulk and cut propagators. At the level of the time integrand, a wavefunction coefficient is a product of bulk-to-bulk propagators dressed with external mode functions. The essential idea is that we are combining this integrand with the analogous integrand where the bulk-to-bulk propagator has been replaced with the anti-bulk-to-bulk propagator. This is (up to a sign) the way we have defined the conjugate wavefunction.
Therefore, we are combining the wavefunction and the conjugate wavefunction in such a way that we can factor out all the information about external lines and apply the identity~\eqref{eq:ABCid} repeatedly.\footnote{A related way to think about the cutting rules is that 
given three objects related by
\beq
A_j - B_j = C_j\,,
\label{eq:ABCid2}
\eeq
the following identity is true~\cite{Bourjaily:2020wvq}
\beq
A_1\cdots A_n - B_1\cdots B_n = C_1 B_2\cdots B_n+A_1 C_2 B_3\cdots B_n+\cdots + A_1\cdots A_{n-1} C_n\,.
\label{eq:propid}
\eeq
In our context, $A={\cal G}$, $B = -\bar{\cal G}$, and $C = \widetilde {\cal G}$ and the cutting rules systematize the application of the
identity~\eqref{eq:propid}, giving a relation between the wavefunction and conjugate wavefunction.} 
The rest is just book-keeping---we need to ensure that the final answer can be written back in terms of the original wavefunction at analytically continued energies. To do this, we use the relation~\eqref{eq:unitarityflat}:
\beq
\bar\psi_n(k_1,\cdots, k_n) = \psi^*_n(-k_1,\cdots, -k_n)\,,
\eeq
which is true in a  unitary theory, where the analytic continuation to negative energies corresponds to a clockwise rotation in the complex plane. Two things are worth emphasizing: first,   the cutting rules are true already at the level of the integrand (similar to the Cutkosky-like rules derived for amplitudes in~\cite{Bourjaily:2020wvq}), and,  second,  they do not rely on unitarity per se---there are cutting identities satisfied by any theory that involve both the wavefunction and conjugate wavefunction. These identities depend only on the structure of the propagators, and can therefore be thought of as a consequence of locality.
The additional input of unitarity is the fact that only in a unitary theory can the relations be written in a way that only involves the wavefunction. (This is also true in the case of scattering amplitudes.)

\vskip4pt
In order to prove the cutting formula~\eqref{eq:corrcutting} at tree level we proceed via induction. In order to streamline the argument, it is useful to introduce some diagrammatic notation. First, we denote the sum of all nontrivial cuts of a diagram (those that intersect at least one internal line) as
\beq
\sum_{\rm cuts} \psi \,= 
\raisebox{-15.5pt}{
\begin{tikzpicture}[line width=1. pt, scale=2]
\draw[fill=lightgray] (0,0) circle (.15cm);
\draw[fill=white] (.15,0) circle (.03cm);
\draw[fill=black] (-.15,0) circle (.03cm);
%\draw[fill=white] (.2,0) circle (.03cm)
\draw[red3, line width=1.5pt,opacity=.85] (0,-.3) -- (0,0.3);
\end{tikzpicture} 
}\,,
\label{eq:cutpicture}
\eeq
where the shaded circle represents an arbitrary tree-level graph.
For example, this would correspond to the sum of diagrams with dashed lines in the above identities. In this diagrammatic notation, we can write the cutting rule~\eqref{eq:corrcutting} as
\beq
\raisebox{-6.5pt}{
\begin{tikzpicture}[line width=1. pt, scale=2]
\draw[fill=lightgray] (0,0) circle (.15cm);
\draw[fill=black] (.15,0) circle (.03cm);
\draw[fill=black] (-.15,0) circle (.03cm);
\end{tikzpicture} 
}
+
\raisebox{-6.5pt}{
\begin{tikzpicture}[line width=1. pt, scale=2]
\draw[fill=lightgray] (0,0) circle (.15cm);
\draw[fill=white] (.15,0) circle (.03cm);
\draw[fill=white] (-.15,0) circle (.03cm);
\draw[line width=1pt] (-.15,.2) -- (.15,.2);
\end{tikzpicture} 
}\,
=-
\raisebox{-15.5pt}{
\begin{tikzpicture}[line width=1. pt, scale=2]
\draw[fill=lightgray] (0,0) circle (.15cm);
\draw[fill=white] (.15,0) circle (.03cm);
\draw[fill=black] (-.15,0) circle (.03cm);
\draw[red3, line width=1.5pt,opacity=.85] (0,-.3) -- (0,0.3);
\end{tikzpicture} 
}\,,
\label{eq:blobcuttingrule}
\eeq
which relates the sum of cuts of a diagram to the wavefunction and its conjugate, as defined above.

\vskip4pt
As we have explained earlier, the bulk-to-bulk and anti-bulk-to-bulk propagators play an essential role, so it is useful to introduce the following graphical notation for them: 
\begin{align}
{\cal G}(k; t_1,t_2) &= 
\raisebox{2pt}{
\begin{tikzpicture}[line width=1. pt, scale=2]
\draw[green3] (0,0) -- (1,0);
\end{tikzpicture} 
}\\
\bar{\cal G}(k; t_1,t_2)  &= 
\raisebox{2pt}{
\begin{tikzpicture}[line width=1. pt, scale=2]
\draw[blue3,line width=1.6pt] (0,0) -- (1,0);
\end{tikzpicture} 
}\,.
\end{align}
In this pictorial form, the identity~\eqref{eq:ABCid} is simply
\beq
\raisebox{2pt}{
\begin{tikzpicture}[line width=1. pt, scale=2]
\draw[green3] (0,0) -- (1,0);
\end{tikzpicture} 
}
~+~
\raisebox{2pt}{
\begin{tikzpicture}[line width=1. pt, scale=2]
\draw[blue3,line width=1.6pt] (0,0) -- (1,0);
\end{tikzpicture} 
}
~-~
\raisebox{-9.5pt}{
\begin{tikzpicture}[line width=1. pt, scale=2]

\draw[fill=black,dashed] (0,0) -- (1,0);
\draw[red3, line width=1.5pt,opacity=.85] (.5,-.2) -- (.5,0.2);
\end{tikzpicture}
}=0\,.
\label{eq:3propidapp}
\eeq
Having introduced this notation, we are ready to proceed with the proof. As mentioned before, the argument is inductive. We are going to take advantage of the fact that any tree graph can be built iteratively by successively adding internal lines. 
Consider adding a single additional internal line that connects some general graph to an additional vertex. Take this vertex to be the farthest ``right" vertex and assume that the graph it connects to satisfies the cutting formula~\eqref{eq:corrcutting}.  We want to show that the graph obtained by adding the extra line then also satisfies~\eqref{eq:corrcutting}.

\vskip4pt
We begin by considering the sum of internal cuts of a general graph, which is denoted pictorially by~\eqref{eq:cutpicture}. At the integrand level, we then multiply by the left-hand side of~\eqref{eq:3propidapp}, and imagine connecting the propagators to a white vertex (but evaluated for {\it positive} energy). Since we are multiplying by zero, we get
\beq
\raisebox{-15.5pt}{
\begin{tikzpicture}[line width=1. pt, scale=2]
\draw[fill=lightgray] (0,0) circle (.15cm);
\draw[fill=black] (-.15,0) circle (.03cm);
\draw[red3, line width=1.5pt,opacity=.85] (0,-.3) -- (0,0.3);
\draw[green3] (.15,0) -- (1,0);
\draw[fill=white] (1,0) circle (.03cm);
\draw[fill=white] (.15,0) circle (.03cm);
\end{tikzpicture} 
}
~+~
\raisebox{-15.5pt}{
\begin{tikzpicture}[line width=1. pt, scale=2]
\draw[fill=lightgray] (0,0) circle (.15cm);
\draw[fill=black] (-.15,0) circle (.03cm);
\draw[red3, line width=1.5pt,opacity=.85] (0,-.3) -- (0,0.3);
\draw[blue3,line width=1.6pt] (.15,0) -- (1,0);
\draw[fill=white] (1,0) circle (.03cm);
\draw[fill=white] (.15,0) circle (.03cm);
\end{tikzpicture} 
}
~-~
\raisebox{-15.5pt}{
\begin{tikzpicture}[line width=1. pt, scale=2]
\draw[fill=lightgray] (0,0) circle (.15cm);
\draw[fill=black] (-.15,0) circle (.03cm);
\draw[red3, line width=1.5pt,opacity=.85] (0,-.3) -- (0,0.3);
\draw[fill=black,dashed] (.15,0) -- (1,0);
\draw[fill=white] (1,0) circle (.03cm);
\draw[fill=white] (.15,0) circle (.03cm);
\draw[red3, line width=1.5pt,opacity=.85] (.575,-.2) -- (.575,0.2);
\end{tikzpicture}
}
= 0\,.
\label{eq:appform1}
\eeq
The goal is to arrange this into the appropriate cutting formula for the new graph with this additional line. First, we notice that the middle term (with the blue internal line) corresponds to almost all of the cuts of this graph. In particular, it is a sum over all cuts except for the one where we cut the additional line, because we are instructed to use the anti-bulk-to-bulk propagator for lines to the right of cuts. We then want to arrange the other two terms in a form that produces the last cut, along with the trivial cuts to the right and left of the wavefunction. We can accomplish this by using the cutting rule for the grey blob~\eqref{eq:blobcuttingrule} to write the following diagrammatic identities
\begin{align}
\raisebox{-15.5pt}{
\begin{tikzpicture}[line width=1. pt, scale=2]
\draw[fill=lightgray] (0,0) circle (.15cm);
\draw[fill=black] (-.15,0) circle (.03cm);
\draw[red3, line width=1.5pt,opacity=.85] (0,-.3) -- (0,0.3);
\draw[green3] (.15,0) -- (1,0);
\draw[fill=white] (1,0) circle (.03cm);
\draw[fill=white] (.15,0) circle (.03cm);
\end{tikzpicture} 
}
~&=~-
\raisebox{-6.5pt}{
\begin{tikzpicture}[line width=1. pt, scale=2]
\draw[fill=lightgray] (0,0) circle (.15cm);
\draw[fill=black] (-.15,0) circle (.03cm);
\draw[green3] (.15,0) -- (1,0);
\draw[fill=white] (1,0) circle (.03cm);
\draw[fill=black] (.15,0) circle (.03cm);
\end{tikzpicture} 
}
~-~
\raisebox{-6.5pt}{
\begin{tikzpicture}[line width=1. pt, scale=2]
\draw[line width=1pt] (-.15,.2) -- (.15,.2);
\draw[fill=lightgray] (0,0) circle (.15cm);
\draw[fill=white] (-.15,0) circle (.03cm);
\draw[green3] (.15,0) -- (1,0);
\draw[fill=white] (1,0) circle (.03cm);
\draw[fill=white] (.15,0) circle (.03cm);
\end{tikzpicture}
}  \,, \\
\raisebox{-15.5pt}{
\begin{tikzpicture}[line width=1. pt, scale=2]
\draw[fill=lightgray] (0,0) circle (.15cm);
\draw[fill=black] (-.15,0) circle (.03cm);
\draw[red3, line width=1.5pt,opacity=.85] (0,-.3) -- (0,0.3);
\draw[fill=black,dashed] (.15,0) -- (1,0);
\draw[fill=white] (1,0) circle (.03cm);
\draw[fill=white] (.15,0) circle (.03cm);
\draw[red3, line width=1.5pt,opacity=.85] (.575,-.2) -- (.575,0.2);
\end{tikzpicture}
}
~&=~-
\raisebox{-9pt}{
\begin{tikzpicture}[line width=1. pt, scale=2]
\draw[fill=lightgray] (0,0) circle (.15cm);
\draw[fill=black] (-.15,0) circle (.03cm);
\draw[fill=black,dashed] (.15,0) -- (1,0);
\draw[fill=white] (1,0) circle (.03cm);
\draw[fill=black] (.15,0) circle (.03cm);
\draw[red3, line width=1.5pt,opacity=.85] (.575,-.2) -- (.575,0.2);
\end{tikzpicture}
}
~-~
\raisebox{-9pt}{
\begin{tikzpicture}[line width=1. pt, scale=2]
\draw[line width=1pt] (-.15,.2) -- (.15,.2);
\draw[fill=lightgray] (0,0) circle (.15cm);
\draw[fill=white] (-.15,0) circle (.03cm);
\draw[fill=black,dashed] (.2,0) -- (1,0);
\draw[fill=white] (1,0) circle (.03cm);
\draw[fill=white] (.15,0) circle (.03cm);
\draw[red3, line width=1.5pt,opacity=.85] (.575,-.2) -- (.575,0.2);
\end{tikzpicture}
}\,,
\label{eq:Bidapp}
\end{align}
and then use the identity~\eqref{eq:3propidapp} in the last term of~\eqref{eq:Bidapp} to write
\beq
\raisebox{-9pt}{
\begin{tikzpicture}[line width=1. pt, scale=2]
\draw[line width=1pt] (-.15,.2) -- (.15,.2);
\draw[fill=lightgray] (0,0) circle (.15cm);
\draw[fill=white] (-.15,0) circle (.03cm);
\draw[fill=black,dashed] (.15,0) -- (1,0);
\draw[fill=white] (1,0) circle (.03cm);
\draw[fill=white] (.15,0) circle (.03cm);
\draw[red3, line width=1.5pt,opacity=.85] (.575,-.2) -- (.575,0.2);
\end{tikzpicture}
}
~=~
\raisebox{-5.5pt}{
\begin{tikzpicture}[line width=1. pt, scale=2]
\draw[line width=1pt] (-.15,.2) -- (.15,.2);
\draw[fill=lightgray] (0,0) circle (.15cm);
\draw[fill=white] (-.15,0) circle (.03cm);
\draw[green3] (.15,0) -- (1,0);
\draw[fill=white] (1,0) circle (.03cm);
\draw[fill=white] (.15,0) circle (.03cm);
\end{tikzpicture}
}
~+~
\raisebox{-5.5pt}{
\begin{tikzpicture}[line width=1. pt, scale=2]
\draw[line width=1pt] (-.15,.2) -- (.15,.2);
\draw[fill=lightgray] (0,0) circle (.15cm);
\draw[fill=white] (-.15,0) circle (.03cm);
\draw[blue3,line width=1.6pt] (.15,0) -- (1,0);
\draw[fill=white] (1,0) circle (.03cm);
\draw[fill=white] (.15,0) circle (.03cm);
\end{tikzpicture}
}\,.
\eeq
Putting all of this together, various terms cancel and we can write~\eqref{eq:appform1} as
\beq
-
\raisebox{-6pt}{
\begin{tikzpicture}[line width=1. pt, scale=2]
\draw[fill=lightgray] (0,0) circle (.15cm);
\draw[fill=black] (-.15,0) circle (.03cm);
\draw[green3] (.15,0) -- (1,0);
\draw[fill=white] (1,0) circle (.03cm);
\draw[fill=black] (.15,0) circle (.03cm);
\end{tikzpicture} 
}
~+~
\raisebox{-6pt}{
\begin{tikzpicture}[line width=1. pt, scale=2]
\draw[line width=1pt] (-.15,.2) -- (.15,.2);
\draw[fill=lightgray] (0,0) circle (.15cm);
\draw[fill=white] (-.15,0) circle (.03cm);
\draw[blue3,line width=1.6pt] (.15,0) -- (1,0);
\draw[fill=white] (1,0) circle (.03cm);
\draw[fill=white] (.15,0) circle (.03cm);
\end{tikzpicture}
}
~+~
\raisebox{-15.5pt}{
\begin{tikzpicture}[line width=1. pt, scale=2]
\draw[fill=lightgray] (0,0) circle (.15cm);
\draw[fill=black] (-.15,0) circle (.03cm);
\draw[red3, line width=1.5pt,opacity=.85] (0,-.3) -- (0,0.3);
\draw[blue3,line width=1.6pt] (.15,0) -- (1,0);
\draw[fill=white] (1,0) circle (.03cm);
\draw[fill=white] (.15,0) circle (.03cm);
\end{tikzpicture} 
}
~+~
\raisebox{-9.5pt}{
\begin{tikzpicture}[line width=1. pt, scale=2]
\draw[fill=lightgray] (0,0) circle (.15cm);
\draw[fill=black] (-.15,0) circle (.03cm);
\draw[fill=black,dashed] (.15,0) -- (1,0);
\draw[fill=white] (1,0) circle (.03cm);
\draw[fill=black] (.15,0) circle (.03cm);
\draw[red3, line width=1.5pt,opacity=.85] (.575,-.2) -- (.575,0.2);
\end{tikzpicture}
} = 0\,.
\label{eq:cutappform3}
\eeq
This is nearly the desired formula: Notice that in the first term the rightmost vertex is $\circ$ instead of $\bullet$. However, these two vertices only differ by a minus sign ($i$ versus $-i$), so we can write
\beq
-
\raisebox{-6pt}{
\begin{tikzpicture}[line width=1. pt, scale=2]
\draw[fill=lightgray] (0,0) circle (.15cm);
\draw[fill=black] (-.15,0) circle (.03cm);
\draw[green3] (.15,0) -- (1,0);
\draw[fill=white] (1,0) circle (.03cm);
\draw[fill=black] (.15,0) circle (.03cm);
\end{tikzpicture} 
}
~=~
\raisebox{-6pt}{
\begin{tikzpicture}[line width=1. pt, scale=2]
\draw[fill=lightgray] (0,0) circle (.15cm);
\draw[fill=black] (-.15,0) circle (.03cm);
\draw[green3] (.15,0) -- (1,0);
\draw[fill=black] (1,0) circle (.03cm);
\draw[fill=black] (.15,0) circle (.03cm);
\end{tikzpicture} 
}\,.
\eeq
In addition, the last two terms of~\eqref{eq:cutappform3} combine into the full set of internal cuts of the tree diagram, which implies
\beq
\psi+ \bar\psi = -\sum_{\rm cuts} \psi\,,
\eeq
as desired. In order to complete the argument, we need one additional step, which is to note that the fact that the sum of cuts of a graph vanishes for any ordering of its vertices right to left, where we sum over all cuts according to the prescription described in Section~\ref{app:cutrules}. This can be seen starting from the cutting formula~\eqref{eq:corrcutting} and on the right hand side (in the sum over internal cuts) using the identity ${\cal G}+\bar{\cal G} = \tl{\cal G}$ to transform any subgraph built from all bulk-to-bulk propagators into one using all anti-bulk-to-bulk propagators (or vice versa). This will necessarily also introduce a number of new cuts to the graph, effectively ordering the vertices in a different way. As an example, we can transform the two ways of cutting the three-site graph into each other via this procedure. With this information in hand, we can effectively build any tree graph we want iteratively using~\eqref{eq:appform1}, where the added line attaches to the rightmost vertex. Then, since we have proven the cutting rules explicitly in the base two-site case, they then follow for a general graph.

\subsubsection*{Single cuts}
The cutting rules that we have described are useful because they apply to general graphs. However, there is another identity satisfied by the bulk-to-bulk propagator that is useful in some cases. Specifically, the bulk-to-bulk propagator satisfies
\beq
\bar{\cal G}(k; t_1,t_2)  = -{\cal G}(-k; t_1,t_2)\,.
\label{eq:ktmkbb}
\eeq
We can use this relation to make the cut propagator appear in expressions involving the wavefunction, without having to introduce the conjugate wavefunction. For example, the two-site graph satisfies
\beq
\psi_{(2)}(x_1,x_2,s_{12})-\psi_{(2)}(x_1,x_2,-s_{12}) = - P(s_{12})\, \underset{s_{12}}{\tl{\rm Disc}}[\psi_{(1)}(x_1+s_{12})]\, \underset{s_{12}}{\tl{\rm Disc}}[\psi_{(1)}(-x_2+s_{12})]\,,
\eeq
which follows from combining~\eqref{eq:ktmkbb} and~\eqref{eq:ABCid}. The extension to more complicated tree graphs is immediate, we just take the difference between a wavefunction coefficient and itself with one of its internal energies flipped, which has the interpretation of cutting this internal line.
Identities of this kind are somewhat similar to single discontinuities in the context of scattering amplitudes. A drawback of these cuts is that they do not readily extend to loop level because they require shifting internal energies, which are integrated over in a loop. Nevertheless, they may be useful in helping to constrain correlator ansatzes by demanding that all of their individual cuts reproduce the correct lower-point objects. See~\cite{Goodhew:2021oqg} for some examples along these lines.

\subsection{From Trees to Loops}
\label{sec:loops}
So far, we have focused on cuts of tree-level graphs. In the context of cosmological correlators, these relations already provide useful structural information. However, in the case of flat-space scattering amplitudes, it is really at loop level that unitarity methods are the most powerful. It stands to reason that cuts will also prove to be useful at loop level for correlators. Fortunately, much of our previous discussion can be adapted to consider loop diagrams.

\vskip4pt
A useful feature of the cutting formulas proved in the previous section is that they hold at the level of the {\it integrand}, 
because they are essentially identities satisfied by products of bulk-to-bulk propagators. From this perspective, there is not so much difference between a tree diagram and a loop---to turn a tree integrand into a loop integrand, we just identify two of the vertices.\footnote{Interestingly, an avatar of this relation between trees and loops persists even after doing the time integrals, as was pointed out in~\cite{Benincasa:2018ssx}.} Consequently, we can re-interpret our tree-level cutting formulas as ones involving loop diagrams.

\vskip4pt
There are two important subtleties when carrying out this procedure. The first is that we have to be careful about vertex factors---for the most part we will focus on theories with polynomial interactions, so that this only requires tracking factors of $i$ which depend on the number of actual vertex insertions. For more complicated interactions involving nontrivial vertex factors, it should not be difficult to factorize the problem by combining vertex factors for the loop diagram and then casting the time integral as a tree integrand with identifications. However, we will not pursue this systematically. The second important subtlety relates to loop momenta---in a loop diagram there are unconstrained momenta running in the loops that have to be integrated over. After cutting loop diagrams, it is important to still integrate over the undetermined loop momenta. We will not attempt to actually perform the loop momentum integrals, so the cutting formulas that we present should be read as relations between the time integrals involved in loop calculations.

\vskip4pt
Since loops are not our main focus, we will just illustrate the procedure for a couple simple examples, and leave a systematic exploration to the future. See also~\cite{Melville:2021lst} for a detailed discussion of cutting loops.

\subsubsection*{One-site loop}
The simplest loop diagram involves a single vertex, where the loop connects the vertex to itself: 
\beq
\hat\psi_{(1)}^\text{1-loop} = ~
\raisebox{-24pt}{
\begin{tikzpicture}[line width=1. pt, scale=2]
\draw[fill=none] (0,0) circle (.2cm);
\draw[fill=black] (0,-.2) circle (.03cm);
\node[scale=1] at (0,-.35) {$x$};
\node[scale=1] at (0,.3) {$y$};
\end{tikzpicture} 
}\,,
\eeq
where we include a hat on $\hat\psi_{(1)}^\text{1-loop}$ to emphasize that this is not the full wavefunctional coefficient, but is rather the loop integrand.
Since this diagram
only has a single exchange, we can adapt the tree-level two-site cutting rule~\eqref{eq:app2sitecutpic} to this graph by identifying the two vertices, which leads to the identity
\beq
\raisebox{-24pt}{
\begin{tikzpicture}[line width=1. pt, scale=2]
\draw[fill=none] (0,0) circle (.2cm);
\draw[fill=black] (0,-.2) circle (.03cm);
\node[scale=1] at (0,-.35) {$x$};
\end{tikzpicture} 
}
~-~
\raisebox{-24pt}{
\begin{tikzpicture}[line width=1. pt, scale=2]
\draw[fill=none] (0,0) circle (.2cm);
\draw[fill=white] (0,-.2) circle (.03cm);
\node[scale=1] at (0,-.35) {$-x$};
\end{tikzpicture} 
}
~-~
\raisebox{-24pt}{
\begin{tikzpicture}[line width=1. pt, scale=2]
\draw[fill=none,dashed] (0,0) circle (.2cm);
\draw[fill=black] (0,-.2) circle (.03cm);
\draw[red3, line width=1.5pt,opacity=.85] (0,.05) -- (0,0.35);
\node[scale=1] at (0,-.35) {$x$};
\end{tikzpicture} 
} ~ = 0\,,
\label{eq:1siteloopcutrule}
\eeq
where the minus signs come from the  factors of $i$ and $-i$ that are different for a loop and for the identified tree diagram.  We can translate this graphical identity into the following formula:
\beq
\underset{x}{\tl{\rm Disc}}[\hat\psi_{(1)}^\text{1-loop}] - \lim_{y'\to y}P(y)\,\underset{y}{\tl{\rm Disc}}\big[\underset{y'}{\tl{\rm Disc}}[\hat\psi_{(1)}(x+y+y')]\big] =0\,,
\eeq
where it is important to split the energies $y$ and $y'$ when taking the double discontinuity in order to preserve the correct identification of the two vertices in the tree graph~\eqref{eq:app2sitecutpic}.

\vskip 4pt
We can then check that this relation is indeed satisfied by the one-site loop in flat space, where the explicit wavefunction coefficient is given by
\beq
\hat\psi_{(1)}^\text{1-loop}  = \frac{1}{x(x+2y)}\,.
\label{eq:1siteloopWF}
\eeq
In this case, the shifted one-site wavefunctional coming from the cut is 
\beq
\lim_{y'\to y}P(y)\,\underset{y}{\tl{\rm Disc}}\big[\underset{y'}{\tl{\rm Disc}}[\hat\psi_{(1)}(x+y+y')] = \frac{1}{2y}\left(\frac{1}{x-2y}-\frac{2}{x}+\frac{1}{x+2y}
\right) ,
\label{eq:shifted1siteloopwf}
\eeq
where $\hat\psi_{(1)}(x+y+y') = 1/(x+y+y')$.
Combining~\eqref{eq:shifted1siteloopwf} with~\eqref{eq:1siteloopWF}, we can verify that~\eqref{eq:1siteloopcutrule} is satisfied. As mentioned earlier, it is worth bearing in mind that this is a relation satisfied by the loop integrand, where the integration over the undetermined loop momentum must still be carried out.

\subsubsection*{Two-site loop}
The next-simplest example of a loop consists of two vertices with two internal lines: 
\beq
\hat\psi_{(2)}^\text{1-loop} = \raisebox{-22pt}{
\begin{tikzpicture}[line width=1. pt, scale=2]
\draw[fill=none] (0,0) circle (.2cm);
\draw[fill=black] (.2,0) circle (.03cm);
\draw[fill=black] (-.2,0) circle (.03cm);
\node[scale=1] at (-.35,0) {$x_1$};
\node[scale=1] at (.375,0) {$x_2$};
\node[scale=1] at (0,.3) {$y_a$};
\node[scale=1] at (0,-.3) {$y_b$};
\end{tikzpicture} 
}\,.
\eeq
Since this has the same number of propagators as the three-site graph example at tree level, we can transform that cutting rule into one for this loop graph by identifying vertices $1$ and $3$. Interestingly, as in the case of the three-site tree, there are two ways of cutting this loop diagram. The first comes from the identity~\eqref{eq:3siteID1} and reads pictorially:
\begin{equation}
\raisebox{-10pt}{
\begin{tikzpicture}[line width=1. pt, scale=2]
\draw[fill=none] (0,0) circle (.2cm);
\draw[fill=black] (.2,0) circle (.03cm);
\draw[fill=black] (-.2,0) circle (.03cm);
\end{tikzpicture} 
}
~-~
\raisebox{-10pt}{
\begin{tikzpicture}[line width=1. pt, scale=2]
\draw[fill=none] (0,0) circle (.2cm);
\draw[fill=white] (.2,0) circle (.03cm);
\draw[fill=white] (-.2,0) circle (.03cm);
\end{tikzpicture} 
}
~-~
\raisebox{-19pt}{
\begin{tikzpicture}[line width=1. pt, scale=2]
\draw (.2,0) arc (0:180:.2cm);
\draw[dashed]  (.2,0) arc (0:-180:.2cm);
\draw[fill=white] (.2,0) circle (.03cm);
\draw[fill=white] (-.2,0) circle (.03cm);
\draw[red3, line width=1.5pt,opacity=.85] (0,-.35) -- (0,-.05);
\end{tikzpicture} 
}
~+~
\raisebox{-10pt}{
\begin{tikzpicture}[line width=1. pt, scale=2]
\draw[dashed] (.2,0) arc (0:180:.2cm);
\draw  (.2,0) arc (0:-180:.2cm);
\draw[fill=black] (.2,0) circle (.03cm);
\draw[fill=black] (-.2,0) circle (.03cm);
\draw[red3, line width=1.5pt,opacity=.85] (0,.05) -- (0,0.35);
\end{tikzpicture} 
}
~=0\,,
\end{equation}
which, translated to an equation, is
\beq
\begin{aligned}
\underset{x_1,x_2}{\tl{\rm Disc}} [\hat\psi_{(2)}^\text{1-loop}] &-\lim_{y_b'\rightarrow y_b}P(y_b)\underset{y_b}{\tl{\rm Disc}}\big[ \underset{y_b}{\tl{\rm Disc}} [\hat\psi_{(2)}(-x_1+y_b,-x_2+y_b',y_a)]\big] \\&+\lim_{y_a'\rightarrow y_a}P(y_a)\underset{y_a}{\tl{\rm Disc}}\big[ \underset{y_a}{\tl{\rm Disc}} [\hat\psi_{(2)}(x_1+y_a,x_2+y_a',y_b)]\big] = 0\,,
\end{aligned}
\label{eq:cutloop2}
\eeq
where again the nested ${\tl{\rm Disc}}$ operations have to be treated with some care. 
For example:
\beq
\begin{aligned}
&\lim_{y_a'\rightarrow y_a}\underset{y_a}{\tl{\rm Disc}}\big[ \underset{y_a}{\tl{\rm Disc}} [\hat\psi_{(2)}(x_1+y_a,x_2+y_a',y_b)] = \hat\psi_{(2)}(x_1-y_a,x_2-y_a,y_b)  \\
&\hspace{0.5cm}-\hat\psi_{(2)}(x_1-y_a,x_2+y_a,y_b) -\hat\psi_{(2)}(x_1+y_a,x_2-y_a,y_b)+\hat\psi_{(2)}(x_1+y_a,x_2+y_a,y_b)\,.
\end{aligned}
\eeq
To check that the identity (\ref{eq:cutloop2}) is actually satisfied in flat space, we use the explicit expression for the two-site loop~\cite{Arkani-Hamed:2017fdk}
\beq
\hat\psi_{(2)}^\text{1-loop}  = \frac{1}{(x_1+x_2)(x_1+y_a+y_b)(x_2+y_a+y_b)}\left[\frac{1}{x_1+x_2+2y_a}+\frac{1}{x_1+x_2+2y_b}\right] ,
\eeq
as well as~\eqref{eq:1and2sitegraphs} for the two-site graph at tree level.

\vskip4pt
We can also imagine manipulating the identity~\eqref{eq:2nd3sitecuts} at the level of the integrand into a loop identity. This different way of cutting the three-site graph leads to the following diagrammatic identity: 
\begin{equation}
\raisebox{-10pt}{
\begin{tikzpicture}[line width=1. pt, scale=2]
\draw[fill=none] (0,0) circle (.2cm);
\draw[fill=black] (.2,0) circle (.03cm);
\draw[fill=black] (-.2,0) circle (.03cm);
\end{tikzpicture} 
}
~-~
\raisebox{-10pt}{
\begin{tikzpicture}[line width=1. pt, scale=2]
\draw[fill=none] (0,0) circle (.2cm);
\draw[fill=white] (.2,0) circle (.03cm);
\draw[fill=white] (-.2,0) circle (.03cm);
\end{tikzpicture} 
}
~-~
\raisebox{-10pt}{
\begin{tikzpicture}[line width=1. pt, scale=2]
\draw[dashed] (.2,0) arc (0:180:.2cm);
\draw  (.2,0) arc (0:-180:.2cm);
\draw[fill=white] (.2,0) circle (.03cm);
\draw[fill=white] (-.2,0) circle (.03cm);
\draw[red3, line width=1.5pt,opacity=.85] (0,.05) -- (0,0.35);
\end{tikzpicture} 
}
~-~
\raisebox{-19pt}{
\begin{tikzpicture}[line width=1. pt, scale=2]
\draw (.2,0) arc (0:180:.2cm);
\draw[dashed]  (.2,0) arc (0:-180:.2cm);
\draw[fill=white] (.2,0) circle (.03cm);
\draw[fill=white] (-.2,0) circle (.03cm);
\draw[red3, line width=1.5pt,opacity=.85] (0,-.35) -- (0,-.05);
\end{tikzpicture} 
}
~+~
\raisebox{-16pt}{
\begin{tikzpicture}[line width=1. pt, scale=2]
\draw[fill=none,dashed] (0,0) circle (.2cm);
\draw[fill=white] (.2,0) circle (.03cm);
\draw[fill=black] (-.2,0) circle (.03cm);
\draw[red3, line width=1.5pt,opacity=.85] (0,-.3) -- (0,0.3);
\end{tikzpicture} 
}
~=0\,,
\end{equation}
which it is also straightforward to verify given the explicit formulas. 

\vskip4pt
As mentioned before, the fact that there are multiple different ways of cutting diagrams suggests that there is a more invariant formulation of the problem. Additionally, it would be interesting to connect these ideas to the formulation of Steinmann relations found in~\cite{Benincasa:2020aoj}.

\newpage
%=======================================
\section{Complex Deformations and Residues}
\label{app:Residues}
%=======================================

In Section~\ref{sec:gluing}, we constructed wavefunction coefficients recursively from lower-point correlators and scattering information by deforming their kinematic arguments into the complex plane.
These deformed wavefunction coefficients have a number of poles at complex momenta, whose residues can be summed to obtain the physical (i.e.~the undeformed) wavefunction. In this appendix, we find a general formula for these residues---at any pole order---in terms of the Laurent expansion of the wavefunction coefficient around its singularities.

\subsection{Deformation and Singularities}

Imagine we want to construct a wavefunction coefficient, $\psi(K_a)$, with an arbitrary number of internal and external lines, whose energies we collectively call $K_a  \equiv \{k_a,|\vec{k}_{a_1}+\ldots+\vec{k}_{a_m}|\}$, where $\vec{k}_{a_1}+\ldots+\vec{k}_{a_m}$ denotes a basis for the exchanged momenta in a given process.
We will assume that $\psi(K_a)$ is a rational function of these kinematic arguments, and that it has a pole of order $n$ when a certain linear combination of the energy variables, which we will denote by $p_j(K_a)$, goes to zero, where $j$ is an index that labels the different poles. We can then write 
\beq
\psi(K_a) = \frac{1}{p_j(K_a)^n}\, F_j(K_a) \, ,
\eeq
where $F_j$ is the correlator after factoring out the pole. We could  similarly factor out all of the poles, but we just show the $j$th pole for illustration.
As explained in Section~\ref{sec:gluing}, we can isolate a subset of the poles by performing 
 a complex linear shift in the energy variables: 
\beq 
\label{shift}
K_a \,\mapsto\, \hat K_a(z) = \{ k_a+c_a z, |\vec{k}_{a_1}+\ldots+\vec{k}_{a_m}| +d_{a_1\ldots a_m}z  \} \, .
\eeq
After doing this shift, we can write $p_j$ as
\beq
p_j(K_a)\, \mapsto\, p_j(\hat K_a) = A (z-z_j) \, ,
\label{eq:complexpolep}
\eeq
where $z_j\neq 0$ is a linear combination of energy variables and $A$ a numerical factor that we will ignore in the following, as it can always be absorbed in the definition of $F_j$. Equation~\eqref{eq:complexpolep} merely expresses that the zero of $p_j(\hat K_a)$ has been moved into the complex plane by the deformation, and now sits at $z = z_j$.

\vskip4pt
As we described in Section~\ref{sec:gluing}, we can write a formal formula for the original undeformed wavefunction coefficient as
\beq
\psi(0) = \frac{1}{2\pi i}\underset{{\cal C}}{\oint}\rd z\, \frac{\psi(z)}{z}\,,
\eeq
by integrating around a small circle ${\cal C}$ centered around $z=0$. By assumption, all of the singularities of $\psi(z)$ are poles, so we can write a residue formula for the original wavefunction by deforming the contour
\beq
\psi(0) = -\sum_{j} \underset{z=z_j}{\rm Res} \left(\frac{\psi(z)}{z}\right) ,
\label{eq:appresidueform}
\eeq
where we are assuming the absence of a pole at infinity.

\subsection{Residues and Laurent Coefficients}
In order to use the formula~\eqref{eq:appresidueform}, we need to know the residues of $\psi(z)/z$ at each pole $z_j$.
Using the general residue formula, we can write
\begin{align}
\underset{z=z_j}{\rm Res} \left(\frac{\psi(z)}{z}\right)&= \frac{1}{(n-1)!} \lim_{z\to z_j} \frac{\rd^{n-1}}{\rd z^{n-1}} \left[ (z-z_j)^n \frac{\psi(z)}{z} \right]  \nonumber \\
	&= \frac{1}{(n-1)!} \lim_{z\to z_j} \frac{\rd^{n-1}}{\rd z^{n-1}} \left[ \frac{1}{z} F_j (z) \right]  \nonumber \\
 &= - \sum_{l=1}^{n} \frac{1}{(-z_j)^l} \frac{1}{(n-l)!} \lim_{z\to z_j} \frac{\rd^{n-l}}{\rd z^{n-l}} F_j (z) \, .  \label{genres}
\end{align}
We want to relate this expression to the Laurent series of $\psi$. To do this, we note that we can write the singular terms of the Laurent series of the deformed function $\psi(z)$ around $z=z_j$ as
\begin{align}
	\left. \lim_{z\to z_j} \psi(z) \vphantom{\frac{1}{x}} \right|_{\text{sing}} &= \sum_{m=-n}^{-1} (z-z_j)^m \,\, \frac{1}{2\pi i} \oint_{{\cal C}_j} \rd w \frac{\psi(w)}{(w-z_j)^{m+1}}  \nonumber \\
	&= \sum_{m=-n}^{-1} (z-z_j)^m \,\, \text{Res} \left[ \frac{\psi(w)}{(w-z_j)^{m+1}}  , w=z_j \right] \nonumber \\
	&= \sum_{m=-n}^{-1} (z-z_j)^m \,\, \frac{1}{(n+m)!} \lim_{w\to z_j} \frac{\rd^{n+m}}{\rd w^{n+m}} \left[ (w-z_j)^n \psi(w) \right] \nonumber \\
	&= \sum_{m=-n}^{-1} (z-z_j)^m \,\, \frac{1}{(n+m)!} \lim_{w\to z_j} \frac{\rd^{n+m}}{\rd w^{n+m}} F_j(w) \, ,
\end{align}
where ${\cal C}_j$ is a small circle centered around $w=z_j$. Relabelling the summation index $m\to -l$ yields
\beq \label{eq:Laurent}
\left. \lim_{z\to z_j} \psi(z) \vphantom{\frac{1}{x}} \right|_{\text{sing}} = \sum_{l=1}^{n} \frac{R^{(l)}(K_a)}{(z-z_j)^l}\,, \quad\text{with}\quad R^{(l)}(K_a) \equiv \frac{1}{(n-l)!} \lim_{w\to z_j} \frac{\rd^{n-l}}{\rd w^{n-l}} F_j(w) \, .
\eeq
Comparing this with \eqref{genres}, we obtain a general expression relating the residue to the coefficients in the Laurent expansion of $\psi(z)$:
\begin{tcolorbox}[colframe=white,arc=0pt,colback=greyish2]
%\vspace{-0.4cm}
	\beq \label{genfor}
	\underset{z=z_j}{\rm Res} \left(\frac{\psi(z)}{z}\right)= - \sum_{l=1}^{n} \frac{R^{(l)}(K_a)}{(-z_j)^l} \, .
	\eeq
\end{tcolorbox}
\noindent
In Section~\ref{sec:gluing}, we use our knowledge of $\psi(K_a)$ around its singularities---i.e.~knowledge of the Laurent series~\eqref{eq:Laurent}---to compute the residues and reconstruct the full function. In the case of partial energy singularities, the cutting equation allows us to compute all the singular terms in~\eqref{eq:Laurent}. For the total energy singularity, we only know the most singular term of the series, which is given by the high-energy limit of the corresponding scattering amplitude. The subleading total energy singularities can however be fixed by consistency conditions, as they are not in general independent from the partial energy singularities.

\subsubsection*{Shifting a single variable}
The relevant formulas simplify if
the complex deformation affects only one of the energy variables, e.g.~$k_1 \mapsto k_1 + z$.
In this case, we have 
\beq
p_j(K_a) = k_1- \theta_j(K_a)  \quad \implies\quad z_j = \theta_j-k_1 \, ,
\eeq
and 
\begin{align}
	\left. \lim_{k_1 \to \theta_j} \psi(K_a) \vphantom{\frac{1}{x}} \right|_{\text{sing}} &= \sum_{m=-n}^{-1} (k_1-\theta_j)^m \,\, \frac{1}{(n+m)!} \lim_{w\to \theta_j} \frac{\rd^{n+m}}{\rd w^{n+m}} \left[ (w-\theta_j)^n \psi(w,k_2,\ldots) \right] \nonumber \\
	&= \sum_{m=-n}^{-1} (-z_j)^m \,\, \frac{1}{(n+m)!} \lim_{w\to \theta_j} \frac{\rd^{n+m}}{\rd w^{n+m}} \psi_j(w,k_2,\ldots) \nonumber \\
	&= \sum_{l=1}^{n} \frac{1}{(-z_j)^l} \frac{1}{(n-l)!} \lim_{w\to \theta_j} \frac{\rd^{n-l}}{\rd w^{n-l}} \psi_j(w,k_2,\ldots) \, ,
\end{align}
which coincides exactly with \eqref{genres}. 
The residue at $z_j=\theta_j-k_1$ then takes the simple form
\begin{tcolorbox}[colframe=white,arc=0pt,colback=greyish2]
%\vspace{-0.4cm}
	\beq \label{eq:Res1var}
		\underset{z=z_j}{\rm Res} \left(\frac{\psi(z)}{z}\right) = - \left. \lim_{k_1\to \theta_j} \psi(K_a) \vphantom{\frac{1}{x}} \right|_{\text{sing}} \, .
	\eeq
\end{tcolorbox}
\noindent
To sum up, when only one of the energy variables is shifted, the residue is just equal to minus the singular terms of the Laurent series of the undeformed wavefunction coefficient.

\subsubsection*{Simple poles}

If all poles are simple poles, as is the case for flat-space correlators, then the formula \eqref{genfor} for each pole simplifies even further and reduces to
\begin{tcolorbox}[colframe=white,arc=0pt,colback=greyish2]
%\vspace{-0.3cm}
	\beq
	\underset{z=z_j}{\rm Res} \left(\frac{\psi(z)}{z}\right)= \frac{1}{z_j} \lim_{z\to z_j} \left[ (z-z_j) \psi (z) \right]  .
	\eeq
\end{tcolorbox}
\noindent
Furthermore, if we shift only one energy variable, then the formula  involves only the unshifted wavefunction coefficient:
\begin{tcolorbox}[colframe=white,arc=0pt,colback=greyish2]
%\vspace{-0.3cm}
	\beq \label{eq:ResSim1}
	\underset{z=z_j}{\rm Res} \left(\frac{\psi(z)}{z}\right) = - \frac{1}{k_1-\theta_j} \lim_{k_1\to \theta_j} \left[ (k_1-\theta_j) \psi (K_a) \right]  .
	\eeq
\end{tcolorbox}
\noindent
This implies that for simple poles, and deforming only one of the energy variables, the residues of the deformed wavefunction are simply given by a limit of the (undeformed) wavefunction coefficient. We typically know expressions for these singular limits of the wavefunction from other principles, like bulk locality.

\newpage
%=======================================
\section{Yang--Mills Correlators}
\label{app:YMapp}
%=======================================

In this appendix, we construct correlators involving non-Abelian currents using the strategies presented in the main text. In particular, we construct the correlators corresponding to non-Abelian Compton scattering, $\psi_{J\vp J\vp}$, and the Yang--Mills four-point function, $\psi_{JJJJ}$. Pure Yang--Mills is Weyl invariant in four dimensions, and we will consider a coupling to a conformally coupled scalar, so that these correlators will be the same in both flat space and de Sitter. This makes these correlation functions particularly simple, and we expect that these techniques can fruitfully be applied to construct higher-multiplicity correlators.

\subsection{Non-Abelian Compton}
We first consider the non-Abelian Compton correlator between two massless spin-1 particles and two conformally coupled scalar. This is quite similar to the Abelian Compton scattering example considered in Section~\ref{sec:gluing}, but with the addition of gluon exchange in the $u$-channel. We perform the same complex deformation as in the Abelian case
\beq
k_1\mapsto k_1+z,
\label{eq:jojodeformation}
\eeq
which now accesses singularities located at
\beq
z_E = - E\,,\quad
z_s = - E_L^{(s)}\,,\quad
z_t =- E_L^{(t)}\,,\quad
z_u =- E_L^{(u)}\,.
\label{eq:jojopoles}
\eeq
The residues of these singularities can be computed from the known three-point amplitudes and correlators
\be
A_{J\vp\vp}&= 2T^A_{ab}(\vec\xi_1\cdot\vec k_2)\,,  \\[6pt]
A_{JJJ} &= f^{ABE}\left[ (\vec{\xi}_1\cdot\vec{\xi}_2)(\vec{\xi}_3\cdot\vec{\alpha}) +2(\vec{\xi}_2\cdot\vec{\xi}_3)(\vec{\xi}_1\cdot\vec{k}_2) -2(\vec{\xi}_1\cdot\vec{\xi}_3)(\vec{\xi}_2\cdot\vec{k}_1) \right]  ,\\
\psi_{J\vp\vp}&=\frac{A_{J\vp^2}}{K}\,,\quad \psi_{JJJ} = \frac{A_{JJJ}}{K} ,
\label{eq:JJJ3pt}
\ee
where, as before $\vec\alpha\equiv \vec k_1-\vec k_2$, and $T^A_{ab}$ and $f^{ABC}$ are the couplings of the non-Abelian vector to scalars and its self-coupling, respectively, which satisfy the relation $[T^A, T_B]_{ab} = f^{ABC}T^C_{ab}$.
We also require the four-point amplitude
\beq
\begin{aligned}
A_{J\vp J\vp}= ~&T^A_{ac}T^B_{cb}\,\frac{4(\epsilon_1\cdot p_2)(\epsilon_3\cdot p_4)}{S}+ T^B_{ac}T^A_{cb}\,\frac{4(\epsilon_1\cdot p_4)(\epsilon_3\cdot p_2)}{T}\\
&+(T^A T^B-T^BT^A)_{ab} \frac{1}{U}\left[-UP_1\left(1+\frac{2S}{U}\right)(\epsilon_1\cdot \epsilon_3) + 4\epsilon_1\circ\epsilon_3\right]\\
&-(T^{A}T^{B}+T^{B}T^{A})_{ab} (\epsilon_1\cdot \epsilon_3)\,,
\end{aligned}
\eeq
where $\epsilon_1\circ\epsilon_3 =  (\epsilon_1\cdot p_2)(\epsilon_3\cdot p_4)-(\epsilon_1\cdot p_4)(\epsilon_3\cdot p_2)$ is defined as in~\eqref{eq:epsiloncirc}.

Using this information, we can construct the residues of the singularities in~\eqref{eq:jojopoles}:
\begin{align}
-\underset{z=z_E}{\rm Res}\left(\frac{\psi(z)}{z}\right) &=  4T^A_{ac}T^B_{cb}\,\frac{(\vec\xi_1\cdot \vec k_2)(\vec\xi_3\cdot \vec k_4)}{(k_{34}-s)(k_{34}+s)}+4T^B_{ac}T^A_{cb}\,\frac{(\vec\xi_1\cdot \vec k_4)(\vec\xi_3\cdot \vec k_2)}{(k_{23}-t)(k_{23}+t)}\,\nonumber \\
 & ~~~~+(T^A T^B-T^BT^A)_{ab} \frac{1}{(k_{24}-u)(k_{24}+u)}\left[\tl{\cal P}^{(u)}_1(\vec\xi_1\cdot \vec\xi_3) + 4\vec\xi_1\circ\vec\xi_3\right]\\\nonumber
&~~~~- (T^{A}T^{B}+T^{B}T^{A})_{ab} (\vec\xi_1\cdot \vec\xi_3)\,,\\[4pt]
-\underset{z=z_s}{\rm Res}\left(\frac{\psi(z)}{z}\right)&=A_{J\varphi \varphi} \cdot \tl \psi_{\varphi J  \varphi} =- T^A_{ac}T^B_{cb}   \frac{4(\vec \xi_1 \cdot \vec k_2)(\vec \xi_3 \cdot \vec k_4)}{ (k_{34}+s) (k_{34}-s)}\, ,\\[4pt]
-\underset{z=z_t}{\rm Res}\left(\frac{\psi(z)}{z}\right)  &= A_{J\varphi \varphi} \cdot \tl \psi_{\varphi J  \varphi} = -T^B_{ac}T^A_{cb} \frac{4(\vec \xi_1 \cdot \vec k_4)(\vec \xi_3 \cdot \vec k_2) }{ (k_{23}+t) (k_{23}-t)}\, ,\\[4pt]
-\underset{z=z_u}{\rm Res}\left(\frac{\psi(z)}{z}\right)&=A_{JJJ} \otimes \tl \psi_{J\varphi   \varphi} = -(T^A T^B-T^BT^A)_{ab} \frac{u^2 \Pi^{(u)}_{1,1}\, \vec\xi_1\cdot\vec\xi_3 + 4\vec\xi_1\circ\vec\xi_3}{ (k_{24}+u)(k_{24}-u)},
\end{align}
where, as in Sections~\ref{sec:gluing} and~\ref{sec:bootstrapping}, the residues do not depend on $k_1$ and we have introduced the deformed polarization sum as in~\eqref{eq:GenLegendre}:
\beq
\tl{\cal P}_1^{(u)}  = \tl\Pi_{1,1}-(-k_{24}+u)E^{(u)}_R\tl\Pi_{1,0}\,.
\label{eq:appptilde}
\eeq
The partial energy residues could also have been obtained by taking limits of the cuts of correlators.
Summing up all the residues according to~\eqref{eq:residue}, and using the identity
\beq
\frac{1}{E(k_{34}+s)(k_{34}-s)}-\frac{1}{(k_{12}+s)(k_{34}+s)(k_{34}-s)} = -\frac{1}{EE_LE_R},
\eeq
along with the analogous expressions in the other channels, we obtain the final expression for the non-Abelian Compton correlator
\begin{tcolorbox}[colframe=white,arc=0pt,colback=greyish2]
\beq
\begin{aligned}
\psi_{J\varphi J\varphi} = ~&-(T^AT^B)_{ab} \frac{4(\vec \xi_1 \cdot \vec k_2)(\vec \xi_3 \cdot \vec k_4)}{EE_LE_R} -(T^BT^A)_{ab} \frac{4(\vec \xi_1 \cdot \vec k_4)(\vec \xi_3 \cdot \vec k_2)}{EE_L^{(t)}E_R^{(t)}} \\
&~~- [T^A,T^B]_{ab}\frac{1}{EE_L^{(u)}E_R^{(u)}} \left[ {\cal P}_1^{(u)} \,\vec \xi_1\cdot\vec \xi_3 + 4\vec\xi_1\circ\vec\xi_3 \right] \\
&~~ -(T^{A}T^{B}+T^{B}T^{A})_{ab} \frac{\vec\xi_1\cdot \vec\xi_3}{E}\, ,
\end{aligned}
\label{eq:NAcompton}
\eeq
\end{tcolorbox}
\noindent
where ${\cal P}_1^{(u)}$ is the polarization sum~\eqref{eq:P1polsum} adapted to the $u$-channel.
This agrees with the correlator obtained by other methods in~\cite{Baumann:2020dch}. It is straightforward to check that this expression is consistent with the cutting rules discussed in Section~\ref{sec:cuttingsec}. In principle, there is the freedom in the above derivation to add terms that are regular in all of the total energy and partial energy limits, but these terms are not necessary to produce the correct answer.

\subsection{Yang--Mills}
Next, we consider the correlator associated to pure Yang--Mills scattering. As in the previous case, we perform the complex deformation~\eqref{eq:jojodeformation}, which can access the singularities~\eqref{eq:jojopoles}. We can compute the partial energy residues from the three-point wavefunction~\eqref{eq:JJJ3pt} which allows us to construct the cut of the correlator (in the $s$-channel for concreteness):
\beq \label{eq:YMscut}
\begin{aligned}
	 \text{Disc}[\psi^{(s)}_{JJJJ}] &= -2s \tl{\psi}_{JJJ}(k_{12}\mp s)\otimes\tl{\psi}_{JJJ}(-k_{34}\mp s) \vphantom{\frac{1}{2}} \\
	& \hspace{-1cm}= - c_s  \, 2s \, \frac{ (\vec{\xi}_1\cdot\vec{\xi}_2)(\vec{\xi}_3\cdot\vec{\xi}_4) \tl{\Pi}_{1,1} +4(\vec{\xi}_1\cdot\vec{\xi}_2)(\vec{\xi}_3\circ\vec{\xi}_4) +4(\vec{\xi}_1\circ\vec{\xi}_2)(\vec{\xi}_3\cdot\vec{\xi}_4) +4\mathcal{Z}^{(s)} }{(k_{12}^2-s^2)(k_{34}^2-s^2) } \, ,
\end{aligned}
\eeq
with $\mathcal{Z}^{(s)} \equiv \big[ (\vec{\xi}_1\cdot \vec{k}_2) \vec{\xi}_2 - (\vec{\xi}_2\cdot \vec{k}_1) \vec{\xi}_1 \big] \cdot \big[ (\vec{\xi}_3\cdot \vec{k}_4) \vec{\xi}_4 - (\vec{\xi}_4\cdot \vec{k}_3) \vec{\xi}_3 \big]$. 
The color factor $c_s \equiv f^{ABE}f^{CDE}$ guarantees Bose symmetry, with $\{A,B,C,D\}$ labeling the external legs with momenta $\{k_1,k_2,k_3,k_4\}$, respectively. We can obtain the residues of the partial energy singularities by taking the $E_L\to 0$ limit of this equation, as was done in Section~\ref{sec:gluing}. Similar formulas hold for the $t$ and $u$-channels after permutation.

\vskip4pt
The residue of the total energy singularity can be obtained from the scattering amplitude:
\begin{align}
iA_{JJJJ} =  \frac{c_s}{S} \bigg[ &(\epsilon_1\cdot\epsilon_2)(\epsilon_3\cdot\epsilon_4) S P_1\left( 1+\frac{2U}{S} \right) -4(\epsilon_1\cdot\epsilon_2)(\epsilon_3\circ\epsilon_4) -4(\epsilon_1\circ\epsilon_2)(\epsilon_3\cdot\epsilon_4) -4Z^{(s)} \bigg] \nonumber \\
& + c_s \Big[ (\epsilon_1\cdot\epsilon_4)(\epsilon_2\cdot\epsilon_3) - (\epsilon_1\cdot\epsilon_3)(\epsilon_2\cdot\epsilon_4) \Big] + \left( \text{$t$ and $u$-channels} \right)  , \label{eq:YMsc} 
\end{align}
with $P_1(x)$ the Legendre polynomial and
\beq \label{eq:Zstrdef}
Z^{(s)} \equiv \big[ (\epsilon_1\cdot p_2) \epsilon_2^{\mu} - (\epsilon_2\cdot p_1) \epsilon_1^{\mu} \big] \cdot \big[ (\epsilon_3\cdot p_4) \epsilon_4^{\mu} - (\epsilon_4\cdot p_3) \epsilon_3^{\mu} \big]\,  .
\eeq
The $t$ and $u$-channel contributions can be obtained by permuting $2\leftrightarrow 4$ and $2\leftrightarrow 3$, respectively. The full amplitude is gauge invariant as a consequence of the Jacobi identity, $c_s+c_t+c_u = 0$.
Using this information we can construct the partial energy singularity residues as:
\be 
-\underset{z=z_s}{\rm Res}\left(\frac{\psi(z)}{z}\right) = c_s \frac{ (\vec{\xi}_1\cdot\vec{\xi}_2)(\vec{\xi}_3\cdot\vec{\xi}_4) \tl{\Pi}_{1,1} +4(\vec{\xi}_1\cdot\vec{\xi}_2)(\vec{\xi}_3\circ\vec{\xi}_4) +4(\vec{\xi}_1\circ\vec{\xi}_2)(\vec{\xi}_3\cdot\vec{\xi}_4) +4\mathcal{Z}^{(s)} }{E_L(k_{34}^2-s^2)} \, .
\ee
The $z_t$ and $z_u$ residues are simply obtained by permutation. The residue at $z_E$ is simply related to the scattering amplitude \eqref{eq:YMsc}:
\begin{align} \label{eq:ResEYM}
	-\underset{z=z_E}{\rm Res}\left(\frac{\psi(z)}{z}\right) = & \, - c_s \frac{ (\vec{\xi}_1\cdot\vec{\xi}_2)(\vec{\xi}_3\cdot\vec{\xi}_4) \tl{{\cal P}}_1^{(s)} +4(\vec{\xi}_1\cdot\vec{\xi}_2)(\vec{\xi}_3\circ\vec{\xi}_4) +4(\vec{\xi}_1\circ\vec{\xi}_2)(\vec{\xi}_3\cdot\vec{\xi}_4) +4\mathcal{Z}^{(s)} }{E(k_{34}^2-s^2)} \nonumber \\
	&   + \frac{c_s}{E} \left[ (\vec{\xi}_1\cdot\vec{\xi}_4)(\vec{\xi}_2\cdot\vec{\xi}_3) - (\vec{\xi}_1\cdot\vec{\xi}_3)(\vec{\xi}_2\cdot\vec{\xi}_4) \right] + \left( \text{$t$ and $u$-channels} \right) , 
\end{align}
where $\tl{{\cal P}}_1^{(s)}$ is the $s$-channel version of \eqref{eq:appptilde}. Adding up the four residues one obtains the following after some algebra
\begin{tcolorbox}[colframe=white,arc=0pt,colback=greyish2]
\vspace{-0.25cm}
	\beq \label{eq:fullYMcorr}
	\begin{aligned}
		\psi_{JJJJ} = & \, c_s \, \frac{ (\vec{\xi}_1\cdot\vec{\xi}_2)(\vec{\xi}_3\cdot\vec{\xi}_4) {\cal P}_1 +4(\vec{\xi}_1\cdot\vec{\xi}_2)(\vec{\xi}_3\circ\vec{\xi}_4) +4(\vec{\xi}_1\circ\vec{\xi}_2)(\vec{\xi}_3\cdot\vec{\xi}_4) +4\mathcal{Z}^{(s)} }{EE_LE_R} \\
		& + \frac{c_s}{E} \left[ (\vec{\xi}_1\cdot\vec{\xi}_4)(\vec{\xi}_2\cdot\vec{\xi}_3) - (\vec{\xi}_1\cdot\vec{\xi}_3)(\vec{\xi}_2\cdot\vec{\xi}_4) \right]  +  \left( \text{$t$ and $u$-channels} \right) .
	\end{aligned} 
	\eeq
\end{tcolorbox}
\noindent
It is straightforward to check both that this correlator satisfies the correct cutting rule, to which only the partial energy singularities contribute, and that it matches the result of an explicit bulk calculation~\cite{Albayrak:2018tam}.

\newpage
%=======================================
\section{Correlators in the EFT of Inflation}
\label{app:EFTapp}
%=======================================

As another application of the tools introduced in the main text, in this appendix we consider correlation functions in the effective field theory of inflation~\cite{Creminelli:2006xe,Cheung:2007st}. Since our goal is not to be comprehensive, but merely illustrate the techniques,  we will focus on the correlators arising from a particular vertex in the EFT. However, we anticipate that this approach will help to streamline more general computations in the EFT of inflation.

\subsection{Goldstone Action}
The effective field theory of inflation views inflation as a process of spontaneous symmetry breaking, where the fact that inflation ends signals the presence of some ``clock" that spontaneously breaks time reparametrization invariance by picking a preferred space-like foliation of the inflationary spacetime.  From this perspective, fluctuations of the inflaton are the Goldstone modes of this symmetry breaking. In general, these perturbations mix with the graviton and form a single multiplet in the infrared. However, we can isolate the dynamics of the Goldstone field by taking the decoupling limit $M_{\rm Pl} \to \infty$, $\dot H\to 0$, while holding the combination $f^4_\pi/H^4 = 2c_s M_{\rm Pl}^2 \lvert \dot H\rvert/H^4$ fixed, where $c_s$ is the speed of sound of the Goldstone fluctuation. This latter scale, $f_\pi$, has the interpretation as the scale of symmetry breaking, where the Goldstone field becomes relevant. In the decoupling limit, the action governing the dynamics of the Goldstone fluctuations at leading order in derivatives and up to cubic order in fields is~\cite{Cheung:2007st,Cheung:2007sv}:
\beq
S = \int\rd^4x\,a^3\, \frac{f_\pi^4}{c_s^3}\left(\frac{1}{2}\dot\pi^2 - \frac{c_s^2}{2a^2}(\vec\nabla\pi)^2+\frac{(1-c_s^2)}{2}\left(1+\frac{2c_3}{3}\right)\dot\pi^3 +\frac{(c_s^2-1)}{2}\dot\pi(\vec\nabla\pi)^2+\cdots
\right),
\label{eq:eftinf}
\eeq
where $c_3$ is a free parameter and $a(t) = e^{Ht}$ is the scale factor of the de Sitter spacetime.
We see that there are two inequivalent cubic couplings---the coefficient of $\dot\pi^3$ can be tuned independently of the quadratic action, while the coefficient of the $\dot\pi(\vec\nabla\pi)^2$ is fixed by the quadratic action.
Correlators involving only the $\dot\pi^3$ interaction fall into the class considered in \S\ref{sec:dotphi3bbreaking}. In this appendix, we instead focus on the correlators arising from the universal interaction $\dot\pi(\vec\nabla\pi)^2$.

\vskip4pt
Our goal is to derive the trispectrum coming from the $\dot\pi(\vec\nabla\pi)^2$ interaction following the logic of transmuting the flat-space exchange correlator. This contribution was computed explicitly in~\cite{Chen:2009bc}. The computation was rather involved, and we will see that we are able to reproduce the answer with substantially less effort.

\subsection{Two and Three-Point Functions}
In order to construct the four-point function, we will need to know its cut, which will require the three-point function as an input. Fortunately, this is relatively straightforward to compute. The power spectrum coming from the quadratic part of the action~\eqref{eq:eftinf} is
\beq
P_\pi(k) = \frac{H^2}{f_\pi^4}\,\frac{1}{2k^3}\,,
\eeq
while the three-point wavefunction coefficient coming from the $\dot\pi(\vec\nabla\pi)^2$ vertex is~\cite{Cheung:2007sv}
\beq
\psi_{\dot\pi(\vec\nabla\pi)^2} = \frac{2f_\pi^4}{H^2}\left( \frac{1}{c_s^2}-1 \right) \left[\frac{6 k_1^2k_2^2k_3^2}{K^3}+  \bigg( -\frac{4}{K}\sum_{i>j} k_i^2k_j^2 +\frac{2}{K^2}\sum_{i\neq j}k_i^2k_j^3 +\frac{1}{2}\sum_i k_i^3  \bigg) \right].
\label{eq:eft3pt}
\eeq
Note that in the way that we are organizing the calculation, we are thinking of~\eqref{eq:eft3pt} as given, remaining agnostic as to how it was obtained, either from a bulk calculation~\cite{Cheung:2007sv} or from symmetry and consistency requirements directly at late times~\cite{Pajer:2020wxk,Jazayeri:2021fvk}.
We now want to use the three-point function~\eqref{eq:eft3pt} to construct the cut of the four-point function.

\subsection{Four-Point Function}
Our strategy to compute the four-point function arising from the $\dot\pi(\vec\nabla\pi)^2$ term will be similar to the approach we took in \S\ref{sec:dotphi3bbreaking}. The cut of the four-point function is
\beq
{\rm Disc}[\psi_4] = \frac{1}{P_\pi(s)} \,\tl\psi_{\dot\pi(\vec\nabla\pi)^2}(k_1,k_2,s) \, \tl\psi_{\dot\pi(\vec\nabla\pi)^2}(k_3,k_4,s)\,,
\label{eq:cuteft4pt}
\eeq
where the shifted wavefunctions appearing in this expression are
\beq
\tl\psi_{\dot\pi(\vec\nabla\pi)^2} (k_1,k_2,s) \equiv  P_\pi(s)\left(\psi_{\dot\pi(\vec\nabla\pi)^2} (k_1,k_2,-s)-\psi_{\dot\pi(\vec\nabla\pi)^2} (k_1,k_2,s)\right) ,
\eeq
and similarly for the right vertex with $\{k_1,k_2\}\mapsto \{k_3,k_4\}$.
Using the shifted three-point wavefunction in flat space, the cut of the four-point function is
\be
{\rm Disc}[\psi_4^{\rm (flat)}] = - \frac{2s}{(k_{12}^2-s^2)(k_{34}^2-s^2)}\,.
\ee
Similarly, the cut of the de Sitter four-point function can be computed from the three-point wavefunction~\eqref{eq:eft3pt}. The result can be related to its flat-space counterpart by a transmutation operator
\beq
{\rm Disc}[\psi_4^{(\rm dS)}] =-\frac{H^2}{f_\pi^4}\left( \frac{1}{c_s^2}-1 \right) {\cal O}_L{\cal O}_R\,{\cal I}_s^{[1]}\,{\rm Disc}[\psi_4^{\rm (flat)}]\,,
\eeq
where we have defined
\beq
{\cal O}_L \equiv k_1^2(s^2-k_1^2){\cal S}_{k_2}^{[1]}+k_2^2(s^2-k_2^2){\cal S}_{k_1}^{[1]}+\frac{1}{2}(s^2-k_1^2-k_2^2){\cal S}_{k_1}^{[1]}{\cal S}_{k_2}^{[1]}\left(\int_{k_{12}}^\infty\rd  k_{12}\right)^2\,,
\eeq
which acts on the left vertex. The analogous operator for the right vertex can be obtained by sending  $\{k_1,k_2\}\mapsto \{k_3,k_4\}$.
These operators have many of the features that we have become accustomed to from other examples; the operators ${\cal O}_{L,R}$ transform the flat-space vertices into the de Sitter boost-breaking $\dot\pi(\vec \nabla\pi)^2$ vertices, and the ${\cal I}_s$ operator transforms the exchanged field into a massless scalar field in de Sitter space.\footnote{One way to understand the various types of terms that appear in ${\cal O}_{L,R}$ comes from looking at the Feynman rule for the $\dot\pi(\vec\nabla\pi)^2$ vertex and inferring what types of operations should be present to generate it from the flat-space vertex. Alternatively, these operators can also be obtained without reference to the bulk by making a systematic ansatz and matching the cut.}

\vskip4pt
Given this operation that transmutes the flat-space cut into the cut of the desired EFT of inflation correlator, we can then act on the full flat-space exchange to obtain the $s$-channel contribution to the correlator:
\begin{tcolorbox}[colframe=white,arc=0pt,colback=greyish2]
\beq
\psi_4^{{\rm (dS)}} = -\frac{H^2}{f_\pi^4}\left( \frac{1}{c_s^2}-1 \right) {\cal O}_L{\cal O}_R\,{\cal I}_s^{[1]} \frac{1}{EE_LE_R}\,.
\eeq
\end{tcolorbox}
\noindent
By construction, this correlation function has the correct cut, and correspondingly has the correct factorization properties on its $E_L$ and $E_R$ singularities. Indeed, one of the nice features of this representation is that it makes the singularity structure of the answer manifest. Note that this expression is for a single permutation of external momenta, corresponding to exchange in the $s$-channel. In order to produce the full answer we should sum over permutations.
This representation of the wavefunction can be checked to agree with an explicit bulk calculation~\cite{Chen:2009bc}.\footnote{Because~\cite{Chen:2009bc} computed the corresponding correlation function in the in-in formalism, the comparison is a bit subtle. First, there are some overall normalization factors involved in going between the wavefunction and correlators. In addition to this, two modifications must be made to the correlator exchange contribution (called $T_{s3}$ in~\cite{Chen:2009bc}) to get the wavefunction: first we must subtract off a disconnected contribution coming from the three-point wavefunction coefficient squared. In addition, the cubic $\dot\pi(\vec\nabla\pi)^2$ term contributes to the fourth-order interaction Hamiltonian. It is therefore necessary to include a contact solution contribution (proportional to $T_{c3}$ in~\cite{Chen:2009bc}).}

\vskip4pt 
We can write the correlator more explicitly by actually performing all the energy integrals, after which we have
\begin{align}
&{\cal O}_L{\cal O}_R\,{\cal I}_s^{[1]} \frac{1}{EE_LE_R} = 
-\tl{\cal O}_L\tl{\cal O}_R\frac{2k_{12}k_{34}+4Es+2s^2}{E^3E_L^2E_R^2}\nonumber \\
&\ \ -\tl{\cal O}_L\left[\frac{(s^2-k_3^2-k_4^2)}{2}\frac{(k_{12}k_{34}+2sE)(k_{34}E+2k_3k_4)+\left[(E+k_{34})E+2k_3k_4\right]s^2}{E^3 E_L^2E_R^2}\right] \nonumber \\
&\ \ -\tl{\cal O}_R\left[\frac{(s^2-k_1^2-k_2^2)}{2}\frac{(k_{12}k_{34}+2sE)(k_{34}E+2k_1k_2)+\left[(E+k_{12})E+2k_1k_2\right]s^2}{E^3 E_L^2E_R^2}\right] \label{eq:simplifiedeftcorr}\\
&\ \ -\frac{(s^2-k_1^2-k_2^2)(s^2-k_3^2-k_4^2)}{4}\bigg[\frac{E_LE_R+sE}{E_L^2E_R^2}\left( \frac{2k_1k_2k_3k_4}{E^3}+\frac{k_{12}k_3k_4+k_{34}k_1k_2}{E^2}+\frac{k_{12}k_{34}}{E}\right) \nonumber \\
&\hspace{5.5cm}+\frac{s^2}{E_L^2 E_R^2}\left(\frac{k_1k_2+k_3k_4}{E}+s+E
\right)\bigg]\,,
\nonumber
\end{align}
where we have defined the operators
\beq
\begin{aligned}
\tl{\cal O}_L &\equiv k_1^2(s^2-k_1^2){\cal S}_{k_2}^{[1]}-k_2^2(s^2-k_2^2){\cal S}_{k_1}^{[1]}\,,\\
\tl{\cal O}_R &\equiv k_3^2(s^2-k_3^2){\cal S}_{k_4}^{[1]}-k_4^2(s^2-k_4^2){\cal S}_{k_3}^{[1]}\,.
\end{aligned}
\eeq
We can see from the expression~\eqref{eq:simplifiedeftcorr} that the correlator has up to $E^{-5}$ total energy singularities and $E_{L,R}^{-3}$ partial energy singularities.

\clearpage
%\setlength{\textheight}{655pt}
%\newpage
\phantomsection
%\enlargethispage{\baselineskip}
%\addtocontents{toc}{\protect\enlargethispage{\baselineskip}}
\addcontentsline{toc}{section}{References}
\bibliographystyle{utphys}
{\linespread{1.075}
\bibliography{Cutting-Refs}
}

\end{document}